\documentclass[12pt]{article}
\pdfoutput=1
\usepackage{amssymb, amsmath,amsfonts}
\usepackage{multirow}
\usepackage{mathrsfs}
\usepackage{array}
\usepackage{cite}
\usepackage{booktabs}
\usepackage{tikz}
\usepackage{pgfplots}
\usepackage{float}
\usepackage{subfigure, graphicx}
\usepackage[pdftex, bookmarks=true,colorlinks,linkcolor=red,urlcolor=blue,citecolor=blue]{hyperref}
\usepackage{cite}
\usepackage{enumerate}
\usepackage{slashed}
\usepackage{tabularx}
\usepackage{enumerate} 

\usepackage{tikz}

\makeatletter\@addtoreset{equation}{section}\makeatother

\def\be{\begin{equation}}
\def\ee{\end{equation}}
\def\bea{\begin{eqnarray}}
\def\eea{\end{eqnarray}}
\newcommand{\comment}[1]{{\bf {\textcolor{blue}{ [#1]}}}}

\def\Dslash{\,\,{\raise.15ex\hbox{/}\mkern-12mu D}}
\def\Dbarslash{\,\,{\raise.15ex\hbox{/}\mkern-12mu {\bar D}}}
\def\delslash{\,\,{\raise.15ex\hbox{/}\mkern-9mu \partial}}
\def\delbarslash{\,\,{\raise.15ex\hbox{/}\mkern-9mu {\bar\partial}}}
\def\pslash{\,\,{\raise.15ex\hbox{/}\mkern-9mu p}}
\def\calDslash{\,\,{\raise.15ex\hbox{/}\mkern-12mu {\cal D}}}

\makeatletter\@addtoreset{equation}{section}\makeatother

\renewcommand{\title}[1]{\vbox{\center\LARGE{#1}}\vspace{5mm}}
\renewcommand{\author}[1]{\vbox{\center#1}\vspace{5mm}}
\newcommand{\address}[1]{\vbox{\center\em#1}}

\def\arXiv#1{\href{http://arxiv.org/abs/#1}{arXiv:#1}}
\def\arXiv#1#2{\href{http://arxiv.org/abs/#1}{arXiv:#1}}

\def\doi#1#2{\href{http://doi.org/#1}{#2}}

\begin{document}

\unitlength = .8mm

\begin{titlepage}
\vspace{.5cm}
 
\begin{center}
\hfill \\
\hfill \\
\vskip 1cm

\title{On AdS$_3$/ICFT$_2$ with\\ a dynamical scalar field located on the brane}
\vskip 0.5cm
{Yan Liu$^{\,a,b}$}\footnote{Email: {\tt yanliu@buaa.edu.cn}}, 
{Hong-Da Lyu$^{\,a,b}$}\footnote{Email: {\tt hongdalyu@buaa.edu.cn}}  and {Chuan-Yi Wang$^{\,a, b}$}\footnote{Email: {\tt by2230109@buaa.edu.cn}}

\address{${}^{a}$Center for Gravitational Physics, Department of Space Science\\ and International Research Institute
of Multidisciplinary Science, \\Beihang University, Beijing 100191, China}

\address{${}^{b}$Peng Huanwu Collaborative Center for Research and Education, \\Beihang University, Beijing 100191, China}

\end{center}
\vskip 1.5cm

We exploit the holographic duality to study the system of a one-dimensional interface contacting two semi-infinite two-dimensional CFTs. Central to our investigation is the introduction of a dynamical scalar field located on the bulk interface brane which breaks the scaling symmetry of the dual interface field theory, along with its consequential backreaction on the system. We define an interface entropy from holographic entanglement entropy, to construct a $g$-function. At zero temperature we construct several illustrative examples and consistently observe that the $g$-theorem is always satisfied. These examples also reveal distinct features of the interface entropy that are intricately linked to the scalar potential profiles. At finite temperature we find that the dynamical scalar field enables the bulk theory to have new configurations which would be infeasible solely with a tension term on the interface brane. 

\vfill

\end{titlepage}

\begingroup 
\hypersetup{linkcolor=black}
\tableofcontents
\endgroup





\section{Introduction}

Interfaces contacting with two distinct systems are ubiquitous in nature. For instance, an impurity or defect within a pristine system can be regarded as an interface bridging two identical systems. Similarly, a quantum dot linked to two interacting quantum wires or a bubble separating the interior and exterior environments are also examples of interfaces. The investigation of such systems 
not only enhances our comprehension of field theories but also illuminates the intricate interplay between disparate domains \cite{Andrei:2018die}. 

We will investigate the system of a one-dimensional interface which can be thought as a quantum dot, contacting two two-dimensional CFTs residing on semi-infinite lines.  When both CFTs are identical, the interface resembles a defect or impurity. However, in general the two CFTs may differ significantly. For instance, one of the CFTs could be trivial, leading to a boundary conformal field theory (BCFT) system. Alternatively, folding the system when the two CFTs are identical also yields a BCFT. 
Naturally, richer physical phenomena are anticipated in the realm of interface conformal field theories (ICFTs), where interfaces dynamically interact with CFTs. This is evidenced by the emergence of novel observables within ICFTs, such as energy transports  \cite{Quella:2006de, Meineri:2019ycm}, i.e. transmission or reflection coefficients quantifying the energy flux across the interface. Additionally, ICFTs exhibit intriguing entanglement structures \cite{peschel, Sakai:2008tt, Calabrese:2009qy}, further deepening our understanding of their intricate dynamics.

In the literature, considerable attention has been devoted to  conformal interfaces that preserves a single copy of the Virasoro algebra. Yet, it is crucial to note that the interfaces may host internal dynamical degrees of freedom, or coupling two CFTs nontrivially, which could break the conformal symmetry of the interface field theory. 
Clearly, this is a wide and interesting field of study, which we expect to contact with the real systems in condensed matter physics.  We are focused on systems involving strongly interacting field theories and general interfaces. In such scenarios, the conventional tools of CFT are not applicable, while  holography could provide invaluable insights into the fundamental physics underlying such systems.

Holographic ICFTs have undergone extensive investigation, ranging from the top-down approaches like intersecting D-branes \cite{Karch:2000gx, Bachas:2001vj, DeWolfe:2001pq}, which elucidate the characteristics of supersymmetric field theories with defects, to bottom-up approaches utilizing holography to explore the properties of interface CFTs. 
One such bottom-up approach involves constructing the holographic model for ICFTs using the Janus solution in the bulk \cite{Bak:2011ga, Bak:2007jm}. In this framework, the interface is conceptualized as a non-local defect that smoothly  dissolves within the gravitational bulk. However, 
solving such systems poses significant challenges, primarily due to the involvement of PDE's. 

A more straightforward approach to studying holographic ICFT 
involves considering a localized interface brane embedded within the gravitational system \cite{Erdmenger:2014xya, Erdmenger:2015xpq, Simidzija:2020ukv, Anous:2022wqh}. 
This method has attracted  significant research attention in recent years.
Various dynamics have been examined in this setups, including the energy transports \cite{Bachas:2020yxv, Baig:2022cnb, Bachas:2022etu} and  the entanglement structures \cite{Anous:2022wqh, Karch:2021qhd, Karch:2022vot}.\footnote{The holographic entanglement entropy has been computed for three-dimensional Janus solutions \cite{Gutperle:2015hcv} and Janus black hole solutions \cite{Nakaguchi:2014eiu}.} Additionally, insights into the island formula for double holography in  holographic BCFT have been provided from the perspective of holographic ICFT \cite{Suzuki:2022xwv, Anous:2022wqh}. 
Moreover, the phases of interfaces in compact CFTs have been investigated in \cite{Bachas:2021fqo}.\footnote{In Euclidean CFTs, a new CFT$_2$ state can be prepared from a CFT$_1$ state through a quenching operation at the interface \cite{Simidzija:2020ukv}. This procedure yields an ICFT in Euclidean spacetime, where the properties of approximate CFT states are explored through AdS/ICFT in \cite{Simidzija:2020ukv}. A similar construction can be applied to CFTs undergoing weak measurement, as discussed in \cite{Sun:2023hlu}.} Other progress in AdS/ICFT can be found in e.g. \cite{Geng:2022slq, Karch:2023evr, Tang:2023chv, Basak:2023bnc, Afrasiar:2023nir, Baig:2023ahz, Shen:2024uja, Azeyanagi:2007qj, Estes:2014hka}.  

In all the aforementioned setups, the additional dynamics on the interface has often been overlooked.  Here, we propose to consider an AdS$_3$/ICFT$_2$ setup with a dynamical interface brane involving a localised scalar field. The inclusion of such additional brane-localized scalar field breaks the conformal invariance of the interface field theory, 
thereby uncovering rich physics of ICFT.
Our study differs from the studies of DCFT in \cite{Erdmenger:2014xya} which is for holographic Kondo effect, where the two CFTs that the defect is contacting were identical and a dynamical brane has been considered  at finite temperature. Instead, we consider a scenario where the dynamical brane is embedded within two distinct AdS spacetimes, i.e. the two CFTs that the defect is contacting could be different. On this interface brane within bulk, we introduce a dynamical scalar field. Our primary objective is to explore the impact of this scalar field on AdS/ICFT. We will study both zero-temperature and finite-temperature configurations. Various new solutions will be constructed in the presence of the scalar field. Moreover, to gain further insights into the interface field theory, we will study the entanglement structure and define an interface entropy.

Our paper is organized as follows. In Sec. \ref{sec2}, we introduce the setups of AdS$_3$/ICFT$_2$ and discuss the BCFT limit as well as the null energy condition on the interface brane.  In Sec. \ref{sec3} 
we will study the aspect of holographic entanglement entropy, including the interface entropy and its BCFT limit. Moreover, we will provide several concrete examples of the zero-temperature configuration to illustrate the behavior of the interface entropy. In Sec. \ref{sec:ft} we turn our attention to the system at finite temperature, exploring the influence of the scalar field on the profile of the interface brane. In Sec. \ref{sec:cd} we conclude our study and discuss the open questions for further exploration. 

\section{Setups of AdS$_3$/ICFT$_2$}
\label{sec2}
In this section we first show the setups of the bulk gravitational theory which are applicable for both zero and finite temperature scenarios. Then we will solve it at zero temperature in Sec. \ref{sec:sts}. We will also discuss the BCFT limit of the zero temperature system in Sec. \ref{subsec:bcftlimit} and the null energy condition on the interface brane in Sec. \ref{sec:nec}. 

The field theory under study has the following structure
\be
\label{eq:actionfieldtheory}
S=S_{\text{CFT}_\text{I}}+S_{\text{CFT}_\text{II}}+S_{\text{FT}_{\text{Q}_0}}
\ee
where on a constant time slice, $S_{\text{CFT}_\text{I}}$ is the action of the CFT$_\text{I}$ living on the left half spatial line,  $S_{\text{CFT}_\text{II}}$ is the action of the CFT$_\text{II}$ living on the right half spatial line, while $S_{\text{FT}_{\text{Q}_0}}$ is the action of the interface field theory living at the contacting point $Q_0$. In general, the interface field theory can be as follows, 
\be \label{eq:actionFTQ}
S_{\text{FT}_{\text{Q}_0}}\supset \lambda \int_{Q_0} dt\, \hat{\mathcal{O}}_\text{I}\hat{\mathcal{O}}_\text{II} +S[\mathcal{O}]+\cdots
\ee
which contains the coupling terms between the operators $\hat{\mathcal{O}}_\text{I},\hat{\mathcal{O}}_\text{II}$ from the left and right CFT's as well as the intrinsic dynamics of the localized fields (generically denoted as $\mathcal{O}$) living at the interface alone. If the field theory on $Q_0$ is conformal invariant, the resulting system is known as ICFT. Such systems have been widely studied in both condensed matter systems and holography, e.g. \cite{Kane:1992zza, Quella:2006de}. Here we are interested in the situation that the field theory on $Q_0$ is away from the conformal fixed point. This could be the situation that the interacting term of $\lambda$ in \eqref{eq:actionFTQ} is not marginal, or the intrinsic action $S[\mathcal{O}]$ is not conformal invariant.  When the whole system is strongly coupled, we expect that we can use the bottom-up holographic duality to explore the properties of such system. 

We consider the following gravitational theory
\be
S_\text{bulk}=S_\text{I} +S_\text{II}+S_Q 
\ee
with
\bea
\begin{split}
S_\text{I}&=\int_{N_\text{I}} d^3x\sqrt{-g_\text{I}}\,\bigg[\frac{1}{16\pi G}\bigg(R_\text{I}+\frac{2}{L_\text{I}^2}\bigg)
    \bigg] , \\
 S_\text{II}   &=\int_{N_\text{II}} d^3x\sqrt{-g_\text{II}}\,\bigg[\frac{1}{16\pi G}\bigg(R_\text{II}+\frac{2}{L_\text{II}^2}\bigg)
    \bigg] ,\\
S_{Q}&=\frac{1}{8\pi G}\int_Q d^2y\sqrt{-h}\, \bigg[\big(K_\text{I}-K_\text{II}\big)-(\partial\phi)^2-V(\phi)\bigg]\,,
\end{split}
\eea
which is the holographic dual of the system \eqref{eq:actionfieldtheory}. 
Note that there is a minus sign in front of the extrinsic curvature scalar $K_\text{II}$ in $S_Q$. This is due to that the extrinsic curvatures are computed using the outward normal vector pointing from $\text{I}$ to $\text{II}$. 
Without loss of generality, we assume that $L_\text{I}\geq L_\text{II}$.  Our setups extend the discussion in \cite{Erdmenger:2014xya} to the scenarios where  different field theories reside on either side of the interface. Different from the previous studies in \cite{Bachas:2020yxv, Bachas:2021fqo,  Anous:2022wqh} where there is a constant tension on the interface brane, we here consider a localized dynamical scalar field residing on the interface brane $Q$, which is dual to a scalar operator located on $Q_0$ in \eqref{eq:actionfieldtheory} that breaks the conformal invariance  .

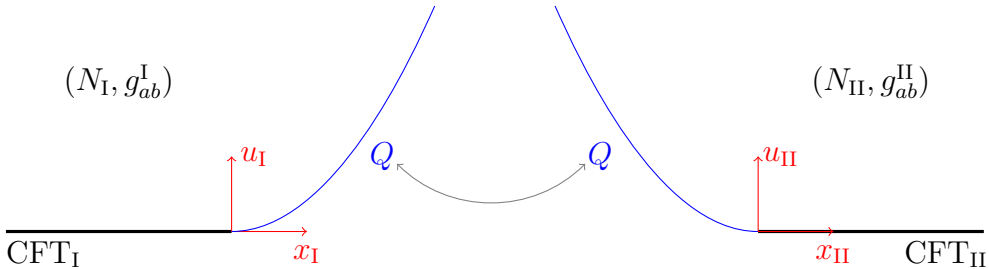
\begin{figure}[h]
\begin{center}

\begin{tikzpicture}
\draw[very thick] plot coordinates {(-3,0)(0,0)};
\node at (-2.5,-0.3) {CFT$_\text{I}$};

\draw[color=red] [->] (0,0) -- (1,0);
\draw[color=red]  [->] (0,0) -- (0,1);

\node[color=red] at (1,-0.3) {$x_\text{I}$};
\node[color=red] at (0.3,1) {$u_\text{I}$};
\draw[color=blue] (0,0) parabola (2.7,3);  
\draw[color=blue] (7,0) parabola (4.3,3);  
\draw[very thick] plot coordinates {(7,0)(10,0)};
\node at (9.5,-0.3) {CFT$_\text{II}$};
\draw[color=red] [->] (7,0) -- (8,0);
\draw[color=red] [->] (7,0) -- (7,1);
\node[color=red] at (8,-0.3) {$x_\text{II}$};
\node[color=red] at (7.3,1) {$u_\text{II}$};
\node at (-1.5,2) {$(N_\text{I},g^\text{I}_{ab})$};
\node at (8.5,2) {$(N_\text{II},g^\text{II}_{ab})$};
\node[color=blue] at (2,1.0) {$Q$};
\node[color=blue] at (4.9,1.0) {$Q$};
\draw[color=gray] [<->] (2.2,0.9)  to [out=-45, in=-135] (4.7,0.9);
\end{tikzpicture}

\end{center}
\vspace{-0.3cm}
\caption{\small Cartoon plot for the setup. The left and right bulk, denoted as   $N_{\text{I}}$ and $N_{\text{II}}$ respectively, are dual to CFT$_{\text{I}}$ and CFT$_{\text{II}}$ respectively. The interface brane $Q$ is in blue.}
\label{fig:cartoonplot}
\end{figure}

The configuration on a constant time slice is shown in Fig. \ref{fig:cartoonplot}. In the left bulk  $N_\text{I}$, we use coordinates $x_\text{I}^a=(t_\text{I}, x_\text{I}, u_\text{I})$ (where $a=0,1,2$) while in the right bulk $N_\text{II}$ we use coordinates $x_{\text{II}}^a=(t_\text{II}, x_\text{II}, u_\text{II})$. On the boundary, CFT$_\text{I}$ occupies the regime $x_\text{I}<0$ while CFT$_\text{II}$ is in the regime $x_\text{II}>0$. The two conformal field theories, 
CFT$_\text{I}$ and CFT$_\text{II}$, contact at the interface point where  $x_\text{I}=x_\text{II}=0$. 
On the interface brane $Q$, we have the intrinsic coordinates $y^\mu=(t, z)$ where $\mu=0,1$.\footnote{In Sec. \ref{sec:ft}, we use intrinsic coordinates $y^\mu=(t, w)$.} 
The embedding equations of $Q$ in $N_\text{I}$ and $N_\text{II}$ are $x_\text{I}^a(y^\mu)$ and $x_\text{II}^a(y^\mu)$, which obeys the following continuous condition on $Q$
\begin{align}
h_{\mu\nu}
= \frac{\partial x^{a}_{\text{I}}}{\partial y^{\mu}}\frac{\partial x^{b}_{\text{I}}}{\partial y^{\nu}}\,g^{\text{I}}_{ab} 
=\frac{\partial x^{a}_\text{II}}{\partial y^{\mu}}\frac{\partial x^{b}_\text{II}}{\partial y^{\nu}}\,g^\text{II}_{ab}\,.
\end{align}
Here $h_{\mu\nu}$ is the induced metric on $Q$, i.e. $ds_Q^2=h_{\mu\nu}dy^\mu dy^\nu$. 

The equations of motion are
\bea
\label{eq:einNI}
R^{\text{I}}_{ab}-\frac{1}{2}g^{\text{I}}_{ab}R^{\text{I}}-\frac{1}{L_\text{\text{I}}^2}g^{\text{I}}_{ab}&=&0\,,\\
\label{eq:einNII}
R^\text{II}_{ab}-\frac{1}{2}g^\text{II}_{ab}R^\text{II}-\frac{1}{L_\text{II}^2}g^\text{II}_{ab}&=&0\,,\\
\Delta K_{\mu\nu}-h_{\mu\nu}\Delta K
+\left[  
(\partial\phi)^2+V(\phi)
\right]
h_{\mu\nu}-2\partial_{\mu}\phi\partial_{\nu}\phi&=&0\,, \label{eq:Q1} \\
2\partial_{\mu}(\sqrt{-h}h^{\mu\nu}\partial_{\nu}\phi)-\sqrt{-h}\frac{dV(\phi)}{d\phi}&=&0\,,\label{eq:j2} 
\eea
where $\Delta X\equiv X_{\text{I}}-X_\text{II}$ with $X$ as $K_{ab}$ or $K$. Here \eqref{eq:einNI} and \eqref{eq:einNII} are the equations for the metric fields in the left bulk $N_\text{I}$ and the right bulk $N_\text{II}$, while \eqref{eq:Q1} and \eqref{eq:j2} are the equations on the interface brane $Q$.

We can separate \eqref{eq:Q1} into two parts, i.e. the trace part and the traceless part. The trace part is
\begin{align}
V(\phi)=\frac{1}{2}\Delta K\,,
\end{align}
and the traceless part is
\begin{align}
\Delta K_{\mu\nu}-\frac{1}{2} h_{\mu\nu} \Delta K
+(\partial\phi)^2h_{\mu\nu}
-2\partial_{\mu}\phi\partial_{\nu}\phi=0\,.
\end{align}
Now the equations on $Q$  
can be summarized as follows
\bea
g^{\text{I}}_{ab}dx_{\text{I}}^{a}dx_{\text{I}}^{b}\Big{|}_{Q}  
&=&g^\text{II}_{ab}dx_\text{II}^{a}dx_\text{II}^{b}\Big{|}_{Q}  \,,\label{eq:jj0}\\
\Delta K_{\mu\nu}-\frac{1}{2}h_{\mu\nu} \Delta K
+(\partial\phi)^2h_{\mu\nu}
-2\partial_{\mu}\phi\partial_{\nu}\phi&=&0\,,\label{eq:jj1}\\
V(\phi)-\frac{1}{2}\Delta K&=&0\,, \label{eq:jj2}\\
2\partial_{\mu}(\sqrt{-h}h^{\mu\nu}\partial_{\nu}\phi)-\sqrt{-h}\frac{dV(\phi)}{d\phi}&=&0\,.\label{eq:jj3}
\eea

We will concentrate on the static systems, anticipating the existence of a global time within the dual field theory. The above setups are applicable for both zero-temperature and finite-temperature scenarios. In the following we will present the zero temperature solution and then study its properties. We will study the finite temperature solution in Sec. \ref{sec:ft}.

\subsection{Solve the system at zero temperature}
\label{sec:sts}

At zero temperature, the bulk equations \eqref{eq:einNI} and \eqref{eq:einNII} give the planar AdS$_3$ 
solution 
\begin{align}
\label{eq:poincare}
ds^2_A=\frac{L^2_A}{u^2_A}\,\big[
-dt^2_A+dx^2_A+du^2_A
\big]\,, ~~~~~~A=\mbox{I,\, ~II}\,.
\end{align}
Here $u_A\in (0, \infty)$ with the boundary field theory CFT$_A$ located at the boundary $u_A\to 0$. We consider the case that the spatial coordinates $x_A$ are non-compact.\footnote{It is worth noting that the case of compact spatial directions has been recently investigated in \cite{Bachas:2021fqo}.} CFT$_\text{I}$ lives in the regime $x_\text{I}<0$ while CFT$_\text{II}$ is in the regime $x_\text{II}>0$.  
The central charges for the dual CFT$_\text{I}$ and CFT$_\text{II}$ are \cite{Brown:1986nw}  
\begin{align}
c_\text{I}=\frac{3L_\text{I}}{2G_N}\,,\ \ 
~~~~
c_\text{II}=\frac{3L_\text{II}}{2G_N}\,.
\end{align}
We have assumed that $L_\text{I}\geq L_\text{II}$, thus $c_\text{I}\geq c_\text{II}$.
For convenience, we define 
\be
\nu \equiv\frac{c_\text{II}}{c_\text{I}}=\frac{L_\text{II}}{L_\text{I}}
\ee
where $\nu\in (0,1]$. Then we can parameterize the system using $L_\text{I}, \nu$ instead of $L_\text{I}, L_\text{II}$.

It will be convenient to use the rescaled coordinates $z_A$,
\begin{align}
\label{eq:rescal}
u_\text{I}=\sqrt{\frac{L_\text{I}}{L_\text{II}}}z_\text{I}= \frac{z_\text{I}}{\sqrt{\nu}}\,, ~~~~
u_\text{II}=\sqrt{\frac{L_\text{II}}{L_\text{I}}}z_\text{II}= \sqrt{\nu} z_\text{II}\,.
\end{align}
Since we are interested in static configuration, the brane $Q$ is supposed to be a timelike hypersurface. Assuming it is given by 
\be
x_\text{I}=\psi_\text{I}(z_\text{I})\,~~~~ 
\text{or equivalently} ~~~ x_\text{II}=\psi_\text{II}(z_\text{II})\,, 
\ee
then the continuous condition of the metric on $Q$ reads
\begin{align}
\label{eq:indQ}
\frac{\nu L_\text{I}^2}{z^2_\text{I}}\left[
-dt_{\text{I}}^2
+
\left(\frac{1}{\nu}+\psi'^2_\text{I}(z_\text{I})\right)dz_{\text{I}}^2\,
\right] 
=\frac{\nu L_\text{I}^2}{z^2_\text{II}}\left[
-dt_\text{II}^2+\left(\nu+\psi'^2_\text{II}(z_\text{II})\right)dz_\text{II}^2
\,\right]\,.  
\end{align}

We assume\footnote{In principle, one could consider the solution of $t_\text{I}=\alpha\, t_\text{II}$, $z_\text{I}=\alpha \,z_\text{II}$ and $\frac{1}{\nu}
+\psi'^2_\text{I}(z_\text{I})=\nu+\psi'^2_\text{II}(z_\text{II})$. Here we set $\alpha=1$ such that time is globally well-defined.}
\be t_{\text{I}}=t_\text{II}\equiv t\,,
\ee
 then the continuous condition of the induced metric \eqref{eq:indQ} gives
\be
\label{eq:zrelation}
z_\text{I}=
z_\text{II}\equiv z\,,~~~~
\frac{1}{\nu}
+\psi'^2_\text{I}(z_\text{I})=\nu+\psi'^2_\text{II}(z_\text{II})\,.
\ee
In following calculations, we identify $\psi_\text{I}(z_\text{I}),\psi_\text{II}(z_\text{II})$ with $\psi_\text{I}(z),\psi_\text{II}(z)$.

Let us remark on the parameterisation of a point $p$ on the brane $Q$, which can be expressed in three distinct coordinate systems: the intrinsic coordinates $y^{\mu}=(t,z)$, the $N_\text{I}$ part of AdS 
$
x_\text{I}^a=(t,\psi_\text{I}(z),  z/\sqrt{\nu})$ 
and the $N_\text{II}$ part of AdS 
$
x_\text{II}^a=(t,\psi_\text{II}(z),\sqrt{\nu}z)$. 
We consider the simplest static case $\phi=\phi(z)$\footnote{More generally, one could consider $\phi=\phi(t,z)$ and then the embedding equation as well as the induced metric of $Q$ is time dependent. It would be interesting to further explore this case.}, then from the equations (\ref{eq:jj1}, \ref{eq:jj2}, \ref{eq:jj3})  we obtain 
\begin{align}
\phi'^2=&~\,\frac{L_{\text{I}}}{2z}\frac{-\psi''_{\text{I}}+\nu \psi''_{\text{II}}}{\sqrt{\nu+\psi'^{2}_{\text{II}}}}\,,\label{eq:eq2}\\
 V(\phi(z))= &~\,\frac{\sqrt{\nu}\left( 2(\psi_\text{I}'-\nu\psi_\text{II}') + 2 (\nu\psi_\text{I}'^3-\psi_\text{II}'^3)-z(\psi_\text{I}''-\nu\psi_\text{II}'')\right)}{2L_\text{I}(1+\nu\psi_\text{I}'^2)^{3/2}}
 \,,\label{eq:eq3} \\
\frac{dV(\phi)}{d\phi}=&~
\frac{2z^2\phi{''}}
{L^2_\text{I}(1+\nu\psi'^{2}_\text{I})}-
\frac{2z^2\nu\phi'\psi_\text{I}'\psi_\text{I}{''}}{L^2_\text{I}(1+\nu\psi'^{2}_\text{I})^2}\,,\label{eq:eq4}
\end{align} 
where $\psi_\text{I},\psi_\text{II},\phi$ are all function of $z$ and the prime $'$ is the derivative with respect to $z$.
Note that the above three equations are not independent, i.e. we can derived the third one from the first two. 

The Ricci tensor of the induced metric on the brane is 
\begin{align}\label{eq:Qads2}
R^{Q}_{\mu\nu}=\lambda(z)\,h_{\mu\nu}\,,
\end{align}
where
\begin{align}\label{eq:lam0}
\lambda(z)=-\frac{1+
\nu\psi'_\text{I}\,(\psi'_\text{I}+z \psi''_\text{I})
}{
L_\text{I}^2\, (1+\nu\psi'^{2}_\text{I})^2
}\,.
\end{align}
Obviously when $\psi'_\text{I}=\text{constant}$, we have $\lambda(z)=\text{constant}$. This is consistent with the picture that in the case of trivial scalar field, the brane is a straight line and has the induced metric with pure AdS$_2$ \cite{Suzuki:2022xwv, Anous:2022wqh}.\footnote{One can show that when $\lambda(z)=\text{constant}$, the only allowed profile for the brane is a straight line. Another unphysical solution may exist that the brane terminates at a special point in the bulk and we do not consider such solution.} 

We are interested in the case that the brane is asymptotically AdS$_2$ in UV. Then in the UV limit $z\to 0$, $\lambda(z)$ should be a negative constant. 
One class of such solution is 
\be\label{eq:gensolads}
\psi_\text{I} \to \gamma\, z^n,~~~\text{with}~ n\geq 1\,,
\ee
where $\gamma$ is a constant. With the above solution, from \eqref{eq:zrelation} we have 
\be \psi_\text{II}\to \pm \sqrt{\frac{1}{\nu}-\nu+n^2\gamma^2\delta_{n1}} \,z\,,~~~ \text{when}~ z\to 0\,.
\ee

\subsubsection{Simple examples of solutions}
\label{sec:simpleexsol}

With the above setups, given the potential $V(\phi)$, one can obtain consistent solutions $\phi(z),\psi_{\text{I}}(z),\psi_{\text{II}}(z)$ of the system. In practice, we construct the consistent solution as follows. We start from an ``arbitrary" function $\psi_{\text{I}}(z)$ which should satisfy the null energy condition discussed in Sec. \ref{sec:nec}. Then  $\psi_{\text{II}}(z)$ can be solved from  \eqref{eq:zrelation}. The consistent scalar field  $\phi(z)$ should satisfy \eqref{eq:eq2} which are determined by $\psi_{\text{I}}(z)$ and $\psi_{\text{II}}(z)$, and  $V(\phi)$ satisfies \eqref{eq:eq3}. In the following we will show two simple examples. More examples will be discussed in Sec. \ref{sec:heeexample}. 

The first example is the case that the brane $Q$ is a straight line. In this case we have solutions, 
\begin{align}
\label{eq:straightline}
\begin{split}
\psi_\text{I} &= \gamma z\,,\\
\psi_\text{II}&=\pm\sqrt{\frac{1+\gamma^2\nu - \nu^2}{\nu}} z\,,\\
\phi(z)&=\phi_0\,, \\
V(\phi)&=\frac{\gamma \nu^{3/2} \mp \sqrt{1+\gamma^2\nu-\nu^2}}{L_{\text{I}} \nu \sqrt{1+\gamma^2\nu}}\,,
\end{split}
\end{align}
where $\gamma, \phi_0$ are constants. Note that the sign in front of $\psi_\text{II}$ is related to if the configuration of $Q$ is an acute angle or an obtuse angle in the bulk $N_\text{II}$. 

This case has been studied in \cite{Simidzija:2020ukv, Anous:2022wqh} with $\phi_0=0$ and the tension $T=V$. 
Similar to the result in \cite{Simidzija:2020ukv, Anous:2022wqh}, from \eqref{eq:straightline} the tension is constrained by 
\be
\label{eq:tensionreg0T}
T_\text{min}
<|T| < T_\text{max}\, ,
\ee
where 
\be
\label{eq:tensionregTmaxTmin}
T_\text{min}=\frac{1-\nu}{L_\text{I} \nu}
\, ,\ \ T_\text{max}=\frac{1+\nu}{L_\text{I} \nu}\, .
\ee
For a given value of the tension satisfying \eqref{eq:tensionreg0T}, the profiles of the branes are uniquely determined, as shown in \eqref{eq:straightline}. 

The second example is the case when $\nu=1$, i.e. the two CFT's have the same central charge. In this case, the system can be viewed as the presence of a defect located at the origin of field theory. The equations \eqref{eq:zrelation} become
\begin{eqnarray}
\psi'^{2}_{\text{I}}&=&\psi'^{2}_{\text{II}}\,,
\end{eqnarray}
which have two kinds of solutions. The first one is 
\be\label{eq:e2sol1}
\psi'_{\text{I}}=\psi'_{\text{II}}\,,~~~~\phi=c_1\,,~~~~ V=0\,.\ee
This is a trivial one and the brane does not play any role, i.e. there is no defect at all on the field theory. 
The second one is 
\begin{align}
\begin{split}
\psi_{\text{I}}(z)=&-\psi_{\text{II}}(z)=~\varphi(z)\,,   \\
\phi'^2(z)=&~-\frac{L_{\text{I}}\,\varphi''(z)}{z\sqrt{1+\varphi'^2(z)}}\,, \\
V(\phi)=&~ \frac{
2\varphi'+2\varphi'^{3}-z\varphi''
}{L_{\text{I}}(1+\varphi'^{2})^\frac{3}{2}}\,.
\end{split}  \label{eq:nu=1}
\end{align} 
This could be understood as unfolding a holographic BCFT system. 
We will further comment on this solution in the next subsection.

\subsection{BCFT limit of the equation of motion}
\label{subsec:bcftlimit}
ICFTs are more generic than BCFTs in the sense that BCFT can be viewed as special limit of ICFT. There are two different ways to obtain the BCFT limit from ICFT. The first way is to consider the limit 
\be
\label{eq:bcftlimit}
\nu=
\frac{L_\text{II}}{L_\text{I}}\to 0\,,
\ee
which can be realized by setting $L_\text{II}\to 0$ while $L_\text{I}$ is finite. This means that the central charge of CFT$_\text{II}$
is relatively small and we can ignore it in the whole system. 

To study the limit, it is more convenient to work in the coordinates of $t, u$ on $Q$ where $u=u_\text{I}$. Using the relation in \eqref{eq:rescal}, we can fix $u_\text{II}=\nu u$. The profile of the brane could be parameterized as $x_\text{I}=\tilde{\psi}_\text{I}(u)$, or equivalently  $x_\text{II}=\tilde{\psi}_\text{II}(u)$ and $\phi=\tilde{\phi}(u)$. 
The equations for these fields can be obtained by  
rewriting \eqref{eq:zrelation} and (\ref{eq:eq2}, \ref{eq:eq3}, \ref{eq:eq4}) with 
$$\psi'_\text{I}(z) \to  \frac{\tilde\psi'_\text{I}(u)}{\sqrt{\nu}}\,,~~~ \psi''_\text{I}(z) \to 
\frac{\tilde\psi''_\text{I}(u)}{\nu}\,,~~~ \psi'_\text{II}(z) \to \frac{\tilde\psi'_\text{II}(u)}{\sqrt{\nu}}\,,~~~ \psi''_\text{II}(z) \to \frac{\tilde\psi''_\text{II}(u)}{\nu}\,,
~~~
\phi'(z)\to \frac{\tilde\phi'(u)}{\sqrt{\nu}}\,.$$ Here the prime in functions of $z$ (e.g. $\psi'_\text{I}(z)$) represents the derivative with respect to $z$, while the prime in functions of $u$ represents the derivative with respect to $u$. 
We take the $\nu\to 0$ limit, assuming that in this limit   $\tilde\psi_\text{I}(u) = \varphi(u)$ which is a $\nu$-independent and smooth function,  then we can obtain 
\begin{align}
\label{eq:bcftlimit2}
\begin{split}
\tilde\psi_{\text{I}}(u)=&~\varphi(u)\,,   \\
\tilde\psi'_{\text{II}}(u)=&~\pm\sqrt{1+\varphi'^2(u)}+\mathcal{O}(\nu)\,,   \\
\tilde\phi'^2(u)=&~-\frac{L_{\text{I}}\varphi''(u)}{2u\sqrt{1+\varphi'^2(u)}}+\mathcal{O}(\nu)\,, \\
V(\tilde\phi)=&~ \mp\frac{1}{L_{\text{I}}\nu}
+\frac{
2\varphi'(u)+2\varphi'^{3}(u)-u\varphi''(u)
}{2L_{\text{I}}(1+\varphi'^{2}(u))^\frac{3}{2}}
+\mathcal{O}(\nu)\,.
\end{split}
\end{align} 
Note that there is a divergent term in the potential $V$. This also has been seen in \cite{Simidzija:2020ukv, Anous:2022wqh} where the tension is divergent in the AdS/BCFT limit. For the special case $\phi=0, V=T$, the above system reduced to the results in \cite{Simidzija:2020ukv, Anous:2022wqh} and solution of \eqref{eq:bcftlimit2} gives to a straight line.

We can make a comparison to the case without regime $N_\text{II}$, i.e. the framework of AdS/BCFT. 
In the case of AdS/BCFT with a dynamical scalar field on $Q$, we parameterize $Q$ as $x=\tilde{\psi}_\text{I}(u)=\varphi(u)$. Then  
the equations of motion are
\begin{align}
\begin{split}
\phi'^2(u)=&~-\frac{L_{\text{I}}\,\varphi''(u)}{2u\sqrt{1+\varphi'^2(u)}}\,, \\
V(\phi)=&~ \frac{
2\varphi'(u)+2\varphi'^{3}(u)-u\varphi''(u)
}{2L_{\text{I}}(1+\varphi'^{2}(u))^\frac{3}{2}}\,.
\end{split} \label{eq:eom of bcftu}
\end{align} 
These equations have been studied in \cite{Kanda:2023zse}.  The above equations are the same as \eqref{eq:bcftlimit2} except that they do not have divergent term in the potential $V(\phi)$. 

The second way to obtain a BCFT is to consider the limit $\nu=1$ where we can perform the folding trick\cite{Karch:2021qhd}. 
In this case the dynamical equations in ICFT are listed in \eqref{eq:nu=1}. 
Note that when $\nu=1$ we have $u_\text{I}=u_\text{II}=z=u$.  
Obviously, it has exactly the same form as \eqref{eq:eom of bcftu} after setting 
$\phi'^{2},V(\phi)$ in the ICFT \eqref{eq:nu=1} as twice of those in \eqref{eq:eom of bcftu}.

\subsection{Null energy condition on the brane}
\label{sec:nec}

The energy conditions on the brane is important to constrain the dynamics of the interface brane. Particularly within the framework of AdS/BCFT, the null energy condition (NEC) is widely used.\footnote{For discussions regarding other energy conditions in AdS/DCFT, wherein the NEC is deemed the most fundamental, see e.g. \cite{Erdmenger:2014xya}.} In the following we derive the constraints on the profiles of the interface brane from the NEC for the matter field residing on the brane.  

The coordinate system on $Q$ is $y^{\mu}=(t,z)$. The null vector on $Q$ is
\begin{align}
N^{\mu}=
\left(
\pm
\sqrt{\frac{1}{\nu}+\psi'^{2}_\text{I}}\,,~
1
\right)\,,
\end{align}
and therefore the NEC is
\begin{align}
(\Delta K_{\mu\nu}-h_{\mu\nu}\Delta K)N^{\mu}N^{\nu} =\frac{L_\text{I}}{z}\,\frac{(-\psi''_{\text{I}}+\nu\psi''_\text{II})}{\sqrt{\frac{1}{\nu}+\psi'^{2}_\text{I}}}\ge 0\,.
\label{eq:nec}
\end{align}
From \eqref{eq:eq2}, the NEC is equivalent to $\phi'^2(z)\geq 0$. This means that whenever we have a consistent background solution, then the NEC is satisfied. 

However, as we show in the following, this condition actually constrains the possible choices of $\psi_{\text{I}}$. From the junction condition \eqref{eq:zrelation}, we have
\begin{align}
\psi'_{\text{II}}=\pm\sqrt{
\frac{1}{\nu}-\nu+\psi'^{2}_{\text{I}}
}\,.
\end{align} 
The sign above is related to the tangent direction of the brane is form an obtuse angle or an acute angle. When $0<\nu<1$, the NEC \eqref{eq:nec}, at $z$ where $\psi'^{2}_{\text{I}}\not=\nu-\frac{1}{\nu}$, can be simplified as 
\begin{align}
-\psi''_{\text{I}}+\nu\psi''_\text{II}= 
\left(
-1\pm
\frac{\nu\psi'_{\text{I}}
}{
\sqrt{
\frac{1}{\nu}-\nu+\psi'^{2}_{\text{I}}
}
} 
\right)\psi''_{\text{I}}\ge 0\,,
\end{align} 
or equivalently,
\begin{align}\label{eq:necpsi1}
\psi''_{\text{I}}\le 0
\end{align} 
by noticing that
$
\Big{|}
\frac{\nu\psi'_{\text{I}}
}{
\sqrt{
\frac{1}{\nu}-\nu+\psi'^{2}_{\text{I}}
}
} \Big{|}< 1\,.
$  
When $\nu=1$, i.e. the two CFTs have the same central charge, 
the NEC \eqref{eq:nec} can be simplified as
\begin{align}
\label{eq:necfornu1}
\psi'_{\text{I}}=\psi'_{\text{II}} \ \ ~~~
\text{or}\ \ ~~~
\psi'_{\text{I}}=-\psi'_{\text{II}}\,,\ \ 
\psi''_{\text{I}}\le 0\,.
\end{align}
Note that the first case in \eqref{eq:necfornu1} is trivial in the sense that the brane does not play any role, which has been discussed in \eqref{eq:e2sol1}. Therefore we will focus on the case with $\psi''_{\text{I}}\le 0$ for all the values of $\nu$. 
This condition constraints the allowed configurations of $Q$.

We can use NEC to study the configuration of $\psi_\text{II}$. 
From the junction condition \eqref{eq:zrelation}, we have
\be
\label{eq:psiIrelation}
\psi'_{\text{I}}=\pm\sqrt{
-\frac{1}{\nu}+\nu+\psi'^{2}_{\text{II}}
}\,.\ee
Note that for $\nu\in (0,1]$, the expression in the square root needs to be non-negative and this constraints the possible values of $\psi'_\text{II}$. 
Now the NEC \eqref{eq:nec}, at $z$ where $\psi'^{2}_{\text{II}}\not=\frac{1}{\nu}-\nu$, becomes
\begin{align}
-\psi''_{\text{I}}+\nu\psi''_\text{II}=
\left(
\nu\mp
\frac{\psi'_{\text{II}}
}{
\sqrt{
-\frac{1}{\nu}+\nu+\psi'^{2}_{\text{II}}
}
} 
\right)\psi''_{\text{II}}\ge 0\,,
\end{align} 
where the sign in the bracket is aligned with the sign in \eqref{eq:psiIrelation}. 
When $0<\nu<1$ we have $|\psi'_{\text{II}}|\ge \sqrt{\frac{1}{\nu}-\nu}>0$. In this case the possible constraints from the NEC for the monotonic profiles are 
\begin{enumerate}[(1)]
\item $0<\nu<1$, $\psi'_{\text{II}}\not=\text{constant}$, 
~~~$\psi'_{\text{II}}\ge \sqrt{\frac{1}{\nu}-\nu}$,~~
$\psi'_{\text{I}}=+\sqrt{
-\frac{1}{\nu}+\nu+\psi'^{2}_{\text{II}}
}$,~~~ $\psi''_{\text{II}}\le 0$.      
\item $0<\nu<1$, $\psi'_{\text{II}}\not=\text{constant}$, 
~$\psi'_{\text{II}}\le -\sqrt{\frac{1}{\nu}-\nu}$,~~
$\psi'_{\text{I}}=+\sqrt{
-\frac{1}{\nu}+\nu+\psi'^{2}_{\text{II}}
}$, ~~~$\psi''_{\text{II}}\ge 0$.     
\item $0<\nu<1$, $\psi'_{\text{II}}\not=\text{constant}$,~~ ~
$\psi'_{\text{II}}\ge \sqrt{\frac{1}{\nu}-\nu}$,~~
$\psi'_{\text{I}}=-\sqrt{
-\frac{1}{\nu}+\nu+\psi'^{2}_{\text{II}}
}$, ~~~$\psi''_{\text{II}}\ge 0$.    
\item $0<\nu<1$, $\psi'_{\text{II}}\not=\text{constant}$, 
~$\psi'_{\text{II}}\le -\sqrt{\frac{1}{\nu}-\nu}$,~~
$\psi'_{\text{I}}=-\sqrt{
-\frac{1}{\nu}+\nu+\psi'^{2}_{\text{II}}
}$, ~~~$\psi''_{\text{II}}\le 0$.     
\item $0<\nu<1$, $\psi'_{\text{II}}=\text{constant}$, 
~$|\psi'_{\text{II}}|\ge\sqrt{\frac{1}{\nu}-\nu}$,~~
$\psi'_{\text{I}}=\pm\sqrt{
-\frac{1}{\nu}+\nu+\psi'^{2}_{\text{II}}
}$. 
\end{enumerate}
In the first four cases, we always have $\psi''_{\text{I}}\le 0$ as shown in \eqref{eq:necpsi1}. In the case (5) the interface brane is a straight line, which corresponds to the scenario involving a trivial scalar field. Obviously with a nontrivial scalar field, the brane can bend in different ways. Note that in the cases (1-4), for the specific point in the bulk where $\psi'_{\text{II}}=\pm\sqrt{\frac{1}{\nu}-\nu}$, we have $\psi'_{\text{I}}=0$ which is a turning point for the profile $\psi_{\text{I}}$ of the brane. This indicates that the brane is no longer monotonic. We will only consider the case that the brane is monotonic in the main text.\footnote{In appendix \ref{app:a}, an example of a profile $\psi_\text{I}$ of the brane which is a non-monotonic function is shown in Fig. \ref{fig:nm br} . 
} 

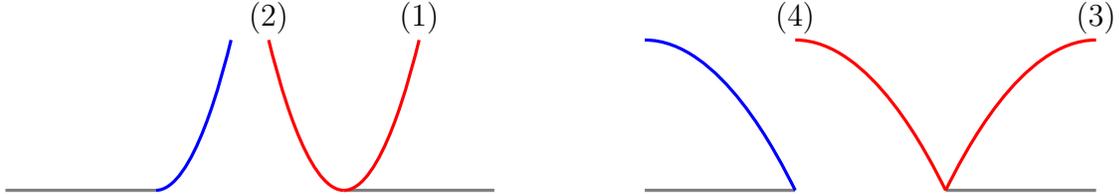
\begin{figure}[h]
\begin{center}
\begin{tikzpicture}

\draw[color=gray,very thick] plot coordinates {(0,0)(2,0)};
\draw[color=blue,very thick] (2,0) parabola (3,2);  

\draw[color=gray,very thick] plot coordinates {(4.5,0)(6.5,0)};
\draw[color=red,very thick] (4.5,0) parabola (3.5,2);  
\draw[color=red,very thick] (4.5,0) parabola (5.5,2);  

\node[color=black] at (3.5,2.3) {(2)};
\node[color=black] at (5.5,2.3) {(1)};

\draw[color=gray,very thick] plot coordinates {(8.5,0)(10.5,0)};
\draw[color=blue,very thick] (8.5,2)  parabola (10.5,0);  

\draw[color=gray,very thick] plot coordinates {(12.5,0)(14.5,0)};
\draw[color=red,very thick]  (10.5,2) parabola (12.5,0);  
\draw[color=red,very thick] (14.5,2)  parabola (12.5,0);  

\node[color=black] at (14.5,2.3) {(3)};
\node[color=black] at (10.5,2.3) {(4)};


\end{tikzpicture}
\end{center}
\vspace{-0.3cm}
\caption{\small Cartoon plot for the four kinds of monotonic configuration of brane profile
when $0<\nu<1$ and $\psi'_{\text{II}}\not=\text{constant}$. These configurations shows clearly about the allowed convexity of the brane. 
}
\label{fig:brane pair}
\end{figure}

We show the cartoon picture of the bending branes in Fig. \ref{fig:brane pair}. We have used the blue line to represent the profile of $Q$ with $\psi_\text{I}$ and the red line for $\psi_\text{II}$. In the cases (1)\&(2) the brane will extend to infinity while in the cases (3)\&(4), the brane can extend to infinity or a finite value of $u$.

When $\nu=1$, we have \eqref{eq:e2sol1} with the relation $\psi'_{\text{I}}=+\psi'_{\text{II}}\,$; or  
\eqref{eq:nu=1} with the relation 
$\psi'_{\text{I}}=-\psi'_{\text{II}}\,, 
\psi''_{\text{II}}\le 0\,,$ i.e. the configurations of (2)\&(3) in Fig. \ref{fig:brane pair}. 

\section{Holographic entanglement entropy}
\label{sec3}

The presence of a nontrivial scalar field on the interface  brane $Q$ leads to more complicated brane profiles, indicating that the ICFT is away from a fixed point.  In this section, we will first study the entanglement entropy of a specific regime within field theory using holographic techniques. Subsequently, we will introduce a $g$-function for the ICFT to quantify the effective degrees of freedom in Sec. \ref{subsec:iegf} and discuss the BCFT limits of holographic entropy in Sec. \ref{subsec:bcftlimitsofhee}. In Sec. \ref{sec:heeexample} we will present several illustrative examples to demonstrate the characteristic features of the system. A consistent observation  is that whenever the null energy condition is satisfied, the $g$-function is monotonically decreasing from UV to IR. We will comment on the entanglement entropy for other intervals in Sec. \ref{sec:comment}.

\begin{figure}[h]
\begin{center}
\begin{tikzpicture}
\draw[very thick] plot coordinates {(-3,0)(0,0)};
\draw[color=red] [->] (0,0) -- (1,0);
\draw[color=red]  [->] (0,0) -- (0,1);
\node[color=red] at (1,-0.3) {$x_\text{I}$};
\node[color=red] at (0.3,1) {$u_\text{I}$};
\draw[color=blue] (0,0) parabola (3,3);  

\draw[color=orange] (5/2,25/12) parabola (-2,0);
\node[color=purple] at (2.6,1.8) {$z_*$};
\node[color=orange] at (0.5,1.9) {$\gamma_1$};
\node[color=purple] at (-2,-0.3) {$-\sigma_\text{I}$};

\draw[color=blue] (7,0) parabola (4,3);  
\draw[very thick] plot coordinates {(7,0)(10,0)};
\draw[color=red] [->] (7,0) -- (8,0);
\draw[color=red] [->] (7,0) -- (7,1);
\node[color=red] at (8,-0.3) {$x_\text{II}$};
\node[color=red] at (7.3,1) {$u_\text{II}$};

\node at (-1.5,2) {$(N_\text{I},g^\text{I}_{ab})$};
\node at (8.5,2) {$(N_\text{II},g^\text{II}_{ab})$};

\node[color=blue] at (2,1.0) {$Q$};
\node[color=blue] at (4.9,1.0) {$Q$};
\draw[color=gray] [<->] (2.2,0.9)  to [out=-45, in=-135] (4.7,0.9);

\draw[color=orange] (4.5,25/12) parabola (9,0);  
\node[color=purple] at (4.4,1.8) {$z_*$};
\node[color=orange] at (6.6,1.9) {$\gamma_2$};
\node[color=purple] at (9,0-0.3) {$\sigma_\text{II}$};

\node[color=purple] at (0,-0.3) {$(0,0)$};
\node[color=purple] at (7,-0.3) {$(0,0)$};
\end{tikzpicture}
\end{center}
\vspace{-0.3cm}
\caption{\small Cartoon plot for the configuration of the extremal surfaces $\gamma_1$ and $\gamma_2$ for chosen the boundary subsystem $[-\sigma_{\text{I}},0]\cup [0,\sigma_{\text{II}}]$.}
\label{fig:exf}
\end{figure}
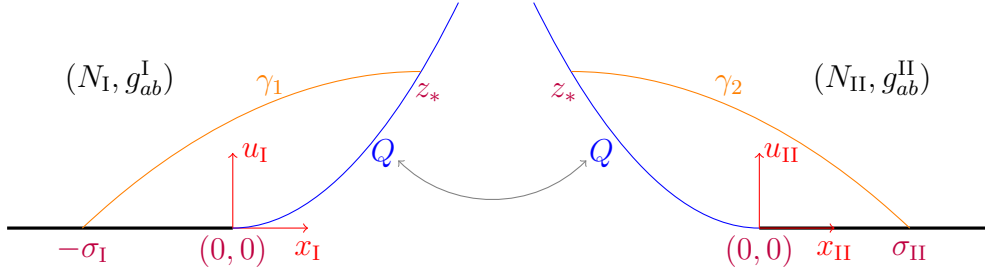

We are interested in the boundary subsystems including the interface. We consider the subsystem $[-\sigma_{\text{I}},0]\cup [0,\sigma_{\text{II}}]$, the corresponding extremal surface contains two parts, denoted by $\gamma_1$ and $\gamma_2$ as shown in Fig. \ref{fig:exf}.
In the coordinate system \eqref{eq:poincare} and using \eqref{eq:rescal},  
at a constant time $t$, $\gamma_1$ and $\gamma_2$ can be written as
\begin{align}
\gamma_1=\left(t,x_{\text{I}}[\xi_1],\frac{z_{\text{I}}[\xi_1]}{\sqrt{\nu}}\right) \,,~~~~~
\gamma_2=(t,x_{\text{II}}[\xi_2],{\sqrt{\nu}} z_{\text{II}}[\xi_2])\,,
\end{align} 
where $\xi_1$ and $\xi_2$ are the parameters of the two curves. The length functional is
\begin{align}
\label{eq:lenfun}
\mathcal{A}=
\int d\xi_1\,\frac{
L_{\text{I}}
\sqrt{\nu}
}{z_{\text{I}}}
\sqrt{
\dot{x}_{\text{I}}^2+\frac{\dot{z}_{\text{I}}^2}{\nu}
}\,
+
\int d\xi_2\, \frac{
L_{\text{I}}
\sqrt{\nu}
}{z_{\text{II}}}
\sqrt{
\dot{x}_{\text{II}}^2+\nu\dot{z}_{\text{II}}^2
}\,,
\end{align} 
where the dots are derivatives with respect to $\xi_1$ or $\xi_2$.

For these two curves, we first perform the variation to obtain the boundary terms and then we choose $\xi_1=z_{\text{I}}$ and $\xi_2=z_{\text{II}}$. 
We get the boundary condition on the brane 
\begin{align}
\frac{
x'_{\text{I}}(z)\psi'_{\text{I}}(z)+\frac{1}{\nu}
}{
\sqrt{
x'^2_{\text{I}}(z)+\frac{1}{\nu}
}
}
+
\frac{
x'_{\text{II}}(z)\psi'_{\text{II}}(z)+\nu
}{
\sqrt{
x'^2_{\text{II}}(z)+\nu
}
}
\Bigg{|}_\text{Q}=0\,, \label{bc:var}
\end{align} 
where the prime is the derivative with respect to $z$. Note that $x_\text{I}(z_\text{I})$ and $\psi_\text{I}(z)$ are curves for $\gamma_1$ and $Q$ respectively, similarly for the other functions.  

Let us give a geometrical interpretation of \eqref{bc:var}. 
The tangent vectors for the curves $\gamma_1$ and $\gamma_2$ are
\begin{align}
\label{eq:vectorvv}
V_1=\left(0,x'_\text{I}[z_{\text{I}}],\frac{1}{\sqrt{\nu}}\right) \, , \ \ \ 
V_2=(0,x'_\text{II}[z_{\text{II}}],\sqrt{\nu})\,,
\end{align} 
where we have chosen the parameter $\xi_1=z_{\text{I}}, ~\xi_2=z_{\text{II}}$. The tangent vectors of the brane $Q$ are
\begin{align}
\label{eq:vectorww}
W_1=\left(0,\psi'_{\text{I}}[z],\frac{1}{\sqrt{\nu}}\right)\, ,\ \ \ 
W_2=(0,\psi'_{\text{II}}[z],\sqrt{\nu})\,.
\end{align} 
The boundary condition \eqref{bc:var} on the brane $Q$ can be rewritten as
\begin{align}
\frac{
g^{\text{I}}_{ab}V^a_1 W^b_1
}{
\sqrt{
g^{\text{I}}_{ab}V^a_1 V^b_1
} \sqrt{
g^{\text{I}}_{ab}W^a_1 W^b_1
}
}+
\frac{
g^{\text{II}}_{ab}V^a_2 W^b_2
}{
\sqrt{
g^{\text{II}}_{ab}V^a_2 V^b_2
} \sqrt{
g^{\text{II}}_{ab}W^a_2 W^b_2
}
}=0\,,
\end{align}
where we have used the continuous condition \eqref{eq:zrelation}. 
The above equation can be formally written as
\begin{align}
\hat{V}_1 \cdot \hat{W}_1
+\hat{V}_2\cdot \hat{W}_2=0\,, \label{eq:angle}
\end{align} 
where the hat means the normalized vector and the dot means the contraction of two vectors by the AdS metric. 
\eqref{eq:angle} indicates that the angles between the tangent vector of $\gamma_1$ and $Q$ and the one of $\gamma_2$ and $Q$ are supplementary, which is consistent with the construction of geodesic in \cite{Anous:2022wqh} when $Q$ is a straight line.

The equations of motion from \eqref{eq:lenfun} give  
\be
\begin{split}\label{eq:eomx}
    x''_{\text{I}} - \frac{x'_{\text{I}}+\nu x'^3_{\text{I}}}{z_{\text{I}} } &=0\,,
    ~~~~~~~
    x''_{\text{II}} - \frac{\nu x'_{\text{II}}+ x'^3_{\text{II}}}{\nu z_{\text{II}} } =0\,,
\end{split}
\ee
where the prime are derivatives with respect to $z_\text{I}$ and $z_\text{II}$. Obviously they take the same form as the ones in AdS/BCFT \cite{Takayanagi:2011zk, Fujita:2011fp}. For the case that the bulk system is a portion of AdS$_3$ \eqref{eq:poincare}, the 
 solutions $x_\text{I}(z_{\text{I}})$ and $x_\text{II}(z_{\text{II}})$ are circle arcs passing through the points marked as $-\sigma_\text{I}$ and $z_*$ as well as $\sigma_\text{II}$ and $z_*$ in Fig. \ref{fig:exf}. 
Plugging these solutions 
into \eqref{bc:var}, the boundary condition can be written as 
\begin{align}
\label{eq:bcc}
    \begin{split}
        &\frac{z^2-\nu(\sigma_{\text{I}}+\psi_{\text{I}})(\sigma_{\text{I}}+\psi_{\text{I}}-2z\psi'_{\text{I}})}{z^2+\nu(\sigma_{\text{I}}+\psi_{\text{I}})^2} + \frac{\nu\left(\nu z^2 - (\sigma_{\text{II}}-\psi_{\text{II}})(\sigma_{\text{II}}-\psi_{\text{II}}+2z\psi'_{\text{II}})\right)}{\nu z^2+ (\sigma_{\text{II}}-\psi_{\text{II}})^2}\Bigg{|}_\text{Q} = 0\,.
    \end{split}
\end{align}
After solving \eqref{eq:bcc}, we obtain the location $z_*(\sigma_\text{I},\sigma_\text{II})$ of the extremal surface intersecting on the interface brane. Then we can obtain the entanglement entropy in terms of the geodesics length
\begin{align}
\label{eq:hee}
\begin{split}
S_\text{E}(\sigma_\text{I},\sigma_\text{II})=&\,
\frac{\mathcal{A}(\sigma_{\text{I}},\sigma_{\text{II}})}{4G}
\\
=&\,
\frac{L_\text{I}}{4G}\,
\left[
\cosh^{-1}\frac{
\frac{z^2}{\sqrt{\nu}}+
\sqrt{\nu}(\sigma_{\text{I}}+\psi_{\text{I}}(z))^2
}
{2z\epsilon_{\text{I}}}
+\nu
\cosh^{-1}\frac{
z^2\nu+(\sigma_{\text{II}}-\psi_{\text{II}}(z))^2
}
{2z\sqrt{\nu}\epsilon_{\text{II}}}
\right]\Bigg{|}_{z=z_*}\\
=&\,\frac{c_{\text{I}}}{6}
\log\frac{
\frac{z^2_*}{\sqrt{\nu}}+
\sqrt{\nu}(\sigma_\text{I}+\psi_{\text{I}}(z_*))^2
}
{z_*\epsilon_{\text{I}}}
+\frac{c_{\text{II}}}{6}
\log\frac{
z^2_*\nu+(\sigma_\text{II}-\psi_{\text{II}}(z_*))^2
}
{z_*\sqrt{\nu}\epsilon_{\text{II}}} \,,
\end{split}
\end{align}
where $z_*=z_*(\sigma_\text{I},\sigma_\text{II})$, $\epsilon_\text{I}$ and $\epsilon_\text{II}$ are UV cutoffs along the radial direction in the left and right AdS$_3$ and we have ignored the terms of order $\mathcal{O}(\epsilon^2_\text{I}),\mathcal{O}(\epsilon^2_\text{II})$.\footnote{If we demand that the two cutoffs are the same on the brane $Q$, then we have $\epsilon_\text{I}=\epsilon_\text{II}/\nu$.} In \eqref{eq:hee}, we have assumed that the extremal curves are smooth and within the bulk. There exists situations that the extremal surface passing the solution of  \eqref{eq:bcc} and the boundary point might be out of bulk of one side, e.g. for $\sigma_\text{I}\gg \sigma_\text{II}$ in the configuration (3) shown in Fig. \ref{fig:brane pair}. In this case, we could use the squashed geodesics \cite{Czech:2016nxc} for the extremal curves out of the bulk, i.e. demanding the extremal curve spanning along the interface brane instead of out of the bulk regime. We will not consider this case in our study.

Another equivalent method to obtain the boundary condition \eqref{eq:bcc} is to minimise the length of the geodesic with respect to the intersecting point between the geodesics and the interface brane.  
Suppose that the intersecting point between the extremal surface and the brane takes the 
intrinsic coordinate $(t,z)$, 
the location 
can be solved from 
\begin{align}
\label{eq:bcc-differential}
\frac{d}{dz}
\left[
\cosh^{-1}\frac{
\frac{z^2}{\sqrt{\nu}}+
\sqrt{\nu}(\sigma_{\text{I}}+\psi_{\text{I}}(z))^2
}
{z\epsilon_{\text{I}}}
+\nu
\cosh^{-1}\frac{
z^2\nu+(\sigma_{\text{II}}-\psi_{\text{II}}(z))^2
}
{z\sqrt{\nu}\epsilon_{\text{II}}}
\right]=0\,,
\end{align} 
which is equivalent to \eqref{eq:bcc}.

\subsection{Interface entropy and $g$-function}
\label{subsec:iegf}

Setting $\sigma_{\text{I}}=\sigma_{\text{II}}=\sigma$, the interface entanglement entropy can be defined for the interface field theory as 
\be\label{eq:sie}
 S_\text{iE}(\sigma)= S_{\text{E}}- \frac{1}{2} S_{\text{E}}^L- \frac{1}{2}S_{\text{E}}^R
\ee
where $S_{\text{E}}$ is the holographic entanglement entropy of the interval $[-\sigma,\sigma]$ from AdS/ICFT \eqref{eq:hee}, while $S_{\text{E}}^L$ and $S_{\text{E}}^R$ are the holographic vacuum entanglement entropy of the same interval that are obtained from the holographic CFT$_{\text{I}}$ and CFT$_{\text{II}}$ respectively, i.e. 
\be
S_{\text{E}}^L=\frac{c_{\text{I}}}{3}\log
\frac{2\sigma}{\epsilon_{\text{I}}}\,,~~~~~
S_{\text{E}}^R=\frac{c_{\text{II}}}{3}\log
\frac{2\sigma}{\epsilon_{\text{II}}}\,.
\ee
Note that similar proposal has also been used in \cite{Simidzija:2020ukv} for AdS/ICFT without scalar field. One nice feature of the above definition \eqref{eq:sie} is that when the brane is completely trivially connecting the two AdS spacetime, i.e. the case of \eqref{eq:e2sol1}, we have $S_\text{iE}(\sigma)=0$. 

Then we can define a $g$-function $g(\sigma)$ as follows 
\begin{align}
\begin{split}
\label{eq:expforg}
\log g(\sigma) =&~ S_\text{iE}(\sigma)\\
=&~\frac{c_{\text{I}}}{6}
\log\frac{
\frac{z^2_*}{\sqrt{\nu}}+
\sqrt{\nu}(\sigma+\psi_{\text{I}}(z_*))^2
}
{2z_*\sigma}
+\frac{c_{\text{II}}}{6}
\log\frac{
z^2_*\nu+(\sigma-\psi_{\text{II}}(z_*))^2
}
{2\sqrt{\nu}z_*\sigma}\,, 
\end{split}
\end{align} 
with $z_*=z_*(\sigma)$ 
determined from \eqref{eq:bcc} by setting $\sigma_{\text{I}}=\sigma_{\text{II}}=\sigma$. The $g$-function 
is finite and independent of the UV cutoffs $\epsilon_{\text{I}}$ and $\epsilon_{\text{II}}$. 

Note that $\log g(\sigma)\in (-\infty,\infty)$. When the brane has a constant tension, i.e. the solution \eqref{eq:straightline}, the interface entropy $S_\text{iE}=\log g(\sigma)$ is a constant so that the above $g$-function is also a constant \cite{Simidzija:2020ukv}.  This is expected as the induced metric on $Q$ is an AdS$_2$ slice. We will show this as the first example in the  section \ref{subsub:solI}. 
Specially, when the profiles satisfy the relations $\psi_{\text{I}}(z)\ge 0$ and $\psi_{\text{II}}(z)\le 0$,
we have $\psi_{\text{I}}(z_*)\ge 0$ and $\psi_{\text{II}}(z_*)\le 0$, then from \eqref{eq:expforg}, $\log g(\sigma)\geq 0$.

The monotonic behavior of $\log g(\sigma)$ is expected to reflect the RG flow the interface-associated degrees of freedom of the interface field theory. In Sec. \ref{sec:UVIRlimits} we will prove $g$-theorem for the profile of the case (2) in Fig. \ref{fig:brane pair}.  In Sec. \ref{sec:heeexample} we will present specific  examples demonstrating that $\log g(\sigma)$ consistently exhibits a monotonic decrease when the NEC is satisfied, which includes other cases in Fig. \ref{fig:brane pair}.\footnote{It is interesting to connect the $g$-function to the defect RG flow in the field theory, see e.g. \cite{Cuomo:2021rkm}.} 

\subsubsection{$g$-theorem}
\label{sec:UVIRlimits}
We will first prove that when the induced metric on the interface $Q$ is asymptotic to AdS$_2$ in both the UV and IR limits,  we have $\lim_{\sigma\to 0}S_\text{iE}(\sigma) \geq \lim_{\sigma\to \infty}S_\text{iE}(\sigma)$. 
Then will give a proof of $g$-theorem for the profile of case (2) in Fig. \ref{fig:brane pair}.

In the UV limit $z\to 0$, the brane is asymptotically AdS$_2$ if we assume that the brane $Q$ has the following asymptotic profile
\begin{align}
\psi_{\text{I}}(z)\simeq \gamma z^n+\cdots\,,
\end{align}
and
\begin{align}
\label{eq:psiIIprofileUV}
\psi_{\text{II}}(z)\simeq \pm
\sqrt{
\frac{1}{\nu}-\nu+\gamma^2 n^2\delta_{1n}
}\ z+\cdots\,,
\end{align}
where the dots are the higher order term in $z$. 
From the above asymptotic behaviors, expanding \eqref{eq:bcc} by assuming that $z_*$ and $\sigma$ are of same order, we have
\begin{align}
    z_* = \sqrt{\frac{\nu}{1+\gamma^2 \nu \delta_{1n}}}\ \sigma + \cdots\,.
\end{align} 
Therefore the interface entropy in the UV limit is
\begin{align}
\label{eq:ieUV}
\begin{split}
S_\text{iE}^\text{UV}=&~ \lim_{\sigma\to 0}\,S_\text{iE}(\sigma)  \\
=&~
 \frac{c_{\text{I}}}{6} \log \left(\gamma\sqrt{\nu}\delta_{1n} + \sqrt{1+\gamma^2\nu\delta_{1n}} \right) \\
&~~~~ + \frac{c_{\text{II}}}{6} \log \left( \frac{
\sqrt{1+\gamma^2 \nu \delta_{1n}}\mp\sqrt{1-\nu^2+\gamma^2 \nu \delta_{1n}}
}
{\nu} \right) + \mathcal{O}(\sigma)\,,
\end{split}
\end{align}
which is a constant in the leading term. The sign in \eqref{eq:ieUV} corresponds to the choice of sign in \eqref{eq:psiIIprofileUV}. 
When $n=1$, $S_\text{iE}^\text{UV}$ is the same as the result with trivial scalar field, which will be written out explicitly in \eqref{eq:loggvac}. When $n>1$, the left brane is perpendicular to the boundary in the near-boundary limit thus the left part contributes nothing to $S_\text{iE}^\text{UV}$, i.e. we have $S_\text{iE}^\text{UV}=\frac{c_\text{II}}{6}\log(\frac{1\mp\sqrt{1-\nu^2}}{\nu})$.

In the IR limit $z\to +\infty$, the brane is asymptotically AdS$_2$ if we assume that the brane $Q$ has the asymptotic profile 
\begin{align}
\psi_{\text{I}}(z)\simeq \psi'_{\text{I}}(+\infty)z+...\,, 
\end{align}
and
\begin{align}
\label{eq:psiIIprofileIR}
\psi_{\text{II}}(z)\simeq \pm
\sqrt{
\psi'^{2}_{\text{I}}(\infty)+\frac{1}{\nu}-\nu
}\,
z+...\,,
\end{align}
with the dots the subleading terms compared to $z$. 
Note that the above asymptotic behaviors are not the most generic ones while a special one which would lead to the IR geometry asymptotically AdS$_2$. 
From the above asymptotic behavior and \eqref{eq:bcc}, we have 
\begin{align}
\sigma \simeq 
\sqrt{
\frac{1}{\nu}+\psi'^2_{\text{I}}(+\infty)
}\,z+...\,.
\end{align}
The interface entropy in the IR limit is 
\begin{align}
\label{eq:ieIR}
\begin{split}
S_\text{iE}^\text{IR}=&\lim_{\sigma\to+\infty}S_\text{iE}(\sigma)  \\
=&~
\frac{c_{\text{I}}}{6}\log
\left(\psi'_{\text{I}}(+\infty)\sqrt{\nu}+\sqrt{1+\psi'_{\text{I}}(+\infty)^2\nu}
\right)\\
&~~~~
+
\frac{c_{\text{II}}}{6}\log
\frac
{
\sqrt{1+\psi'_{\text{I}}(+\infty)^2\nu}\mp\sqrt{1+\psi'_{\text{I}}(+\infty)^2\nu-\nu^2}
}
{\nu}\,,
\end{split}
\end{align}
where the sign corresponds to the choice of sign in \eqref{eq:psiIIprofileIR}.

When the matter field on $Q$ satisfies the NEC as discussed in Sec. \ref{sec:nec}, i.e. when $\nu\neq 1$, or the condition  $\nu= 1,\psi'_{\text{I}}=-\psi'_{\text{II}}$, we have $\psi''_\text{I}\le 0$. This indicates that $\psi'_{\text{I}}(0)\ge\psi'_{\text{I}}(+\infty)$. 
From \eqref{eq:ieUV} and \eqref{eq:ieIR}, 
we always have 
\begin{align}
\label{eq:sieadsUVIR}
S_\text{iE}^\text{UV}\ge S_\text{iE}^\text{IR}\,.
\end{align} 
When $\nu=1,\psi'_{\text{I}}=\psi'_{\text{II}}$, we have $S_\text{iE}=0$  indicating the interface is trivial. 

Note that \eqref{eq:sieadsUVIR} is a relation between the values of $g$-function at the specific UV and IR fixed points. When we are away from the UV and IR limit, the $g$-function is expected to  satisfy the $g$-theorem 
$
\frac{d}{d\sigma} S_\text{iE}(\sigma)\le 0
$ along the whole RG flow. In Sec. \ref{sec:nec} we have proved that the existence of the profile indicates the NEC. With the assumption of a given profile, now let us analyze the $g$-theorem in detail. 

From \eqref{eq:expforg}, we have 
\be
\label{eq:eqdSdsigma}
\frac{d}{d\sigma} S_\text{iE}(\sigma)=
\frac{c_\text{I}}{6}\left[
\frac{
2\nu (\sigma+\psi_\text{I}(z_*))
}{
z_*^2+\nu (\sigma+\psi_\text{I}(z_*))^2
}
+
\frac{
2\nu (\sigma-\psi_\text{II}(z_*))
}{
z_*^2\nu+(\sigma-\psi_\text{II}(z_*))^2
}
-\frac{1+\nu}{\sigma}
\right] \,  ,
\ee
where we have used $\frac{\partial S_\text{E}}{\partial z_*}=0$. 

When $\nu=1$, we consider \eqref{eq:nu=1}, i.e. $\psi_{\text{I}}(z)=-\psi_{\text{II}}(z)$. Eq. \eqref{eq:eqdSdsigma} can be simplified as\footnote{In this case, the system can be viewed as an unfolding of BCFT. It was proved in \cite{Harper:2024aku, Fujita:2011fp} that for AdS/BCFT when NEC is satisfied, the $g$-theorem $\frac{d}{d\sigma} S_\text{iE}(\sigma)\le 0
$ holds.}  
\be
\frac{d}{d\sigma} S_\text{iE}(\sigma)=-
\frac{
c_\text{I}(z_*^2-\sigma^2+\psi^2_\text{I}(z_*))
}{
3\sigma (z_*^2+(\sigma+\psi_\text{I}(z_*))^2)
}\,.
\ee
The profile of system is case (2) or (3) in Fig. \ref{fig:brane pair}, where we can always find $z^2+\psi^2_\text{I}(z_*)\ge \sigma^2$ using the convex property of the interface brane and the boundary condition \eqref{eq:perpenHEEnu=1} introduced in the next subsection, thus $\frac{d}{d\sigma} S_\text{iE}(\sigma)\le 0$ and $g$-theorem holds.

When $0<\nu<1$, using $\frac{\partial S_\text{E}}{\partial z_*}=0$ we can simplify \eqref{eq:eqdSdsigma} as
\be
\frac{d}{d\sigma} S_\text{iE}(\sigma)=
\frac{c_\text{I}}{6\sigma}\left[
-\frac{
2\nu (\sigma+\psi_\text{I}(z_*))
(\psi_\text{I}(z_*)-z\psi'_\text{I}(z_*))
}{
z_*^2+\nu (\sigma+\psi_\text{I}(z_*))^2
}
+
\frac{
2\nu (\sigma-\psi_\text{II}(z_*))
(\psi_\text{II}(z_*)-z\psi'_\text{II}(z_*))
}{
z_*^2\nu+\nu (\sigma-\psi_\text{II}(z_*))^2
}
\right]  \,  .
\ee
For the case (2) in Fig. \ref{fig:brane pair}, we have $\psi_\text{I}(0)=\psi_\text{II}(0)=0$, $\psi_\text{I}(z)\ge 0,\psi_\text{II}(z)\le 0$, and the NEC constraints $\psi''_\text{I}(z)\le 0,\psi''_\text{II}(z)\ge 0$. Then 
\be
 (\sigma+\psi_\text{I}(z_*))
(\psi_\text{I}(z_*)-z\psi'_\text{I}(z_*))\ge 0\,,\ \ ~~~
(\sigma-\psi_\text{II}(z_*))
(\psi_\text{II}(z_*)-z\psi'_\text{II}(z_*))\le 0 
\, ,
\ee
thus $\frac{d}{d\sigma} S_\text{iE}(\sigma)\le 0$, i.e. the $g$-theorem holds. For other cases in Fig. \ref{fig:brane pair}, it is not obvious to prove the $g$-theorem. 
Moreover, the possible existence of squashed geodesics makes the proof more complicated. Instead we will provide some examples in Sec. \ref{sec:heeexample}, 
to show that it is indeed satisfied for the specific examples we considered.

\subsection{BCFT limits of holographic entanglement entropy}
\label{subsec:bcftlimitsofhee}

The BCFT limit of AdS/ICFT was discussed in Sec. \ref{subsec:bcftlimit} and there are two different BCFT limits, i.e. $\nu\to 0$ and $\nu= 1$. Here we take these two BCFT limits to the entanglement entropy and the interface entropy. 

Let us first study the limit $\nu\to 0$. In this case,  
the BCFT limit of the geodesic can be found by solving the boundary condition \eqref{eq:bcc}. Similar to the discussion in Sec. \ref{subsec:bcftlimit}, it is more convenient to work in the $t, u$ coordinate on $Q$. We first  
rewrite \eqref{eq:bcc} in $u$ coordinate
\begin{align}
\label{eq:bccu}
    \begin{split}
        &\frac{u^2-(\sigma_{\text{I}}+\tilde\psi_{\text{I}})(\sigma_{\text{I}}+\tilde\psi_{\text{I}}-2u\tilde\psi'_{\text{I}})}{u^2+(\sigma_{\text{I}}+\tilde\psi_{\text{I}})^2} + \nu \frac{\nu^2 u^2 - (\sigma_{\text{II}}-\tilde\psi_{\text{II}})(\sigma_{\text{II}}-\tilde\psi_{\text{II}}+2u\tilde\psi'_{\text{II}})}{\nu^2 u^2+ (\sigma_{\text{II}}-\tilde\psi_{\text{II}})^2}\Bigg{|}_\text{Q} = 0\,,
    \end{split}
\end{align}
where the conventions in Sec. \ref{subsec:bcftlimit} have been used and the prime denotes the derivative with respect to $u$.
In the limit $\nu\to 0$, it becomes
\begin{align}
\label{eq:bcbcft}
    \frac{u^2-(\sigma_{\text{I}}+\tilde\psi_{\text{I}})(\sigma_{\text{I}}+\tilde\psi_{\text{I}}-2u\tilde\psi'_{\text{I}})}{u^2+(\sigma_{\text{I}}+\tilde\psi_{\text{I}})^2}\Bigg{|}_\text{Q} = 0\,.
\end{align}

Using the fact that the normalized tangent vector of $\gamma_1$ and the brane take the following form in $u$ coordinate from \eqref{eq:vectorvv} and \eqref{eq:vectorww},  
\be
\hat{V}_1= \frac{2(\tilde\psi_\text{I}+\sigma_{\text{I}})}{(\tilde\psi_\text{I}+\sigma_{\text{I}})^2+u^2} \left(-u, ~\frac{(\tilde\psi_\text{I}+\sigma_{\text{I}})^2-u^2}{2(\tilde\psi_\text{I}+\sigma_{\text{I}})}\right)\,,
~~~~~~
\hat{W}_1=\frac{1}{\sqrt{\tilde\psi'^2_\text{I}+1}} \left(\tilde\psi'_\text{I}, ~1\right)\,,
\ee
it is straightforward to prove that 
\eqref{eq:bcbcft} is equivalent to 
\begin{align}
    \hat{V}_1 \cdot \hat{W}_1  =0\,.
\end{align}
Therefore \eqref{eq:bcbcft} is exactly the boundary condition of minimal surface in AdS/BCFT \cite{Takayanagi:2011zk}. Given the solution $u_*$ of \eqref{eq:bcbcft}, we can calculate the entanglement entropy in AdS/BCFT 
\begin{align}
\label{eq:eebcft}
   S_\text{E} &= \frac{c_{\text{I}}}{6}
\log\frac{
u^2_*+(\sigma_\text{I}+\tilde\psi_{\text{I}}(u_*))^2
}
{u_*\epsilon_{\text{I}}}\,.
\end{align}
The interface entanglement entropy or $g$-function 
takes the following form in AdS/BCFT, 
\begin{align}\label{eq:gfunbcft}
\begin{split}
\log g(\sigma) =&~ S_\text{iE}(\sigma)
=~\frac{c_{\text{I}}}{6}
\log\frac{
u^2_*+
(\sigma+\tilde\psi_{\text{I}}(u_*))^2
}
{2u_*\sigma}\,. 
\end{split}
\end{align} 

In the case of trivial scalar field with $\phi=0, V=T$, the configuration for the ICFT is \eqref{eq:straightline}.  From \eqref{eq:bcftlimit2} the BCFT limit is given by  
\begin{align}
\tilde{\psi}_\text{I} =  \frac{L_\text{I} T }{\sqrt{1-L^2_\text{I} T^2}}\, u\,,~~~~
\tilde{\psi}_\text{II} = \pm
\frac{1}{\sqrt{1-L^2_\text{I} T^2}}\, u
+\mathcal{O}(\nu)
\,.
\end{align}
From \eqref{eq:bcbcft}, \eqref{eq:eebcft} and \eqref{eq:gfunbcft}, the entanglement entropy takes the following form 
\begin{align}
    \begin{split}        
        u_* &= \sqrt{1-L^2_\text{I} T^2} \sigma\,,\\
        S_\text{E} &=\frac{c_{\text{I}}}{6}\log\frac{2 \sigma}{\epsilon_\text{I}}+\log g
        = \frac{c_{\text{I}}}{6} \log 
\left( 
\sqrt{\frac{1+L_\text{I} T}{1-L_\text{I} T} }\,\frac{2 \sigma}{\epsilon_\text{I}}
\right)\,,\\
        \log g &= \frac{c_{\text{I}}}{12} \log \left(\frac{1+L_\text{I} T}{1-L_\text{I} T}  \right)\,.
    \end{split}
\end{align}
This is precisely the result obtained in \cite{Takayanagi:2011zk}. For nontrivial scalar field case, boundary entropy is not a constant anymore \cite{Kanda:2023zse} and the results depend on the detailed profile of the brane $Q$.

Another type of BCFT limit is the $\nu= 1$ limit. Noticing that now we have $z=u_\text{I}=u_\text{II}=u$ and we will use $z$-coordinate for convenient. We set $\sigma_\text{I}=\sigma_{\text{II}}=\sigma$ and from \eqref{eq:nu=1} we also set $\psi_{\text{I}}(z)=-\psi_{\text{II}}(z)=\varphi(z)$. The boundary condition \eqref{bc:var} becomes
\begin{align}\label{eq:bc:nu1}
\frac{
x'_{\text{I}}(z)\varphi(z)+1
}{
\sqrt{
x'^2_{\text{I}}(z)+1
}
}
+
\frac{
-x'_{\text{II}}(z)\varphi(z)+1
}{
\sqrt{
x'^2_{\text{II}}(z)+1
}
}
\Bigg{|}_\text{Q}=0\,. 
\end{align}
Note that when $\nu=1$, from \eqref{eq:eomx} the equations for $x_{\text{I}}$ and $x_{\text{II}}$ are the same, so they should have the same solution. From the $Z_2$ folding symmetry between the left part and right part of the system, we have $x'_{\text{I}}=-x'_{\text{II}}=x'$ on the brane. Then \eqref{eq:bc:nu1} becomes
\begin{align}\label{eq:bcnu1}
2\frac{
x'(z)\varphi(z)+1
}{
\sqrt{
x'^2(z)+1
}
}\Bigg{|}_\text{Q}=0\,. 
\end{align}
Repeating the analysis below \eqref{bc:var}, \eqref{eq:bcnu1} indicates that
\begin{align}
\label{eq:perpenHEEnu=1}
    \hat{V}_1 \cdot \hat{W}_1 =  \hat{V}_2 \cdot \hat{W}_2 =0\,,
\end{align}
which is precisely the perpendicular boundary condition of the minimal surface on the EOW brane in AdS/BCFT. We can also use \eqref{eq:eomx} to write \eqref{eq:bcnu1} in the following form
\begin{align}
    \frac{z^2-(\sigma+\varphi)(\sigma+\varphi-2z\varphi')}{z^2+(\sigma+\varphi)^2}\Bigg{|}_\text{Q} =0\,,
\end{align}
which is just a rewrite of \eqref{eq:bcc} by setting $\nu=1$, $\sigma_\text{I}=\sigma_{\text{II}}=\sigma$ and $\psi_{\text{I}}(z)=-\psi_{\text{II}}(z)=\varphi(z)$. Note that this is also the same as \eqref{eq:bcbcft} in the $\nu\to 0$ limit. Similarly, after solving above equation to obtain the intersecting point $z_*$, we can calculate EE and $g$-function and they just have the form of \eqref{eq:eebcft} and \eqref{eq:gfunbcft} by identifying $u_* \to z_*$ and $\tilde\psi_\text{I}(u_*) \to \varphi(z_*)$.

\subsection{Case studies of several examples}
\label{sec:heeexample}

In this subsection, we present six distinct examples with different monotonic profiles of the interface brane $Q$. An example with non-monotonic profile is shown in appendix \ref{app:a}. With a nontrivial scalar field, the interface field theory is expected to deviate from a fixed point. The bending of the interface brane reflects the RG flow of interface, revealing numerous significant physical phenomena.

Before proceeding, let us summarize our findings based on the specific examples examined: 
\begin{enumerate}[(1)]
\item  In cases where the induced metric on the interface brane $Q$ flows from a UV AdS to an IR AdS, the scalar potential evolves from  
a locally maximal in UV to a globally minimal in IR, as shown in the solutions  II, III, V. In solution I where the scalar field is trivial, the induced metric is a pure AdS$_2$. 
\item When the induced metric on the interface brane $Q$ flows from a UV AdS to an IR flat spacetime, the scalar potential evolves from a locally maximal in UV 
(solution VI or the solution in Appendix \ref{app:a}) or a locally minimal in UV (solution IV with $n>3$) to a globally minimal in IR.     
\item The relation $\partial_\phi V(\phi)|_{\phi_\text{UV}}
=\partial_\phi V(\phi)|_{\phi_\text{IR}}=0$ is always satisfied for all examples. 
\item When the scalar potential exhibits non-monotonic behavior, i.e. the solutions IV and V, multiple extremal surfaces may exist for the interval we studied. Consequently, a ``second order phase transition" occurs for the interface entropy $S_\text{iE}(\sigma)$ as $\sigma$ increases. 
\item  When the induced metric is asymptotically flat in IR, i.e. the solutions VI and IV and the solution in Appendix \ref{app:a}, the interface entropy $S_\text{iE}(\sigma\to \infty)$ goes to $-\infty$. Notably, in such cases, the scalar field could be divergent (or finite) if the interface brane spans to $z\to \infty$ (or a finite value) in the IR limit. 
\item  The $g$-theorem is always consistently satisfied, i.e. $S_\text{iE}(\sigma)$ monotonically decreases when $\sigma$ increases. This is consistent with the analytical proof in Sec. \ref{sec:UVIRlimits}.
\end{enumerate}

\subsubsection{Solution I with trivial scalar field}
\label{subsub:solI}

The simplest example is the case with trivial scalar field, i.e. $\phi=0, V=T$. This  case was studied in \cite{Anous:2022wqh, Simidzija:2020ukv}. Now the interface $Q$ is a straight line and the solution is given by \eqref{eq:straightline} and the tension is bounded by \eqref{eq:tensionreg0T}. The induced metric on the interface brane is AdS$_2$. 

In this case, the intersection point between the minimal surface and the brane is
\begin{align}
    z_* = \sqrt{\frac{\nu}{1+\gamma^2 \nu}}\ \sigma\,.
\end{align}
Then we can calculate the entanglement entropy
\begin{align}\label{eq:Svac}
    S_{\text{E}} = \frac{c_{\text{I}}}{6} \log  
    \frac{2\sigma}{\epsilon_{\text{I}}} 
    + \frac{c_{\text{II}}}{6} \log 
    \frac{2\sigma}{\epsilon_{\text{II}}} 
    +\log g
    \,,
\end{align}
with the $g$-function 
\begin{align}
\label{eq:loggvac}
    \log g = \frac{c_{\text{I}}}{6} \log \left(\gamma\sqrt{\nu}+ \sqrt{1+\gamma^2\nu} \right) + \frac{c_{\text{II}}}{6} \log \left( \frac{
\sqrt{1+\gamma^2 \nu}\mp\sqrt{1+\gamma^2 \nu-\nu^2}
}
{\nu} \right)\,,
\end{align}
where the signs $\mp$ in \eqref{eq:loggvac} corresponds to the choices of sign $\pm$ in \eqref{eq:straightline}. Note that the $g$-function is a constant and independent of the interval length $\sigma$. This is consistent with the fact that the induced metric on the interface $Q$ is an AdS$_2$ and the interface field theory on the boundary is expected to be a CFT. Notably, when $\nu=1$ and $\psi_\text{I}=\psi_\text{II}=\gamma z$ we have $\log g=0$ which is precisely the case that the interface $Q$ does not play any role in the construction. 

We can also perform the coordinate transformation on Poincare AdS coordinate \eqref{eq:poincare} via 
\begin{align}
    u_{\text{A}}=\frac{y_{\text{A}}}{\cosh(\rho_{\text{A}}/L_{\text{A}})}\,, \qquad x_{\text{A}} = y_{\text{A}} \tanh(\rho_{\text{A}}/L_{\text{A}})\,, ~~~\text{A}=(\text{I}, \text{II})
\end{align}
then the interface entropy can be expressed as 
\begin{align}
\label{eq:solIie}
S_{\text{iE}}=\log g(\sigma)=
\frac{1}{4G}\left(\rho^*_{\text{I}}+\rho^*_{\text{II}}\right)\,,
\end{align}
where $\rho^*_{\text{I}},\rho^*_{\text{II}}$ are
\begin{align}
\rho^*_{\text{I}}=L_{\text{I}}\log
(\lambda \sqrt{\nu}+\sqrt{1+\lambda^2 \nu}), \ \ \ 
\rho^*_{\text{II}}=L_{\text{II}}\log
\frac
{
\sqrt{1+\lambda^2 \nu}\mp\sqrt{1+\lambda^2 \nu-\nu^2}
}
{\nu}.
\end{align}
which parameterize the positions of the branes. This is the same as the result in \cite{Anous:2022wqh, Simidzija:2020ukv}. 

\subsubsection{Solution II}
\label{subsub:sol2}

\begin{figure}[h!]
\begin{center}
\includegraphics[width=0.4\textwidth]{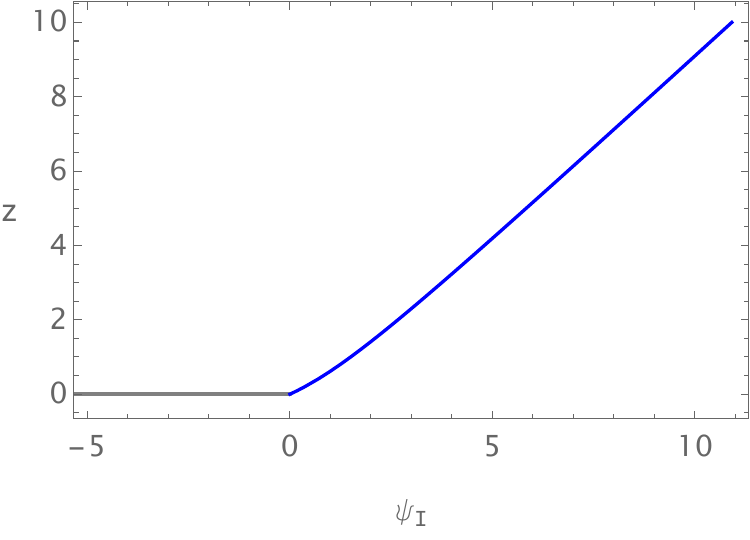}
~~~~~
\includegraphics[width=0.4\textwidth]{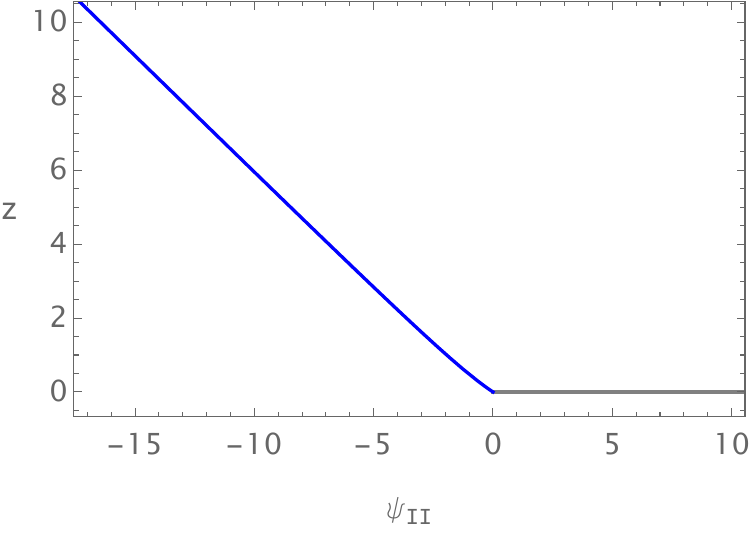}
\end{center}
\vspace{-0.3cm}
\caption{\small The profiles of the interface brane $Q$ for the example \eqref{eq:psiIsol2} with $a=2$, $b=1$, $L_\text{I}=1$, and $\nu=0.5$. 
}
\label{fig:psiz2}
\end{figure}

The second example of the interface is described by 
\begin{align}
\label{eq:psiIsol2}
\psi_\text{I}(z)=&\frac{az+bz^2}{1+z}\,. 
\end{align}
The NEC \eqref{eq:necpsi1} imposes the constraint $a \geq b$. The solution for $\psi_\text{II}$ can be represented using elliptic integrals, which are quite complicated. Fig. \ref{fig:psiz2} illustrates an example of the brane configuration when $a=2$, $b=1$, $L_\text{I}=1$, and $\nu=1/2$, with the minus sign selected for $\psi_\text{II}$. It belongs to the case (2) classified in Fig. \ref{fig:brane pair}. 

In the UV limit $z\to 0$ and the IR limit $z\to\infty$, $\psi_\text{I}$ behaves as
\begin{align}
\begin{split}
\psi_\text{I}(z)=&~az+(b-a)z^2+\mathcal{O}(z^3)\,,\qquad (z\to 0)\,,\\
\psi_\text{I}(z)=&~bz-(b-a)+\mathcal{O}\left( \frac{1}{z} \right)\,,\qquad
(z\to \infty)\,.
\end{split}
\end{align}
Therefore, we can approximate it with a straight line in both the UV and IR regimes, and it exhibits asymptotic AdS$_2$ behavior in both the UV and IR. 
Therefore the profile can be though of a precise example in Sec. \ref{sec:UVIRlimits}. 
The nature of fixed points can also be confirmed by examining equation \eqref{eq:lam0}, where we find
\begin{align}
\begin{split}
\lambda(z)=&-\frac{1}{L^2_\text{I} (1+a^2\nu)}+\mathcal{O}(z) \,,~~~~~\qquad(z\to 0)      \,,\\
\lambda(z)=&-\frac{1}{L^2_\text{I} (1+b^2\nu)}
+\mathcal{O}\left(\frac{1}{z^2}\right)  \,,\qquad (z\to\infty)\,.
\end{split}
\end{align}
The leading term is a negative constant in both the UV and IR regimes, indicating that the interface brane asymptotically approaches AdS$_2$ in these regions. Interestingly, the condition derived from the NEC is equivalent to requiring that the effective AdS$_2$ radius on the interface $Q$ satisfy $L_Q^\text{UV}\geq L_Q^\text{IR}$ for $b>0$.\footnote{The relation is no longer true for $a<0$. Nevertheless, we have checked several examples and find that the $g$-theorem is still satisfied.}  The ICFT dual to the interface is anticipated to flow from a UV fixed point to an IR fixed point. 

The interface entropy at both the UV and IR fixed points can be calculated using equation \eqref{eq:loggvac}
\begin{align}
\begin{split}
    S^\text{UV}_\text{iE} &= \frac{c_{\text{I}}}{6} \log \left(a\sqrt{\nu}+ \sqrt{1+a^2\nu} \right) + \frac{c_{\text{II}}}{6} \log \left( \frac{
\sqrt{1+a^2 \nu}\mp\sqrt{1+a^2 \nu-\nu^2}
}
{\nu} \right)\,, \\
S^\text{IR}_\text{iE} &= \frac{c_{\text{I}}}{6} \log \left(b\sqrt{\nu}+ \sqrt{1+b^2\nu} \right) + \frac{c_{\text{II}}}{6} \log \left( \frac{
\sqrt{1+b^2 \nu}\mp\sqrt{1+b^2 \nu-\nu^2}
}
{\nu} \right)\,.
\end{split}
\end{align} 
For the condition $a\geq b$, we have
\begin{align}
S^\text{UV}_\text{iE}\geq S^\text{IR}_\text{iE}\,,
\end{align}
as discussed already in Sec. \ref{sec:UVIRlimits}. 
A special case is that the $g$-function becomes a constant thus $S^\text{UV}_\text{iE}=S^\text{IR}_\text{iE}$ when $a=b$ and the brane becomes $\psi_\text{I}(z)=az$ which is precisely the example discussed in Sec. \ref{subsub:solI}.

\begin{figure}[h!]
\begin{center}
\includegraphics[width=0.31\textwidth]{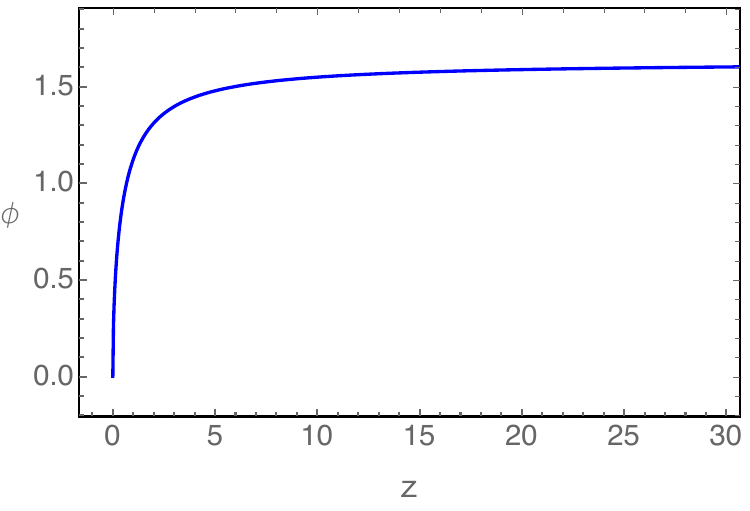}
\includegraphics[width=0.32\textwidth]{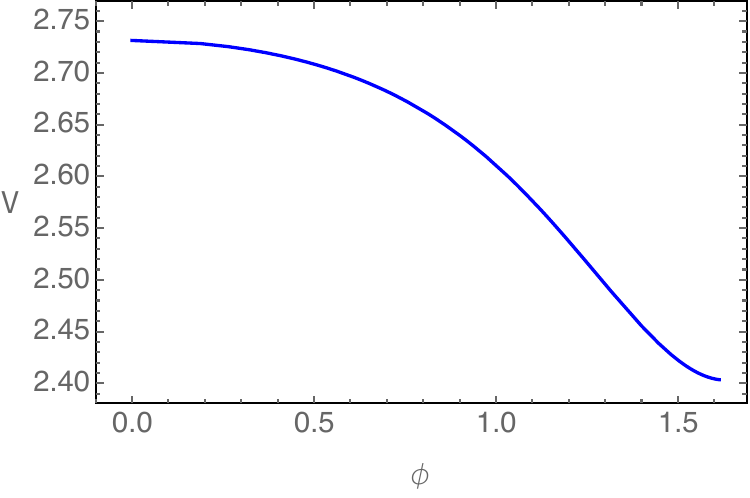}
\includegraphics[width=0.325\textwidth]{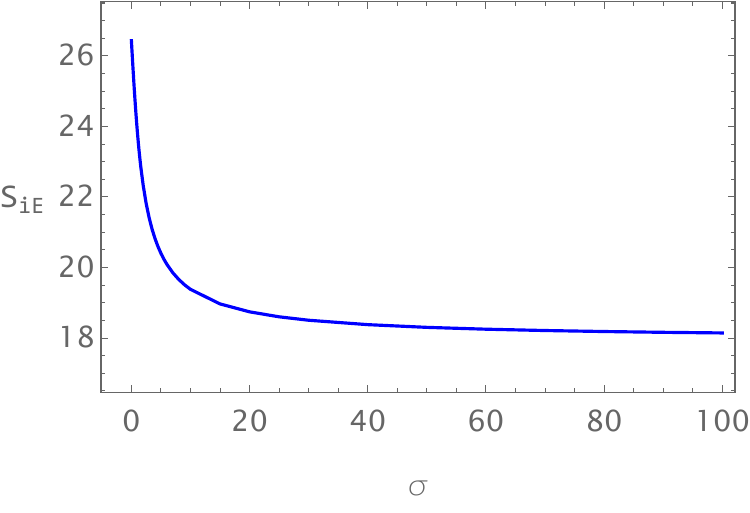}
\end{center}
\vspace{-0.3cm}
\caption{\small 
The $\phi(z)$, $V(\phi)$ and the 
interface entropy for the profile \eqref{eq:psiIsol2}. Here we set $a=2\,, b=1\,, L_\text{I}=1\,, \nu=0.5$. 
}
\label{fig:case2-2}
\end{figure}

We show an example of this solution in Fig. \ref{fig:case2-2}, where the parameters are set to be  $a=2$, $b=1$, $L_\text{I}=1$, and $\nu=1/2$. The left two plots are for $\phi(z)$ and $V(\phi)$. The analytical expressions for the scalar field and its potential in the IR and UV limits are shown in appendix \ref{app:b}. The right plot is for the interface entropy. Obviously the $g$-theorem is satisfied. 

\subsubsection{Solution III}
\label{subsub:sol3}

\begin{figure}[h!]
\begin{center}
\includegraphics[width=0.42\textwidth]{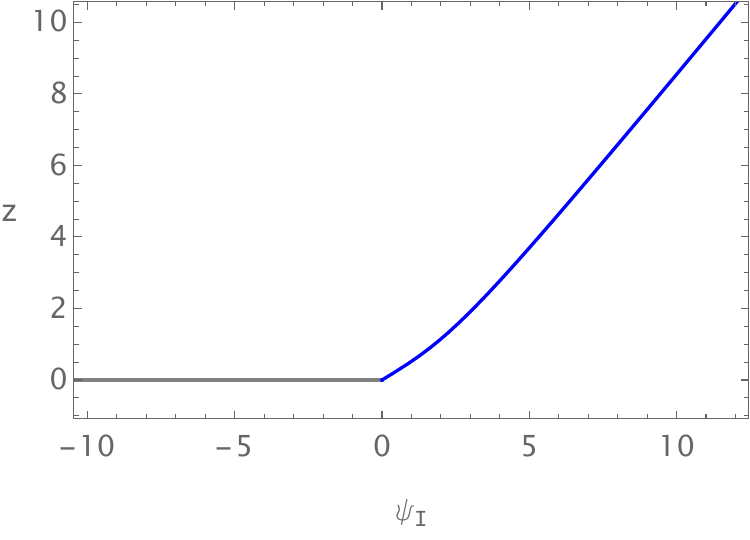}
~~
\includegraphics[width=0.42\textwidth]{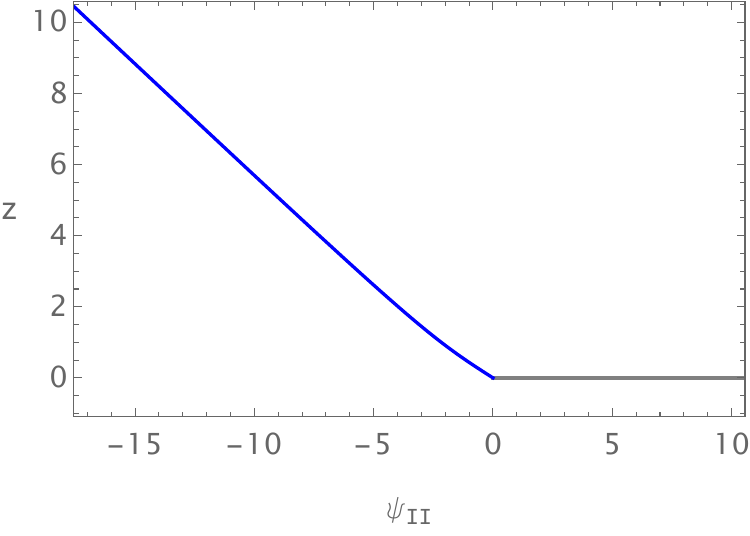}
\end{center}
\vspace{-0.3cm}
\caption{\small The profile of the interface brane $Q$ for the example \eqref{eq:psiIsol3}. Here we choose $a= b=c=1\,, \nu=0.5\,, L_\text{I}=1$. 
}
\label{fig:psitan}
\end{figure}

The third example of the interface brane is described by 
\begin{align}
\label{eq:psiIsol3}
\psi_\text{I}(z)=a\arctan (bz)+cz\,,
\end{align}
where $a,b,c$ are constants. We choose $b>0$ since its sign can be moved out. 
The NEC \eqref{eq:necpsi1} demands 
$
ab\ge 0.
$ 
The solution for $\psi_\text{II}$ can only be obtained numerically. 
In Fig. \ref{fig:psitan}, we show an example of the brane configuration. Similar to the discussion of the previous example, the minus sign for $\psi_\text{II}$ is chosen and it belongs to the case (2) classified in Fig. \ref{fig:brane pair}.

$\psi_\text{I}(z)$ has the following asymptotic behaviour,
\begin{align}
\begin{split}
\psi_\text{I}(z)=&~ (ab+c)z-\frac{ab^3}{3}z^3+\mathcal{O}(z^5)\,,~~~~~~~~~~~~~~ (z\to 0)     \,  ,\\
\psi_\text{I}(z)=&~  cz+\frac{\pi a}{2}-\frac{a}{bz}+\frac{a}{3b^3z^3}+\mathcal{O}\left(\frac{1}{z^5}\right)\,,~~~~ (z\to\infty)\,.
\end{split}
\end{align}
The induced metric on the brane and the Ricci tensor satisfy the relation $R_{Q\mu\nu}=\lambda(z) h_{\mu\nu}$, where
\begin{align}
\begin{split}
\lambda(z)=&~-\frac{1}{(1+(ab+c)^2\nu)L^2_\text{I}}+\mathcal{O}(z^4) \,,~~~  (z\to 0)      \, ,\\
\lambda(z)=&~-\frac{1}{(1+c^2\nu)L^2_\text{I}}+\mathcal{O}\left(\frac{1}{z^2}\right)  \,,~~~~~~~~ (z\to\infty)\,.
\end{split}
\end{align}
This indicates that the brane is also asymptotically AdS in the UV and IR limits.

The entanglement entropy can also computed from \eqref{eq:loggvac}. Following the discussion in Sec. \ref{sec:UVIRlimits}, once the NEC is satisfied, we have 
\be
S^\text{UV}_\text{iE}\ge S^\text{IR}_\text{iE}\,.
\ee
A special case is that the $g$-function becomes a constant thus $S^\text{UV}_\text{iE}=S^\text{IR}_\text{iE}$ when $ab=0$ and the brane becomes $\psi_\text{I}(z)=cz$, i.e. the solution in Sec. \ref{subsub:solI}.

\begin{figure}[h!]
\begin{center}
\includegraphics[width=0.31\textwidth]{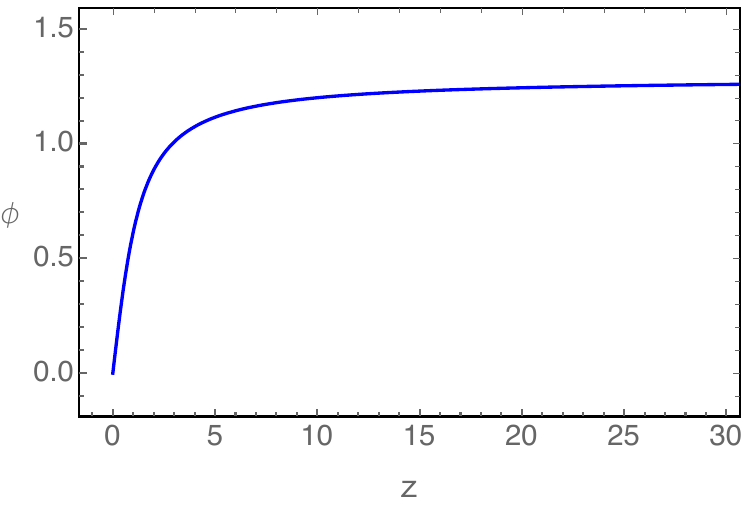}
\includegraphics[width=0.32\textwidth]{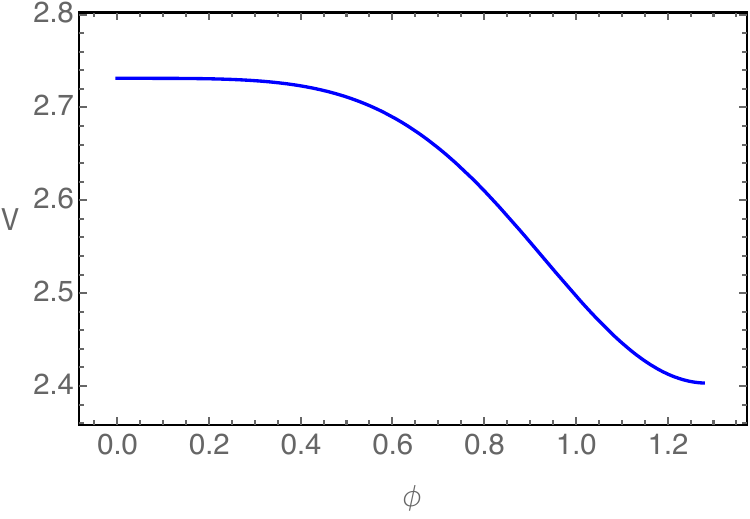}
\includegraphics[width=0.33\textwidth]{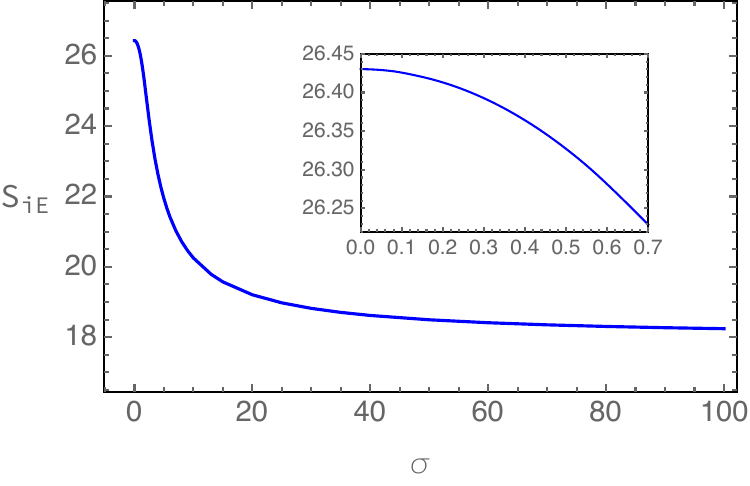}
\end{center}
\vspace{-0.3cm}
\caption{\small The $\phi(z)$, $V(\phi)$ and the interface entropy for the profile \eqref{eq:psiIsol3}. Here we chose
$a= b=c=1\,, \nu=0.5\,, L_\text{I}=1$
and we have $\phi_0\simeq 1.28$.
}
\label{fig:Stan}
\end{figure}

In Fig. \ref{fig:Stan}, we show an example of the scalar field, the potential, as well as 
the interface entropy $S_\text{iE}$. Obviously the $g$-theorem holds. Different from the behavior of $S_\text{iE}$ in Fig. \ref{fig:case2-2}, here when  $\sigma\to 0$, we have $\partial_\sigma S_\text{iE}\to 0$. 

\subsubsection{Solution IV}
\label{subsub:sol4}

We have considered the curved interface branes that can be approximated as straight lines in both the UV and IR regimes in Sec. \ref{subsub:sol2} and Sec. \ref{subsub:sol3}. Here, we will study a more general solution that is asymptotically AdS$_2$ in the UV regime, as shown in \eqref{eq:gensolads}.

The profile $\psi_{\text{I}}$ of the interface brane   has the form
\begin{align}
\label{eq:psiIsol4}
    \psi_{\text{I}}=\gamma\  z^n
\end{align}
where $n\geq 1$ and $\gamma$ are constants. When $n=1$, this exactly matches the case we studied in Sec. \ref{subsub:solI} and there is no constraint on $\gamma$ from NEC. In this subsection we focus on the regime $n>1$, where the NEC further constraints $\gamma<0$. Evaluating \eqref{eq:lam0} we obtain 
\begin{align}
    \lambda(z)=&-\frac{1}{L^2_\text{I}(1+\gamma^2 \nu \delta_{n1})}+\cdots \,, \qquad ~~(z\to 0)    \,,\\
\lambda(z)=&-\frac{1}{n(L^2_\text{I}\gamma^2\nu)z^{2(n-1)}}+\cdots  \,,\qquad (z\to\infty)\,.
\end{align}
This implies that the brane asymptotically approaches AdS$_2$ in the UV regime. However, the condition $\lambda(z)\to 0$ as $z\to\infty$ indicates that it becomes flat in the deep IR. 

\begin{figure}[h!]
\begin{center}
\includegraphics[width=0.4\textwidth]{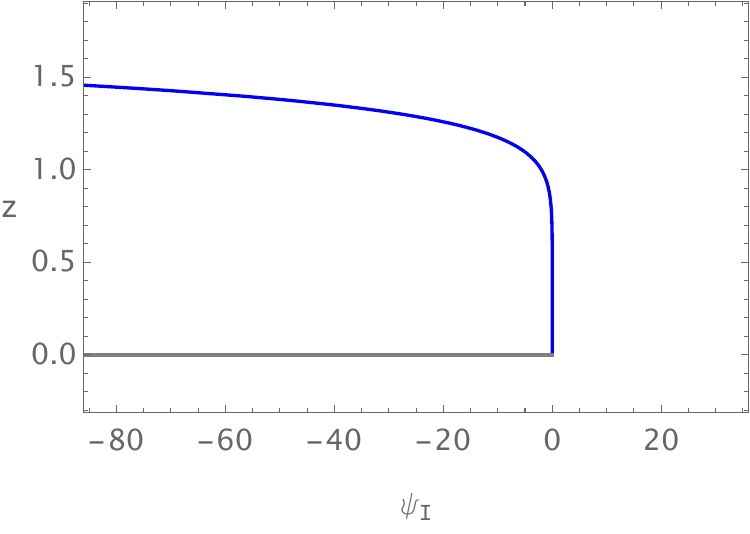}
~~
\includegraphics[width=0.4\textwidth]{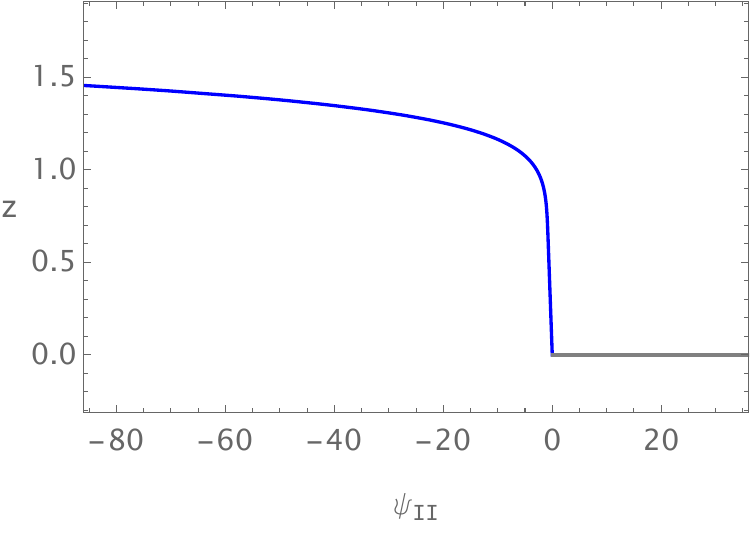}
\end{center}
\vspace{-0.3cm}
\caption{\small  Plots of configurations of $\psi_\text{I}$ of \eqref{eq:psiIsol4} and $\psi_\text{II}$ of \eqref{eq:psiIIsol4}. Here we have set $\gamma=-2$, $n=10$, $L_\text{I}=1$ and $\nu=0.5$. 
}
\label{fig:psiz10}
\end{figure}

The solution of $\psi_{\text{II}}$ can be obtained as
\begin{align}
\label{eq:psiIIsol4}
    \psi_{\text{II}}=\pm \sqrt{\frac{1-\nu^2}{\nu}}\  z \ _2F_1\left( -\frac{1}{2},\  \frac{1}{2n-2},\  \frac{2n-1}{2n-2},\  -\frac{n^2\gamma^2 \nu}{1-\nu^2}\  z^{2n-2} \right)\,,
\end{align}
where $_2F_1$ denotes the hypergeometric function. Fig. \ref{fig:psiz10} illustrates an example of the brane configuration when $\gamma=-2$, $n=10$, $L_\text{I}=1$, and $\nu=1/2$, with the minus sign selected for $\psi_\text{II}$. It belongs to the case (4) classified in Fig. \ref{fig:brane pair}. 

\begin{figure}[h!]
\begin{center}
\includegraphics[width=0.4\textwidth]{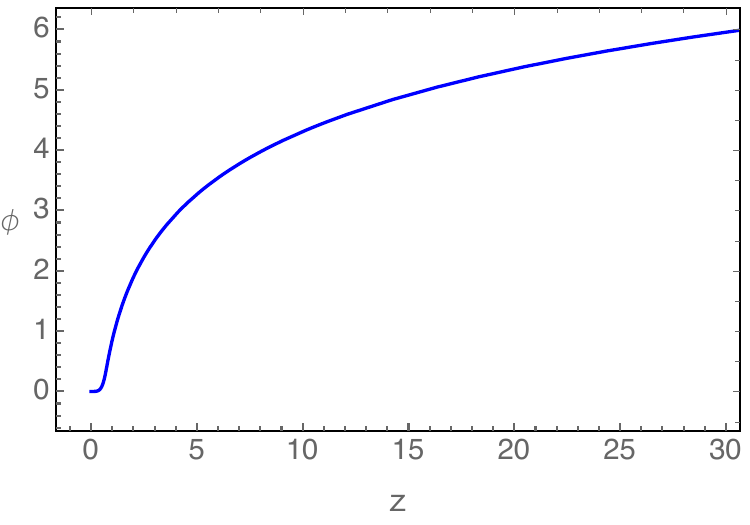}
~~
\includegraphics[width=0.41\textwidth]{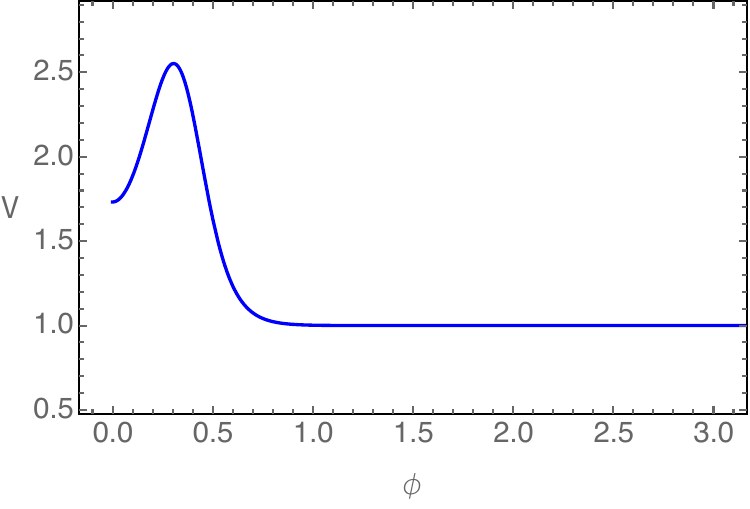}
\end{center}
\vspace{-0.3cm}
\caption{\small The typical configuration of the scalar field $\phi$ and the potential $V$. Here we have set $\gamma=-2$, $n=10$, $L_\text{I}=1$ and $\nu=0.5$.}
\label{fig:z10}
\end{figure}

Fig. \ref{fig:z10} shows a typical example of the scalar field $\phi$ and the potential $V$ with the same parameters in Fig. \ref{fig:psiz10}. The scalar field $\phi$ is divergent when $z\to\infty$, as also shown analytically in \eqref{eq:appB-solIV}. The potential is non-monotonic: $V$ has a local minimum at $\phi=0$ and a global minimum for large $\phi$, while it has a maximum at a finite value of $\phi$.

\begin{figure}[h!]
\begin{center}
\includegraphics[width=0.32\textwidth]{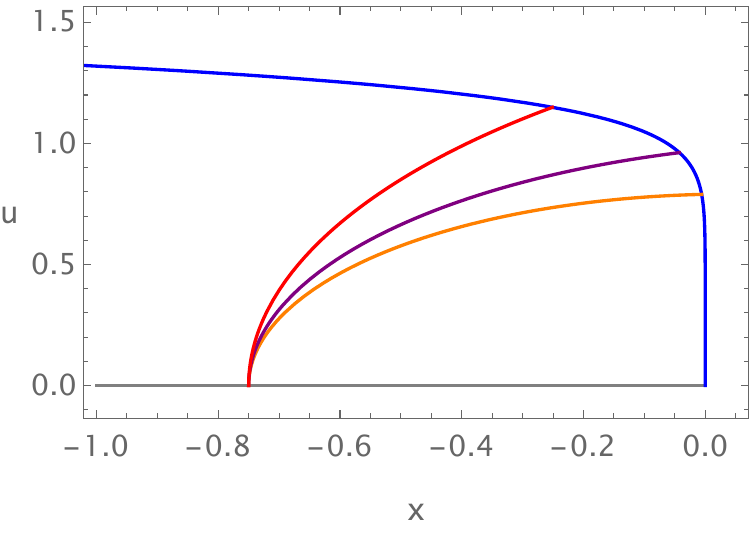}
\includegraphics[width=0.32\textwidth]{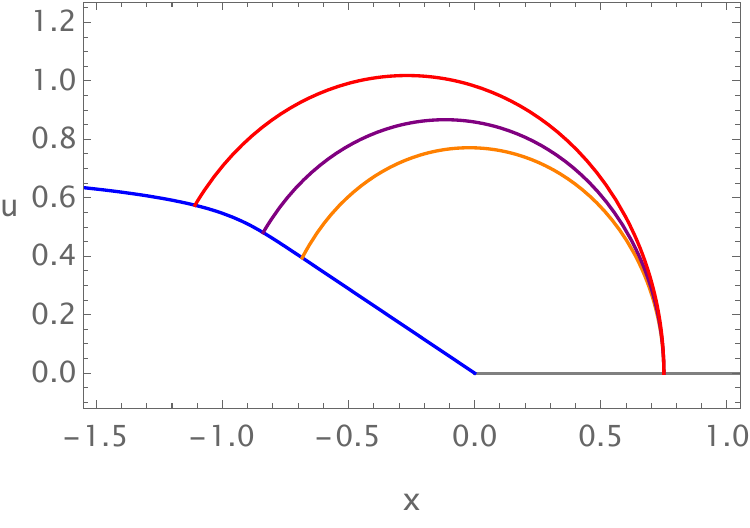}
\includegraphics[width=0.32\textwidth]{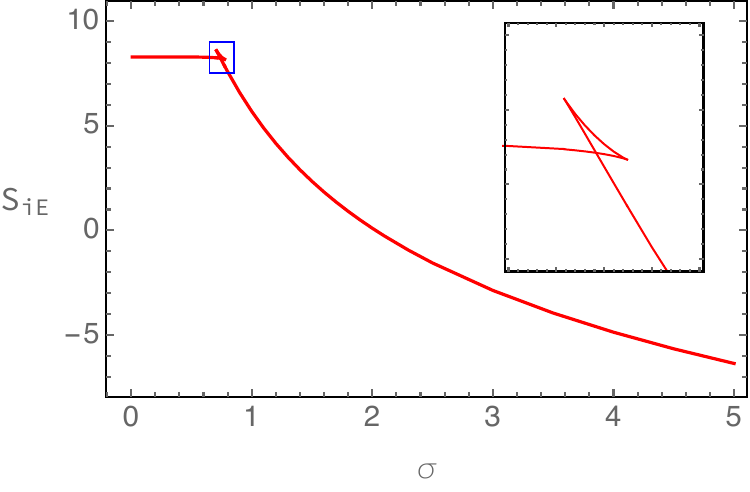}
\end{center}
\vspace{-0.3cm}
\caption{\small The left 
two figures are the typical configurations of the extremal surface for $\sigma=0.75$. The right figure is the 
interface entropy as a function of $\sigma$ for the profile \eqref{eq:psiIsol4} and \eqref{eq:psiIIsol4} with the minus sign selected. Here we have set $\gamma=-2$, $n=10$, $L_\text{I}=1$ and $\nu=0.5$. 
}
\label{fig:Sz4}
\end{figure}

Numerically, we find that the non-monotonic nature of the scalar potential can lead to the existence of multiple extremal surfaces for the specific spatial regime we have considered. More precisely, for the profile $\psi_{\text{II}}<0$ that we analyzed in Fig. \ref{fig:psiz10}, there are three possible intersection points $z_*$ within the interval $(0.708,0.774)$ of $\sigma_{\text{I}}=\sigma_{\text{II}}=\sigma$, as illustrated explicitly in the left two plots of Fig.  \ref{fig:Sz4}. Furthermore, the right plot in Fig. \ref{fig:Sz4} show 
the interface entropy $S_\text{iE}=\log g$ as a function of $\sigma$. A notable observation is that  there is a first-order phase transition for both the entanglement entropy and the boundary entropy when we increase $\sigma$. This transition is a result of the competition among three extremal curves. We anticipate that this result stems from the non-monotonic behavior of the scalar field's potential. Moreover, the interface entropy is monotonically decreasing and satisfies the $g$-theorem. 

\subsubsection{Solution V}
\label{subsub:sol5}

The studies in Sec. \ref{subsub:sol4} demonstrate the occurrence of multiple extremal curves in scenarios where the scalar potential exhibits a non-monotonic behavior. To further support this observation, we will now delve into another intriguing example. We consider 
the following profile for the interface brane 
\begin{align}
\label{eq:psiIsol5}
    \psi_{\text{I}}=\gamma\  a^{-z}-\gamma\,,
\end{align}
where $\gamma, a$ are constants. The NEC requires $\gamma<0$. We set $a>1$ to ensure $\psi_{\text{I}}\geq 0$. After evaluating \eqref{eq:lam0}, we find 
\begin{align}
\begin{split}
    \lambda(z)=&-\frac{1}{L^2_\text{I}(1+\gamma^2 \nu (\log a)^2)}+\cdots\,,~~~  (z\to 0)\,,\\
\lambda(z)=&-\frac{1}{L^2_\text{I}}+\cdots\,,~~~~~~~~~~~~~~~~~~\qquad (z\to\infty)\,.
\end{split}
\end{align}
This implies that the brane asymptotically approaches AdS$_2$ in both the UV and deep IR regions. 

\begin{figure}[h!]
\begin{center}
\includegraphics[width=0.4\textwidth]{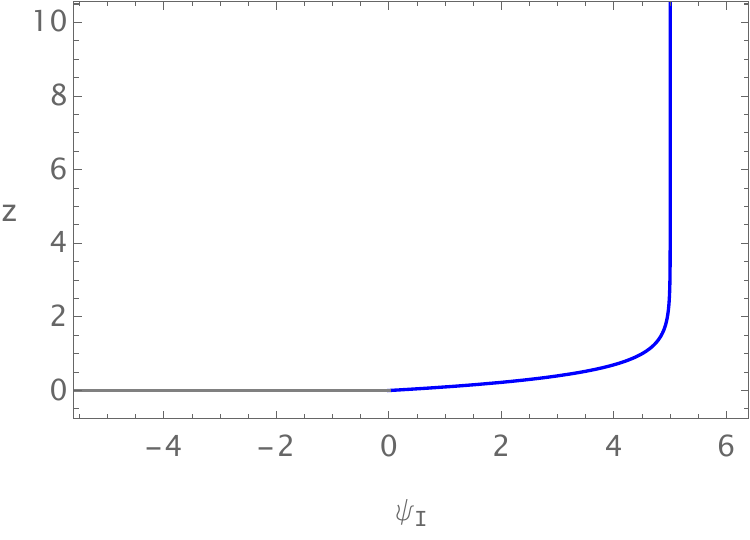}
~~
\includegraphics[width=0.4\textwidth]{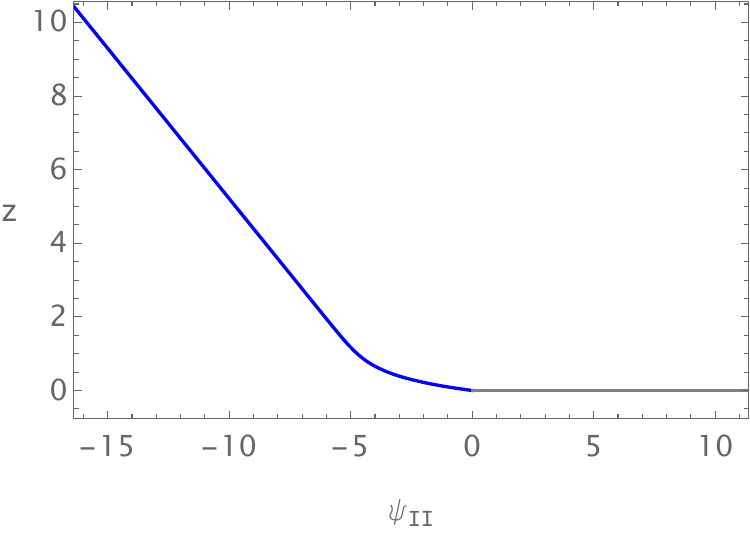}
\end{center}
\vspace{-0.3cm}
\caption{\small  Plots of interface brane for the example \eqref{eq:psiIsol5}. Here we have set $\gamma=-5, a=10, L_\text{I}=1$ and $\nu=0.5$. 
}
\label{fig:psiaz}
\end{figure}

In this case, the solution of $\psi_{\text{II}}$ is quite complicated and its expression will be shown here. In Fig. \ref{fig:psiaz} we show an example of the profiles where $\gamma=-5, a=10, L_\text{I}=1$ and $\nu=0.5$. Note for the left portion of bulk spacetime, the brane only approaches a finite value of $x_\text{I}=-\gamma$.\footnote{This feature reminds us the profiles of the end-of-world (EOW) brane in AdS/BCFT for gapped phases \cite{Liu:2022ezb}.} It belongs to the case (2) classified in Fig. \ref{fig:brane pair}.

\begin{figure}[h!]
\begin{center}
\includegraphics[width=0.4\textwidth]{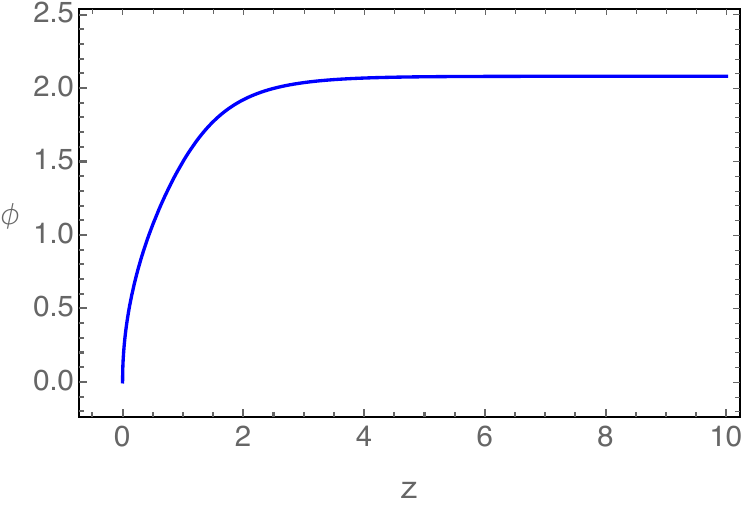}
~~
\includegraphics[width=0.41\textwidth]{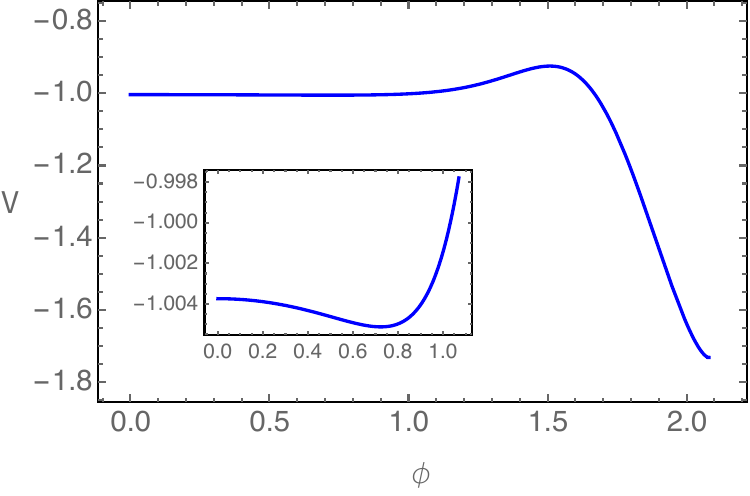}
\end{center}
\vspace{-0.3cm}
\caption{\small Typical configuration of the scalar field $\phi(z)$ and potential $V(\phi)$.  We choose $\gamma=-5, a=10, L_\text{I}=1$ and $\nu=0.5$ and we have $\phi_0=2.09$. 
}
\label{fig:az}
\end{figure}

Obtaining the analytical behavior for scalar field and the potential up to subleading order is challenge.  Here we instead present the numerical plots of $\phi$ and $V$ in Fig. \ref{fig:az}. We observe that the scalar field $\phi$ tends towards a constant value $\phi_0$ as $z$ increases, indicating that the brane asymptotically approaches AdS$_2$ in the deep IR region. Regarding the potential $V$, its behavior is non-monotonic, similar to the potential in Sec. \ref{subsub:sol4}. It exhibits a local maximum at $\phi=0$, corresponding to the UV fixed point, and a global minimum at $\phi=\phi_0$, corresponding to the IR fixed point. Additionally, between these two points, the potential has a local minimum and a local maximum. 

\begin{figure}[h!]
\begin{center}
\includegraphics[width=0.31\textwidth]{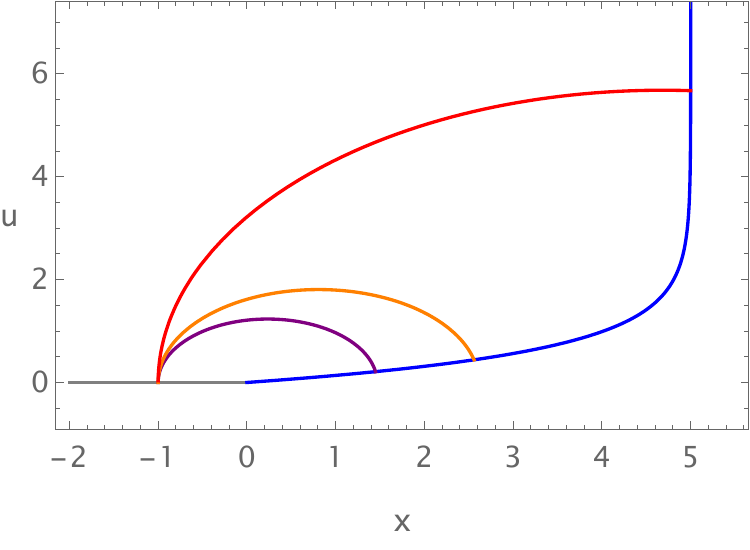}
\includegraphics[width=0.31\textwidth]{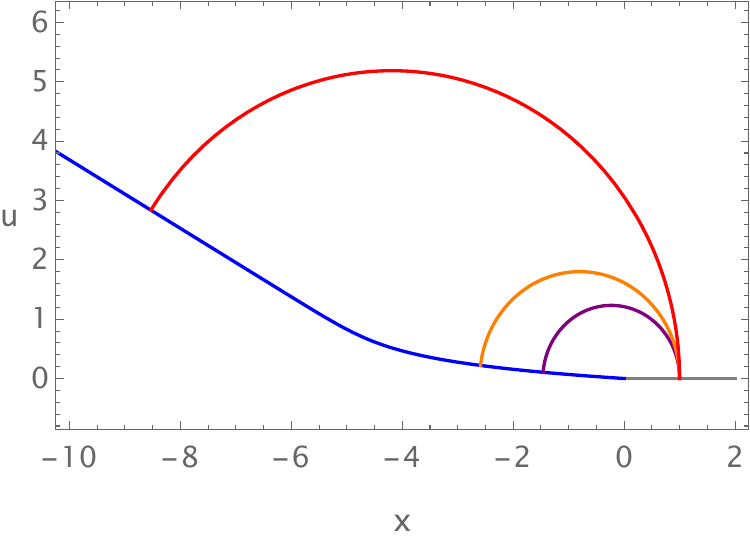}
\includegraphics[width=0.32\textwidth]{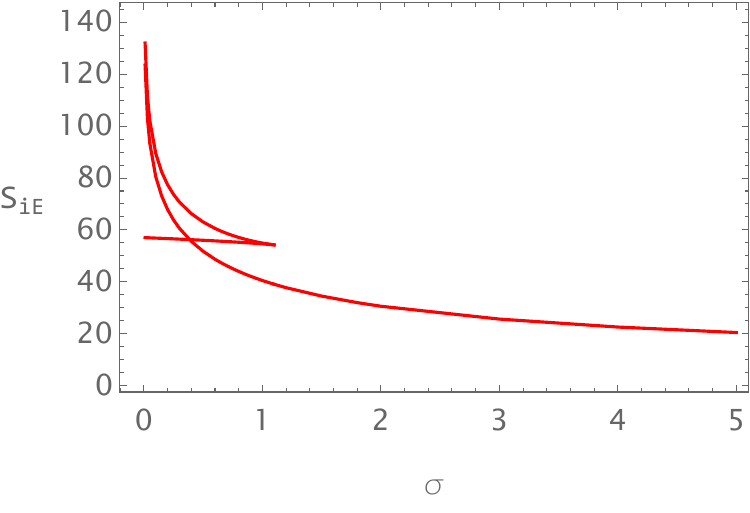}
\end{center}
\vspace{-0.3cm}
\caption{\small The 
left two plots are the multiple extremal surfaces and the right plot is 
the interface entropy $S_{\text{iE}}=\log{g}$ as a function of $\sigma$ for the profile \eqref{eq:psiIsol5}. 
Here we have set $\gamma=-5, a=10, L_\text{I}=1$ and $\nu=0.5$. 
}
\label{fig:Saz5}
\end{figure}

Similar to the observation in Sec. \ref{subsub:sol4}, with a non-monotonic potential, we might have multiple extremal curves. More precisely, as illustrated in the left two plots, Fig. \ref{fig:Saz5}, for the $\psi_{\text{II}}<0$ case we analyzed, there are three possible intersecting points $z=z_*$ within the regime  $(0,1.099)$ of the spatial interval $\sigma_{\text{I}}=\sigma_{\text{II}}=\sigma$. 
In the right figure in Fig. \ref{fig:Saz5}, we show 
the interface  entropy $S_\text{iE}$ as a function of $\sigma$. We observe again there is a first-order phase transition in 
the boundary entropy when we increase $\sigma$. This is due to the competition among the multiple extremal surfaces. 
This observation further confirms 
that 
the potential $V$'s non-monotonic nature
leads to the phase transitions of the $g$-function of ICFT. Here the $g$-theorem is also satisfied. 
Note that the boundary entropy has a lower bound and is always greater than $0$ in this case and this is different from the case in Sec. \ref{subsub:sol4}.

\subsubsection{Solution VI}
\label{subsec:solVI}

In the previous discussions, we have provided several examples where, in the IR regime, the interface brane extend to $z\to \infty$. Here we present a specific example where the interface brane only spans upto a finite regime of $z$.  We consider the following profile of the interface brane 
\begin{align}
\label{eq:ps1vii}
    \psi_\text{I} = \frac{a}{z- b}+\frac{a}{b}\,,
\end{align}
where $a$ and $b$ are constants with $b>0$.  The NEC constraints  $a>0$. Note that when $x$ goes to $-\infty$, $z$ only approaches a finite constant value $b$, therefore the brane exists solely within the finite range $0<z<b$. From \eqref{eq:lam0} we obtain
\begin{align}
\begin{split}
    \lambda(z) &= -\frac{b^4}{L^2_\text{I}(b^4+a^2\nu)}+\mathcal O(z)\,,\quad (z\to 0)\,, \\
    \lambda(z) &=\mathcal O((z-b)^3)\,,~~~~~~~~~~~~~~~\quad (z\to b)\,.
\end{split}
\end{align}
These results suggest that the brane is asymptotic AdS$_2$ near  the UV boundary, yet becomes flat in the vicinity of $z=b$,
similar to the studies in Sec. \ref{subsub:sol4}.

\begin{figure}[h!]
\begin{center}
\includegraphics[width=0.4\textwidth]{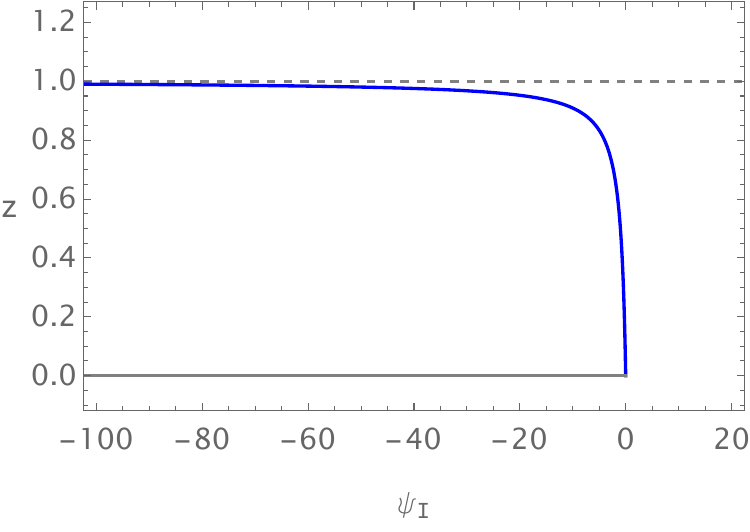}
~~
\includegraphics[width=0.4\textwidth]{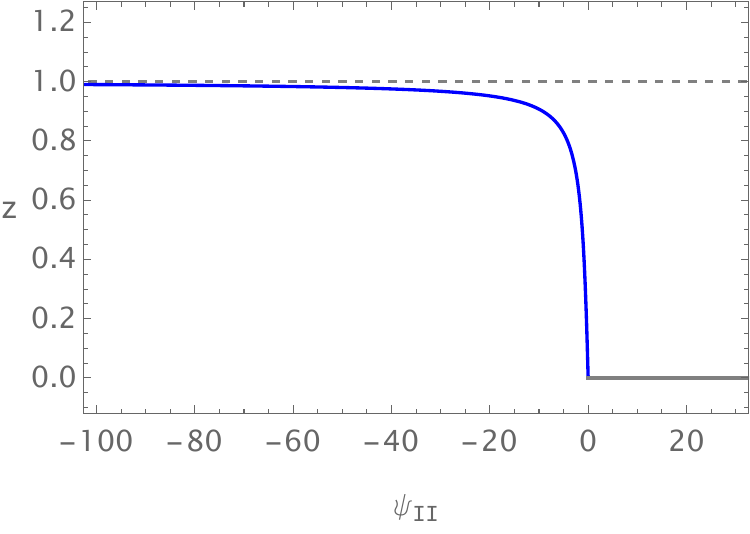}
\end{center}
\vspace{-0.3cm}
\caption{\small  Plots of configurations of $\psi_\text{I}$ in \eqref{eq:ps1vii} and $\psi_\text{II}$ with minus sector in \eqref{eq:ps2vii} when $a=b=1$. We set $L_\text{I}=1$ and $\nu=0.5$.
}
\label{fig:psi1bx}
\end{figure}

The solution for $\psi_\text{II}$ in this scenario is
\begin{align}\label{eq:ps2vii}
\begin{split}
    \psi_\text{II} =&~ \pm \Bigg{(}\frac{a}{b-z}\, _2F_1 \left( -\frac{1}{2}\,, -\frac{1}{4}\,, 
    \frac{3}{4}\,, -\frac{(b-z)^4(1-\nu^2)}{a^2\nu} \right) \\
    &~~~~~~~~~- \frac{a}{b}\, _2F_1 \left( -\frac{1}{2}\,, -\frac{1}{4}\,, \frac{3}{4}\,, 
    -\frac{b^4(1-\nu^2)}{a^2\nu} \right)\Bigg{)}\,,
\end{split}
\end{align}
where $_2F_1$ is a hypergeometric function. Note that \eqref{eq:ps2vii} has a pole at $z=b$ where the brane is asymptotically to and the brane is defined within the interval $0<z<b$. In Fig. \ref{fig:psi1bx} we show an example  configurations of $\psi_\text{I}$ and $\psi_\text{II}$ by setting $a=b=1$. This is a special example of case (3) classified in Fig. \ref{fig:brane pair}. 
When we choose the plus sector in \eqref{eq:ps2vii}, the configuration becomes to case (4) classified in Fig. \ref{fig:brane pair}.  

\begin{figure}[h!]
\begin{center}
\includegraphics[width=0.31\textwidth]{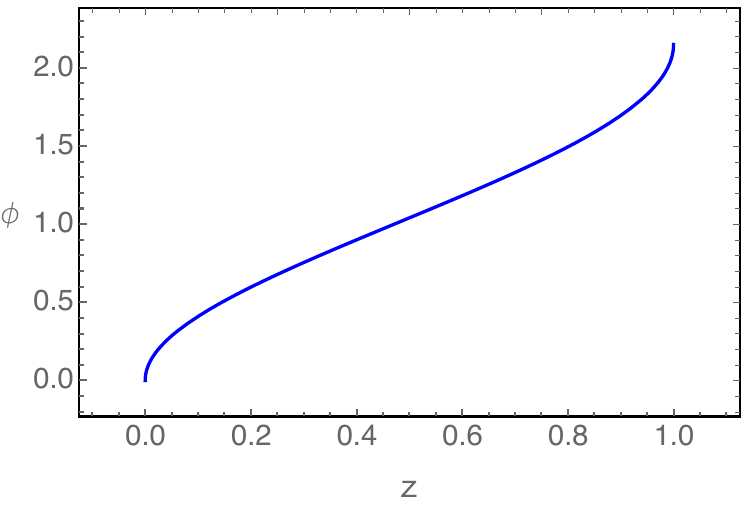}
\includegraphics[width=0.32\textwidth]{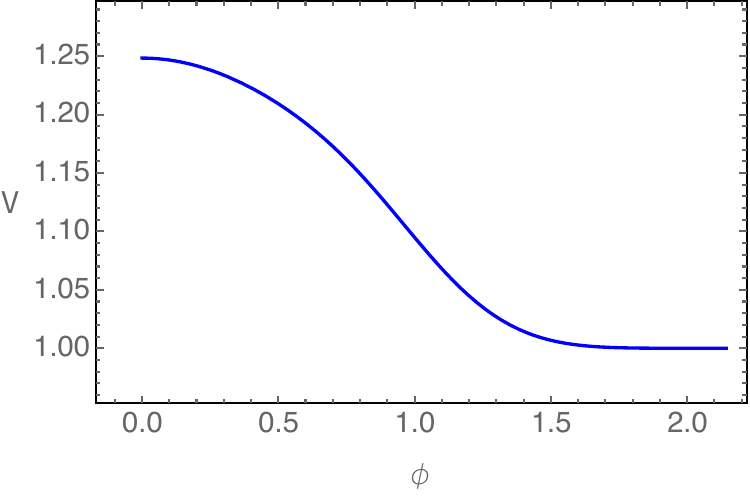}
\includegraphics[width=0.33\textwidth]{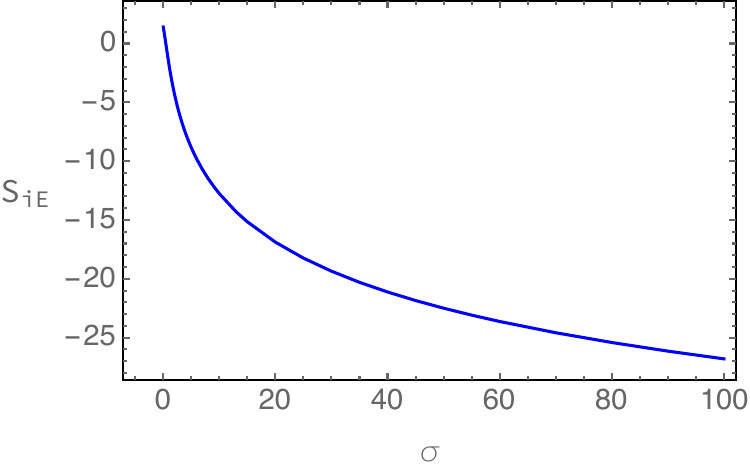}
\end{center}
\vspace{-0.3cm}
\caption{\small The left two figures are plots of scalar field $\phi(z)$ and its potential $V(\phi)$. The right plot is the interface entropy $S_{\text{iE}}$ as a function of $\sigma$. Here we set $a=b=1$, $L_\text{I}=1$,  $\nu=1/2$ and consider the profile with negative sign in \eqref{eq:ps2vii}. 
}
\label{fig:S1bx}
\end{figure}

In Fig. \ref{fig:S1bx}, we choose the minus sign for $\psi_\text{II}$ in \eqref{eq:ps2vii} and set $a=b=L_\text{I}=1$ and $\nu=1/2$. The left two  figures illustrate the configurations of the scalar field $\phi(z)$ and its potential $V(\phi)$. We observe that the potential $V$ monotonically decreases as a function of the scalar field $\phi$ yet remains always positive. Additionally, the scalar field approaches a maximum value of $\phi_0=2.15$ as $z\to b$, corresponding to a local minimum of the potential $V$. The evolution from $\phi=0$ to $\phi=\phi_0$, which moves from a local maximum to a local minimum of the potential $V$, can as thought of a RG flow from UV to IR for ICFT. The right figure of Fig. \ref{fig:S1bx} illustrates  the interface entropy as a function of the subsystem size $\sigma$. We observe  that the boundary entropy $S_\text{iE}$ decreases monotonically, and this decrease does not have a lower bound. Together with the results of solution IV in Fig. \ref{fig:Sz4}, we suspect that this behavior is an intrinsic characteristic of a brane that is asymptotically flat in the IR.

\begin{figure}[h!]
\begin{center}
\includegraphics[width=0.31\textwidth]{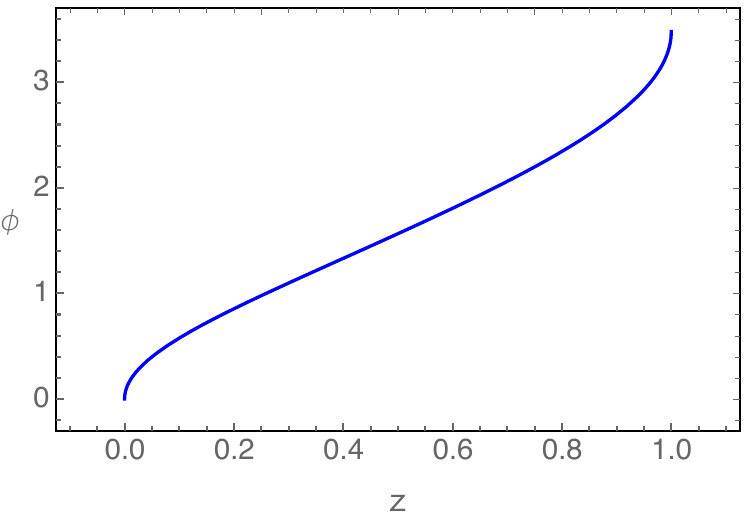}
\includegraphics[width=0.32\textwidth]{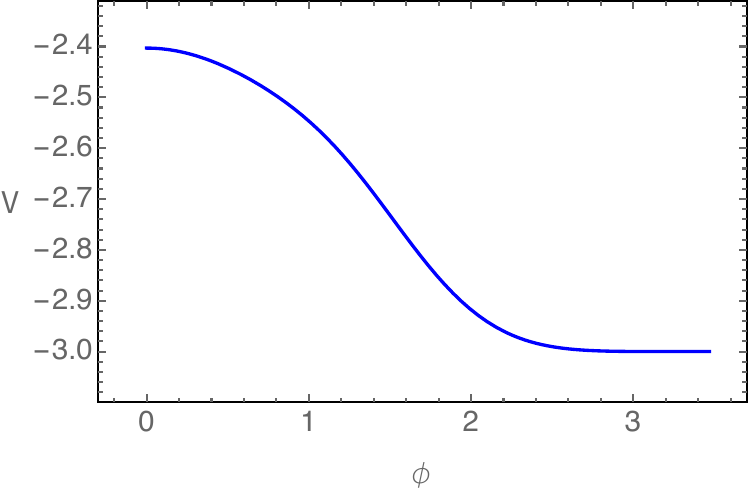}
\includegraphics[width=0.34\textwidth]{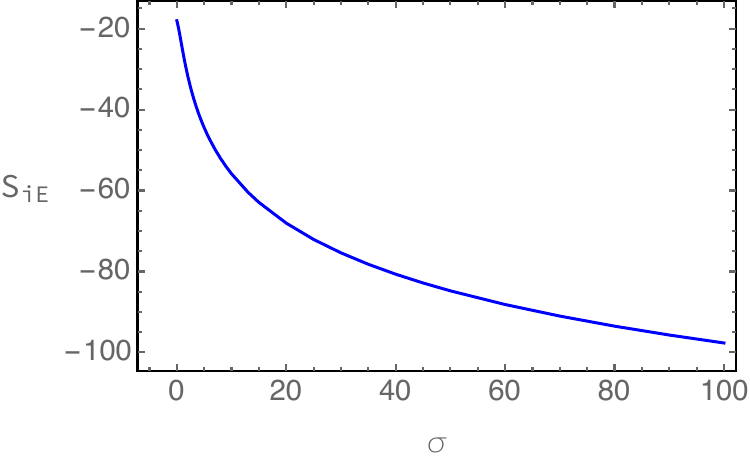}
\end{center}
\vspace{-0.3cm}
\caption{\small  The left two figures are plots of scalar field $\phi(z)$ and its potential $V(\phi)$. The right plot is the 
interface entropy $S_{\text{iE}}$ as a function of $\sigma$. Here set $a=b=1$, $L_\text{I}=1$, $\nu=0.5$ and consider the profile with positive sign in \eqref{eq:ps2vii}.
}
\label{fig:S1bxp}
\end{figure}

In Fig. \ref{fig:S1bxp} we consider the case that $\psi_\text{II}$ in \eqref{eq:ps2vii} is positive. We also set $a=b=L_\text{I}=1$ and $\nu=1/2$. The left two figures in Fig. \ref{fig:S1bxp} illustrate the configurations of the scalar field $\phi(z)$ and its potential $V(\phi)$. Similar to the case with a negative $\psi_\text{II}$, the potential $V$ monotonically decreases as $\phi$ increases, and the maximum value of the scalar field $\phi_0=3.48$ corresponds to the point where the potential reaches its minimum value. However, in this scenario, the potential $V$ is always negative. The right 
figure illustrate the interface entropy as a function of the subsystem size $\sigma$. The interface entropy $S_\text{iE}$ again monotonically decreases from a negative value towards $-\infty$.

\subsection{Comment on the entanglement entropy of various subsystems}
\label{sec:comment}

In comparison to CFT or BCFT, ICFT presents a more intricate entanglement structure, where the entanglement entropy of diverse subsystems offers a detailed exploration of the interface dynamics  \cite{Calabrese:2009qy, {Affleck2009}}. For example, the entanglement entropy serves as a tool to characterize the impact of interaction strength on the defect 
\cite{Calabrese:2009qy}. In \cite{Karch:2021qhd, Karch:2022vot}, the entanglement structure has been studied within a class of holographic ICFT model. Interestingly, a universal relation between the coefficients of divergent terms in the entanglement entropy of various subsystems has been uncovered, in both Janus solutions and RS braneworld models. Compared to the holographic models in\cite{Karch:2021qhd, Karch:2022vot}, due to the presence of a dynamical scalar field on the interface brane the induced metric might no longer be AdS$_2$. However, we can extract information regarding the divergent term in the entanglement entropy from the position of the extremal surface endpoint on the boundary for {\em the specific case (2) classified in Fig. \ref{fig:brane pair}.}: i.e. we assume $\psi'_\text{I}\ge 0,\ \psi'_\text{II}\le 0$. As in previous discussions, we consider the case $\nu\le 1$.

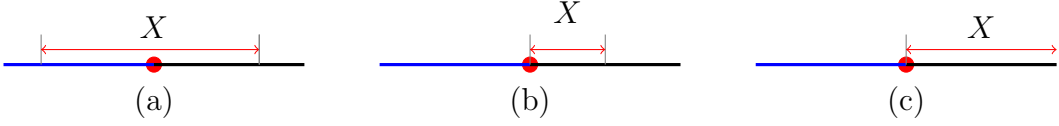
\begin{figure}[h!]
\begin{center}
\begin{tikzpicture}
\draw [fill,color=red] (2,0) circle [radius=0.1];
\draw[very thick,color=blue] plot coordinates {(0,0)(2,0)};
\draw[very thick,color=black] plot coordinates {(2,0)(4,0)};
\draw[color=gray] plot coordinates {(0.5,0)(0.5,0.4)};
\draw[color=gray] plot coordinates {(3.4,0)(3.4,0.4)};
\draw[color=red] [<->] (0.5,0.2) -- (3.4,0.2);
\node[color=black] at (2.0,0.5) {$X$};
\node[color=black] at (2.0,-0.5) {(a)};

\draw [fill,color=red] (7,0) circle [radius=0.1];
\draw[very thick,color=blue] plot coordinates {(5,0)(7,0)};
\draw[very thick,color=black] plot coordinates {(7,0)(9,0)};
\draw[color=gray] plot coordinates {(7,0)(7,0.4)};
\draw[color=gray] plot coordinates {(8,0)(8,0.4)};
\draw[color=red] [<->] (7,0.2) -- (8,0.2);
\node[color=black] at (7.5,0.7) {$X$};
\node[color=black] at (7.0,-0.5) {(b)};

\draw [fill,color=red] (12,0) circle [radius=0.1];
\draw[very thick,color=blue] plot coordinates {(10,0)(12,0)};
\draw[very thick,color=black] plot coordinates {(12,0)(14,0)};
\draw[color=gray] plot coordinates {(12,0)(12,0.4)};
\draw[color=gray] plot coordinates {(3.4,0)(3.4,0.4)};
\draw[color=red] [<->] (12,0.2) -- (14,0.2);
\node[color=black] at (13,0.5) {$X$};
\node[color=black] at (12,-0.5) {(c)};

\end{tikzpicture}
\end{center}
\vspace{-0.3cm}
\caption{\small Cartoon plot for various boundary subsystems. CFT$_\text{I,II}$
are two halves of the straight line and the red dot is the interface. $X$ is the subsystem. (a) the interface is at the interior of the subsystem and the subsystem is finite; (b) the interface is at one end of the subsystem and the subsystem is finite; (c) the interface is at one end of the subsystem and the subsystem is infinite.
}
\label{fig:various subsystems}
\end{figure}

Let us first consider the subsystem (a) in Fig. \ref{fig:various subsystems}, which is the union of $-\sigma_{\text{I}}\le x_{\text{I}}\le 0$ and $0\le x_{\text{II}}\le \sigma_{\text{II}}$. Suppose the intersecting point $S$ between the brane and the extremal surface is $z_*$. Denote the left, right end points of the extremal surface by $A,B$. Then the left Poincare coordinates of $A,S$ are
$
(0,-\sigma_{\text{I}},\epsilon_{\text{I}})\,,$ and $  
(0,\psi_{\text{I}}(z_*),\frac{z_*}{\sqrt{\nu}})
$
and the right Poincare coordinates of $S,B$ are
$(0,\sigma_{\text{II}},\epsilon_{\text{II}})$ and 
$(0,\psi_{\text{II}}(z_*),\sqrt{\nu}z_*)$.   
Then from \eqref{eq:hee} the entanglement entropy is
\begin{align}
\label{eq:vbheeA}
\begin{split}
S_\text{E}=&~
\frac{L_{\text{I}}}{4G}
\cosh^{-1}
\frac
{
(\sigma_{\text{I}}+\psi_{\text{I}}(z_*))^2+
\frac{z_*^{2}}{\nu}
}
{
2\frac{1}{\sqrt{\nu}}z_*
\epsilon_{\text{I}}
}
+
\frac{L_{\text{I}}\nu}{4G}
\cosh^{-1}
\frac
{
(\sigma_{\text{II}}-\psi_{\text{II}}(z_*))^2+
\nu z_*^{2}
}
{
2\sqrt{\nu}z_*
\epsilon_{\text{II}}
}   \\
=&~\frac{c_\text{I}+c_\text{II}}{6}\log\frac{(\sigma_{\text{I}}+\sigma_{\text{II}})}{\epsilon}
+\text{(regular terms)}\,.
\end{split}
\end{align} 
In the last line, we have chosen   $\epsilon_{\text{I}}=\frac{\epsilon_{\text{II}}}{\nu}=\epsilon$.  

Next we consider the subsystem (b) in Fig. \ref{fig:various subsystems}, which is $0\le x_{\text{II}}\le \sigma_{\text{II}}$. For the case we considered, 
the extremal surface does not cross the brane. Suppose the two end points of the extremal surface $\gamma_{b}$ are
$
A:\,(t_{\text{II}},x_{\text{II}},u_{\text{II}})
=(0,\sigma_{\text{II}},\epsilon)$ and $  
B:\,(t_{\text{II}},x_{\text{II}},u_{\text{II}})
=(0,0,\epsilon).
$ 
The entanglement entropy is
\begin{align}
\label{eq:vbheeB}
S_\text{E}=\frac{L_{\text{I}}\nu}{4G}\cosh^{-1}
\frac
{
\sigma_{\text{II}}^2+2\epsilon^2
}
{
2\epsilon^2
}  =\frac{c_\text{II}}{3}\log\frac{\sigma_{\text{II}}}{\epsilon}
+\text{(regular terms)}\,.
\end{align}

Finally, let us consider the subsystem (c) in Fig. \ref{fig:various subsystems}, which is $+\infty\ge x_{\text{II}}\ge 0$. Noticing that $c_\text{I}\ge c_\text{II}$, then the extremal surface $\gamma_{c}$ is $x_{\text{II}}=0$ and the induced metric on $\gamma_c$ is
\begin{align}
ds^2_{\gamma}=L^2_{\text{I}}\nu^2
\frac{dy^2_{\text{II}}}{y^2_{\text{II}}}.
\end{align} 
The entanglement entropy is
\begin{align}
\label{eq:vbheeC}
S_\text{E}=\frac{L_{\text{I}}\nu}{4G}
\int^{\ell_{IR}}_{\epsilon}
\frac{dy_{\text{II}}}{y_{\text{II}}}  
=\frac{c_2}{6}
\log\frac{\ell_{\text{IR}}}{\epsilon}\,,
\end{align}
where $\epsilon$ is the UV cutoff and $\ell_{\text{IR}}$ is the IR cutoff.

From the divergence term in \eqref{eq:vbheeA}, \eqref{eq:vbheeB} and \eqref{eq:vbheeC}, we can obtain an effective central charge that satisfies the universal relation proposed in \cite{Karch:2021qhd, Karch:2022vot}. In \cite{Karch:2021qhd} it was suggested that the endpoint of a finite interval could characterize the underlying field theory, thus $c_\text{II}$ can be viewed as an effective central charge of the defect, satisfying the upper bound proposed in \cite{Karch:2023evr}. However, as it only appears within divergent terms, it appears to lack  information regarding the dynamics of the defect along RG flow, merely reflecting the coupling properties between the defect and CFT. A deeper  understanding on the effective central charge from the perspective of field theory, particularly when the scaling symmetry of interface is broken, is necessary.

\section{Finite temperature}
\label{sec:ft}

We have explored the zero temperature solution and its associated interface entropy. In this section, we extend our previous study to the finite temperature case, focusing on the configurations of brane solutions. We will first glue two BTZ black holes along a brane in Sec. \ref{sec:g2b}, then glue a thermal AdS solution with a BTZ black hole in Sec. \ref{sec:gab}. We mostly concern with the impact of the interface-located scalar field on the configuration.

\subsection{Gluing two BTZ black holes}
\label{sec:g2b}

The planar BTZ black hole metric is given by
\begin{align}
\label{eq:btzbh}
ds^2=\frac{L^2_\text{A}}{u^2_\text{A}}\left[
-f_\text{A}(u_\text{A})dt^2_\text{A}+\frac{du^2_\text{A}}{f_\text{A}(u_\text{A})}+dx^2_\text{A}
\right]\,,\ \  ~~~\text{A=\,I\,,~II}\,,
\end{align}
where
\begin{align}
f_\text{A}(u_\text{A})=1-\frac{u^2_\text{A}}{(u^H_\text{A})^2}\,,\ \  ~~~\text{A=\,I\,,~II}\,.
\end{align}
Similar to the zero temperature case, $L_A$ is the AdS radius. 
Note that $u=u^H_\text{A}$ is the location of the horizon. We consider the case that $x_\text{A}$ is non-compact. The Hawking temperatures of the black holes are\footnote{We use $\Theta$ for temperature to avoid possible confusion with the tension $T$.} 
\be
\label{eq:finiteTtem}
\Theta_\text{I}=\frac{1}{2\pi u^H_\text{I}}\,,~~~~\Theta_\text{II}=\frac{1}{2\pi u^H_\text{II}}\,.~~~~
\ee
The dual field theory lives at the conformal boundary $u\to 0$. 

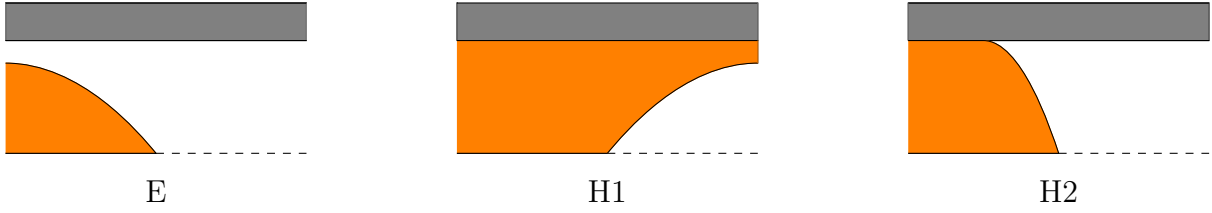
\begin{figure}[h!]
\begin{center}
\begin{tikzpicture}

\draw [fill=gray] (0,1.5)--(4,1.5) -- (4,2) --(0,2)--(0,1.5);
\draw[color=gray] plot coordinates {(0,1.5)(0,2)};
\draw[color=gray] plot coordinates {(4,1.5)(4,2)};
\draw[color=black] plot coordinates {(0,1.5)(4,1.5)};

\node[color=black] at (2.0,-0.5) {E};

\draw[color=black] plot coordinates {(0,2)(4,2)};
\draw[color=black] plot coordinates {(0,0)(2,0)};
\draw[color=black, dashed] plot coordinates {(2,0)(4,0)};
\draw[color=orange] (0,1.2) parabola (2,0);  
\draw [fill=orange]  (0,1.2) parabola (2,0)--(0,0);

\draw [fill=gray] (6,1.5)--(10,1.5) -- (10,2) --(6,2)--(6,1.5);
\draw[color=gray] plot coordinates {(6,1.5)(6,2)};
\draw[color=gray] plot coordinates {(10,1.5)(10,2)};

\node[color=black] at (8.0,-0.5) {H1};

\draw[color=black] plot coordinates {(6,2)(6,2)};
\draw[color=black] plot coordinates {(6,1.5)(10,1.5)};
\draw[color=black] plot coordinates {(6,0)(8,0)};
\draw[color=black, dashed] plot coordinates {(8,0)(10,0)};
\draw[color=orange] (10,1.2) parabola (8,0);  
\draw [fill=orange]  (6,1.5)--(10,1.5)--(10,1.2) parabola (8,0)--(6,0);
\draw[color=orange] plot coordinates {(10,1.5)(10,1.2)};

\draw [fill=gray] (12,1.5)--(16,1.5) -- (16,2) --(12,2)--(12,1.5);
\draw[color=gray] plot coordinates {(12,1.5)(12,2)};
\draw[color=gray] plot coordinates {(16,1.5)(16,2)};

\node[color=black] at (14.0,-0.5) {H2};

\draw[color=black] plot coordinates {(12,2)(16,2)};
\draw[color=black] plot coordinates {(12,1.5)(16,1.5)};
\draw[color=black] plot coordinates {(12,0)(14,0)};
\draw[color=black,dashed] plot coordinates {(14,0)(16,0)};
\draw[color=orange] (13,1.5) parabola (14,0);  
\draw [fill=orange]  (12,1.5)--(13,1.5) parabola (14,0)--(12,0);  

\end{tikzpicture}
\end{center}
\vspace{-0.3cm}
\caption{\small Cartoon plot for the configuration of different types of spacetime slice. The gray region is the interior of the black hole. The orange region is one part of the bulk. ``E" means there is no horizon in the bulk and ``H" means there is a complete or a part of horizon in the bulk.
}
\label{fig:Qconfig}
\end{figure}
The boundary field consists of two semi-infinite CFTs intersecting at the interface. 
Same as the zero temperature case, CFT$_\text{I}$ lives in the regime $x_\text{I}<0$ while CFT$_\text{II}$ is in the regime $x_\text{II}>0$. The left (or right) portion of the bulk has three possible different configurations for the brane $Q$ related to one side of the system $N_\text{I}$ (or $N_\text{II}$), as shown in Fig. \ref{fig:Qconfig}. The gray regime denotes the black hole interior, whereas the orange regime is the bulk  system $N_\text{I}$ (or $N_\text{II}$). In the left figure ``E", there is no horizon for the bulk geometry. In the middle figure ``H1", there is a complete  horizon for the bulk geometry. In the right figure ``H2", there is only a portion of horizon for the bulk geometry. These notations follow those in \cite{Bachas:2021fqo}.

Suppose the intrinsic coordinate system on the  interface brane $Q$ is $(t,w)$ and the brane is parameterized as 
\be
t_\text{I}(t,w)=t_\text{II}(t,w)=t\,,~~ 
\ee
and $u_\text{I}(w), u_\text{II}(w), x_\text{I}(w), x_\text{II}(w)$. 
Assuming the induced metric on $Q$ takes the form
\be
\label{eq:fTinducemetric}
ds^2_Q=-F(w) dt^2+G(w)dw^2\,,
\ee
satisfying 
\bea
\label{eq:finiteTu12}
\begin{split}
&u_\text{I}(w)=\frac{u_\text{I}^HL_\text{I}}{
\sqrt{L_\text{I}^2+(u_\text{I}^H)^2 F(w)}}\,,~~~~
u_\text{II}(w)=\frac{u_\text{II}^HL_\text{II}}{
\sqrt{L_\text{II}^2+(u_\text{II}^H)^2 F(w)}}\,,~~~~
\\
&\left(
\frac{dx_\text{I}}{dw}
\right)^2= \frac{(u_\text{I}^H)^2 \left( 4(L_\text{I}^2-(u_\text{I}^H)^2 F(w)) F(w)G(w) +L_\text{I}^2 (u_\text{I}^H)^2(\frac{dF}{dw})^2\right)}{4L_\text{I}^2F(w)\left(L_\text{I}^2+(u_\text{I}^H)^2 F(w)\right)^2}\,,\\
&\left(
\frac{dx_\text{II}}{dw}
\right)^2= \frac{(u_\text{II}^H)^2 \left( 4(L_\text{II}^2-(u_\text{II}^H)^2 F(w)) F(w)G(w) +L_\text{II}^2 (u_\text{II}^H)^2(\frac{dF}{dw})^2\right)}{4L_\text{II}^2F(w)\left(L_\text{II}^2+(u_\text{II}^H)^2 F(w)\right)^2}\,.
\end{split}
\eea

One convenient parameterization is that 
\bea 
\begin{split}
\label{eq:para}
F(w)&=w\,,\\
G(w)&= \frac{L^2_\text{I}(u^H_\text{I})^2}
{
4 L^2_\text{I} w+4(u^H_\text{I})^2w^2
}+\eta^{2}_\text{I}  
= \frac{L^2_\text{II}(u^H_\text{II})^2}
{
4 L^2_\text{II} w+4(u^H_\text{II})^2w^2
}+\eta^{2}_\text{II}  \,. 
\end{split}
\eea
We will solve the system under this assumption. The expression for $u_A$ and  $x_A$ becomes
\bea
\begin{split}
&u_\text{I}(w)=\frac{u_\text{I}^HL_\text{I}}{
\sqrt{L_\text{I}^2+(u_\text{I}^H)^2 w}}\,,~~~~~~
~~~~~~~~~~~
u_\text{II}(w)=\frac{u_\text{II}^HL_\text{II}}{
\sqrt{L_\text{II}^2+(u_\text{II}^H)^2 w}}\,,~~~~
\\
\label{eq:xIxII}
& x_\text{I}(w)=
-\int^{+\infty}_w
\frac{
u^H_\text{I}\eta_\text{I}(\sigma)}{
\sqrt{
L_\text{I}^2+(u^H_\text{I})^2\sigma}
}d\sigma\,,~~~~
x_\text{II}(w)=
-\int^{+\infty}_w
\frac{
u^H_\text{II}\eta_\text{II}(\sigma)}{
\sqrt{
L_\text{II}^2+(u^H_\text{II})^2\sigma}
}d\sigma\,.
\end{split}
\eea

Note that the parameter $w$ ranges from $w_0$ to $\infty$, where $w_0$ is a non-negative integration constant determined by \eqref{eq:xIxII}. When $w\to \infty$, both $u_\text{I}$ and $u_\text{II}$ tend to zero, marking the boundary of bulk geometry, which corresponds to the UV limit. 

On the other hand, when $\omega_0\to 0$, $u_\text{I}$ and $u_\text{II}$ converge to finite values $u_\text{I}^H$ and $u_\text{II}^H$ at specific, finite values of $x_\text{I}$ and $x_\text{II}$. In this case, the brane should touch the horizon, i.e. the configuration ``H2". Consequently, as $w$ decreases towards zero, $w\to 0, x_\text{I}, x_\text{II}$ are expected to approach finite values. Lastly, when $w_0$ is a positive nonzero number, the brane should only approach a finite value $u_A$, correponding to either the  ``E" or ``H1" configuration. In this case, as $w\to w_0$, it is expected that $x_\text{I}\to \pm \infty, x_\text{II}\to \pm \infty $. 
Obviously, from the possible choice of $w_0$, 
the configurations [E, H2],  [H1, H2], [H2, E] and [H2, H1] are  not allowed.\footnote{We use [E, H2] to refer that the left bulk $N_\text{I}$ is empty while the right bulk $N_\text{II}$ belongs to type H2 shown in Fig. \ref{fig:Qconfig}. The other pairs follow the same convention.}   

With the above setup \eqref{eq:xIxII}, the profile of the interface brane $Q$ in $N_\text{I}$ or $N_\text{II}$ is parameterized by 
\be\label{eq:dxdu}
\frac{dx_\text{A}}{du_\text{A}}=
-2\,\frac{L_\text{A}}{u^2_\text{A}}
\eta_A(\xi)\Bigg{|}_{\xi=
\frac{L^2_\text{A}}{u^2_\text{A}}-\frac{L^2_\text{A}}{(u^H_\text{A})^2}
}\,,~~~~~~~\ \ A=\text{I,~II}\,,
\ee
with the boundary condition $x_\text{A}(u_\text{A}=0)=0.$

For the simplest static case $\phi=\phi(w)$, there are only three independent equations in (\ref{eq:jj0}, \ref{eq:jj1}, \ref{eq:jj2}, \ref{eq:jj3}). Here we write out the explicit expressions as follows
\begin{align}
\begin{split}
\label{eq:q-finiteT}
& 0=\frac{L^2_\text{I}(u^H_\text{I})^2}
{
4 L^2_\text{I} w+4(u^H_\text{I})^2w^2
}+\eta^{2}_\text{I}  
- \frac{L^2_\text{II}(u^H_\text{II})^2}
{
4 L^2_\text{II} w+4(u^H_\text{II})^2w^2
}-\eta^{2}_\text{II}   \, ,\\
&\frac{V(\phi)}{2}= -
\frac{ (L^2_\text{I}+2w (u^H_\text{I})^2) \eta_\text{I} (L^2_\text{I} (u^H_\text{I})^2+2w (L^2_\text{I}+w (u^H_\text{I})^2) \eta^2_\text{I})
+ L^2_\text{I} (u^H_\text{I})^2 w
(L^2_\text{I}+(u^H_\text{I})^2 w)\eta'_\text{I}
}{
L_\text{I}u^H_\text{I}
(L^2_\text{I}(u^H_\text{I})^2+4w(L^2_\text{I}+w (u^H_\text{I})^2)\eta^2_\text{I})^{\frac{3}{2}}
}  \\
&~~+
\frac{
 (L^2_\text{II}+2w (u^H_\text{II})^2) \eta_\text{II} (L^2_\text{II} (u^H_\text{II})^2+2w (L^2_\text{II}+w (u^H_\text{II})^2) \eta^2_\text{II})
+ L^2_\text{II} (u^H_\text{II})^2 w
(L^2_\text{II}+(u^H_\text{II})^2 w)\eta'_\text{II}
}{
L_\text{II}u^H_\text{II}
(L^2_\text{II}(u^H_\text{II})^2+4w(L^2_\text{II}+w (u^H_\text{II})^2)\eta^2_\text{II})^{\frac{3}{2}}
}
\, ,\\
&\phi'^2(w)=
-\frac{
L^2_\text{I}(u^H_\text{I})^2+
4(L^2_\text{I}+(u^H_\text{I})^2w) w \eta^2_\text{I}
}{
4(L^2_\text{I}+(u^H_\text{I})^2w)w^2
}\,\bigg{[}
-wV(\phi)  \\
&~~-
\frac{
2w(L^2_\text{I}+(u^H_\text{I})^2w)\eta_\text{I}
}{
L_\text{I}u^H_\text{I}
\sqrt{
L^2_\text{I}(u^H_\text{I})^2+
4(L^2_\text{I}+(u^H_\text{I})^2w) w \eta^2_\text{I}
}
}+
\frac{
2w(L^2_\text{II}+(u^H_\text{II})^2w)\eta_\text{II}
}{
L_\text{II}u^H_\text{II}
\sqrt{
L^2_\text{II}(u^H_\text{II})^2+
4(L^2_\text{II}+(u^H_\text{II})^2w) w \eta^2_\text{II}
}
}
\bigg{]}
\, .
\end{split}
\end{align}
Note that in the above equations only up to first order derivative of $\eta$ is involved, different from the zero temperature case \eqref{eq:eq2}. This is due to the fact that $\eta$ plays the role of the slop of the brane which can be seen in \eqref{eq:dxdu}. The first equation in \eqref{eq:q-finiteT} is from the continuity of the metric field at the interface brane.

For a given potential $V(\phi)$, by solving \eqref{eq:q-finiteT} one can derive  $\eta_\text{I}(\omega), \eta_\text{II}(\omega)$ and $\phi(\omega)$. Subsequently, one can obtain the induced metric \eqref{eq:fTinducemetric} and all the information about the system. 
In practice, we solve the system in a reverse manner, similar to the zero temperature case. More precisely, for a specified $G(w)$, 
we know the form of $\eta_\text{I}$ and $\eta_\text{II}$ from \eqref{eq:para}, i.e. the profile of the system. Then, by solving $\eqref{eq:q-finiteT}$, we obtain the scalar field $\phi$ and its potential $V(\phi)$, ensuring a consistent system.

Note that our study differs from \cite{Bachas:2021fqo}, where the dual field theory was analyzed in a compact spatial direction. In contrast, the system we investigate lacks any additional scale, affording us the freedom to choose the unit for temperature in the dual field theory. Similar to the discussion in Section \ref{sec:sts}, our focus is on a globally static configuration, implying the existence of a global time in the dual field theory. Furthermore, we assume that the periods of the imaginary time are identical. It is crucial to emphasize that the temperature associated with the H1 and H2 configurations is determined by \eqref{eq:finiteTtem}, whereas the temperature of the type E configuration is independent of the parameter $u_H$. Additionally, it is worth noting that for the type E configuration, any potential conifold singularity that arises due to an arbitrary choice of the periodicity of the imaginary time is hidden behind the interface brane.

In all these different configurations, the brane should intersect with the boundary. We can find a class of solutions by imposing the following asymptotical behaviours
\begin{align}
\label{eq:uvassumption}
\text{UV}&\,,~~~ w \to +\infty\,, ~~~\ \ \eta_{\text{I}}(w)\to \frac{C}{w^a}\,,~~~a\geq 1\,.
\end{align}
Note that the intersecting condition only constraints $a>1/2$. However, when we demand that the brane is asymptotically AdS in the UV limit, we should have $a\ge 1$. 
On the other hand, in the deep IR, depending on the possible configures we have different conditions on the functions $\eta_\text{I}$ and $\eta_\text{II}$. 
If the interface brane intersects with the horizon of the BTZ black holes, then 
$\eta_\text{I}(w)/
\sqrt{L_\text{I}^2+(u^H_\text{I})^2 w},~
\eta_\text{II}(w)/
\sqrt{
L_\text{II}^2+(u^H_\text{II})^2 w}
$ must be integrable at $w=0$ and there should be no singularity for $\eta_\text{A}$ when $0<w<\infty$. 
We expect that 
\begin{align}
\label{eq:irh2}
\text{IR}&\,,~~~~  w \to 0\,,~~~~~
\eta_{\text{A}}(w)\to \frac{B}{w^b}\,,~~~~~~~~~~~~~~~~
(b<1) 
\,, ~~~~~~~ \text{for H2}\\
\label{eq:ireh1}
\text{IR}&\,,~~~~w \to w_0\,,~~~~
\eta_{\text{A}}(w)\to \frac{B}{(w-w_0)^b}\,,~~
~~~~(b\geq 1)\,, ~~~~~~ \text{for E or H1}
\end{align}
where $w_0>0$.

In the finite temperature case, the NEC, expressed as $
(\Delta K_{\mu\nu}-h_{\mu\nu}\Delta K)N^{\mu}N^{\nu} \ge 0\,
$
with $N^{\mu}$ bing an arbitrary null vector on the brane, is again equivalent to $\phi'^2(w)\ge 0$. This is similar to the behavior observed in the zero temperature case. The NEC can be simplified as
\begin{align}
\label{eq:necfiniteT}
\begin{split}
&-\frac{
(u^H_\text{I})^4\,\eta_\text{I}
+(L^2_\text{I}+(u^H_\text{I})^2w)
(-2\eta^3_\text{I}+(u^H_\text{I})^2\eta'_\text{I})
}
{
\nu u^H_\text{I}
(L^2_\text{I}(u^H_\text{I})^2+4w(L^2_\text{I}+(u^H_\text{I})^2w)\eta^2_\text{I})^{\frac{3}{2}}
}\,+\\
&~~~~~~~~
+
\frac{
(u^H_\text{II})^4\,\eta_\text{II}
+(L^2_\text{I}\nu^2+(u^H_\text{II})^2w)
(-2\eta^3_\text{II}+(u^H_\text{II})^2\eta'_\text{II})
}
{
u^H_\text{II}
(L^2_\text{I}(u^H_\text{I})^2\nu^2+4w(L^2_\text{I}\nu^2+(u^H_\text{II})^2w)\eta^2_\text{II})^{\frac{3}{2}}
}\ge 0
\,.
\end{split}
\end{align}

In the special case $u^H_\text{I}=u^H_\text{II}$, i.e. the configuration [H1,H1]\footnote{We will show that the NEC will forbid this type of configuration in Sec. \ref{sec:noh1h1}.} 
or [H2,H2], the NEC \eqref{eq:necfiniteT} can be further simplified. 
When  $\nu=1$, from the first equation in \eqref{eq:q-finiteT}, we have $\eta_\text{I}=\pm \eta_\text{II}$. When $\nu=1$ and 
$\eta_\text{I}=\eta_\text{II}$, the NEC holds automatically.  If  $\nu=1$, $\eta_\text{I}=-\eta_\text{II}$, or if $0<\nu<1$ the NEC can be further simplified as
\bea
\label{eq:nech1h1}
\left(
2\eta_\text{I}^2-
\frac{
(u^H_\text{I})^4
}
{(u^H_\text{I})^2w+
L^2_\text{I}
}
\right)\eta_\text{I}
-(u^H_\text{I})^2\,\eta'_\text{I}\ge 0\,,
\eea
or equivalently,
\bea
\label{eq:nech2h2}
\left(
2\eta_\text{II}^2-
\frac{
(u^H_\text{I})^4
}
{(u^H_\text{I})^2w+
L^2_\text{I}\nu^2
}
\right)\eta_\text{II}
-(u^H_\text{I})^2\,\eta'_\text{II}\ge 0\,.
\eea
Note that here it is quite similar to the zero temperature case discussed in Sec. \ref{sec:nec}, i.e. we have nontrivial constraint on the ansatz of the interface brane $Q$.

The Ricci tensor of the induced metric on the brane is
\begin{align}
R^Q_{\mu\nu}=\lambda(w) h^Q_{\mu\nu}\,,
\end{align}
where
\begin{align}
\label{eq:lamfin}
\lambda(w) =
\frac{
-L^2_\text{I}(u^H_\text{I})^4
+4(
L^2_\text{I}+(u^H_\text{I})^2 w)^2
(
\eta^2_\text{I}+2w\eta_\text{I}\eta'_\text{I}
)
}{
\left[
L^2_\text{I}(u^H_\text{I})^2
+4w(
L^2_\text{I}+(u^H_\text{I})^2w)\eta^2_\text{I}
\right]^2
}\,.
\end{align}
From \eqref{eq:uvassumption}, the condition that the brane is asymptotically AdS in the UV limit (i.e. $w\to \infty$) is $a\ge 1$ in \eqref{eq:uvassumption}. 
When $a=1$, we have the following asymptotic behaviour 
\bea
\begin{split}
\eta_{\text{II}}(w)&\to  \pm
\frac{
\sqrt{
4 C^2+L^2_\text{I}(1-\nu^2)
}
}{
2w
}
\,,\\
\phi'(w)&\to 0\,,\\
V(\phi)&\to 
\frac{
-2C\nu\pm \sqrt{
4C^2+L^2_\text{I}(1-\nu^2)
}
}{
L_\text{I}\nu  \sqrt{
4C^2+L^2_\text{I}
}
}\,,\\
\lambda^\text{UV}&\to -\frac{1}{4C^2+L^2_\text{I}}\,.
\end{split}
\eea
When $a>1$, we have the following asymptotic behaviour
\begin{align}
\begin{split}
\eta_{\text{II}}(w)&\to 
\pm
\frac{
L_\text{I} \sqrt{1-\nu^2}
}{
2w
}
\,,\\
\phi'(w)&\to 0\,,\\
V(\phi)&\to \pm
\frac{
\sqrt{1-\nu^2}
}{
L_\text{I}\nu
}\,,\\
\lambda^\text{UV}&\to-\frac{1}{L^2_\text{I}}\,.
\end{split}
\end{align}

The absolute value of the UV limit of the potential always satisfies the bound for the tension with trivial scalar field 
\begin{align}
T_\text{min}
<|V^\text{UV}|<
T_\text{max}\,,
\end{align}
where $T_\text{min}$ and $T_\text{max}$ are defined in \eqref{eq:tensionreg0T}.

\subsubsection{Permissible configurations at finite temperature}

We have shown above the equations of the systems at finite temperature.  We will solve them and illustrates sample profiles. It turns out when the scalar field is trivial, the only allowed profile of the interface brane is [H2,H2]. However, with the presence of a nontrivial scalar field, we find four permissible  configurations: [E, E], [E, H1], [H1, E] and [H2, H2]. These configurations are summarized in table \ref{tab:allowedprofile}. One common feature among all the profiles we have found is that the scalar potential flows from a global maximal in the UV region to a global minimum in the IR region. Furthermore, in the IR region, the induced metric on the brane exhibits distinct behaviors: it could be asymptotically AdS (in the case of [H2,H2] with a trivial scalar field), dS (for [H2,H2] with a nontrivial scalar field) or flat (for other permissible configurations). Moreover, when the induced metric is asymptotically flat in the IR region the scalar field diverges. 

\begin{table}[h!]
\centering 
\begin{tabular}{|c|c|c|c|}
\hline
 & E & H1 & H2  
 \\ \hline
E & $\surd$ & $\surd$ & $\times$
  \\ \hline
H1  & $\surd$ & $\times$ & $\times$
  \\ \hline
 H2  & $\times$ & $\times$ & $\surd$
  \\ \hline
\end{tabular}
\caption{\small The permissible configurations with a dynamical scalar filed located on the interface brane. The column denotes the profile of the interface brane in $N_\text{I}$ while the row signifies the profile in $N_\text{II}$. 
\label{tab:allowedprofile}}
\end{table}

\subsubsection{The solution with trivial scalar field: [H2,H2]}

Let us first study the case with $\phi=0$ and $V(\phi)=T$. The zero temperature solution has been discussed in Sec. \ref{sec:simpleexsol} and \ref{subsub:solI}. Here we consider the finite temperature generalizations. 

The equations of motion on $Q$ 
(\ref{eq:q-finiteT}) become
\begin{align}
\begin{split}
&\frac{L^2_\text{I}(u^H_\text{I})^2}
{
4L^2_\text{I}
w+4(u^H_\text{I})^2w^2 
}+\eta^{2}_\text{I}
=
\frac{L^2_\text{I}(u^H_\text{II})^2\nu^2}
{
4L^2_\text{I}
\nu^2w+4(u^H_\text{II})^2w^2
}
+\eta^{2}_\text{II}\,,  \\
&\frac{L_\text{I}T}{2}=
\frac{-
((u^H_\text{I})^2w+
L^2_\text{I})
\eta_\text{I}
}{u^H_\text{I}
\sqrt{
L^2_\text{I}(u^H_\text{I})^2
+4w((u^H_\text{I})^2w+L^2_\text{I} 
)
\eta^{2}_\text{I}
}
}+
\frac{
((u^H_\text{II})^2w+
L^2_\text{I}
\nu^2)
\eta_\text{II}
}{
\nu u^H_\text{II}
\sqrt{
L^2_\text{I}(u^H_\text{II})^2\nu^2
+4w((u^H_\text{II})^2w+L^2_\text{I}
\nu^2)
\eta^{2}_\text{II}
}
}\,.
\end{split}
\end{align}

We eliminate $\eta_{\text{II}}$ in above equations and obtain the solution of $\eta_{\text{I}}$ which takes the following form 
\begin{align}
\label{eq:vaceta1}
\eta_\text{I}=\pm\frac{
L_\text{I}u^H_\text{I}\left( L^2_\text{I} \nu^2 ((u^H_\text{II})^2-(u^H_\text{I})^2) + w (u^H_\text{I})^2 (u^H_\text{II})^2
(-1+
(1+L^2_\text{I}T^2)\nu^2
) \right)
}{2
\sqrt{w(L^2_\text{I} + w (u^H_\text{I})^2 )}\,
\sqrt{A (w-w_+)(w-w_-)}
} \,,
\end{align}
where
\begin{align}\label{eq:wpm}
w_{\pm}= \frac{-B\pm \sqrt{B^2-AC}}{A}
\end{align}
and
\begin{align}
\begin{split}
\label{eq:abc}
    A&=(u^H_\text{I})^4\left(u^H_\text{II}\right)^4L_\text{I}^4\nu^4
\left(\left(\frac{1+\nu}{\nu L_\text{I}}\right)^2-T^2\right)\,\left(T^2-\left(\frac{1-\nu}{\nu L_\text{I}}\right)^2\right)\,,\\
    B&= L^2_\text{I} \nu^2 (u^H_\text{I})^2 (u^H_\text{II})^2 \left( (u^H_\text{I})^2 (-1+ (1+L^2_\text{I}T^2)\nu^2 ) + (u^H_\text{II})^2 \left(1+(-1+L^2_\text{I}T^2)\nu^2 \right) \right)\,,\\
    C&= - L^4_\text{I} \left((u^H_\text{I})^2-(u^H_\text{II})^2\right)^2 \nu^4\,.
\end{split}
\end{align}
In the above equations \eqref{eq:abc}, we always have $C\leq 0$ where $C=0$ only occurs when $u^H_\text{I}=u^H_\text{II}$. 
Meanwhile, $A$ has to be positive otherwise the square root in \eqref{eq:vaceta1} is negative and therefore no consistent solution for $\eta_\text{I}$. We should impose the constraint 
\begin{align}
\label{eq:Trangevac}
   \frac{1-\nu}{\nu L_\text{I}} < |T| < \frac{1+\nu}{\nu L_\text{I}}\,. 
\end{align} 
Note that this constraint is exactly the same as the zero temperature result discussed in \eqref{eq:tensionreg0T}.  

Similarly, the solution of $\eta_\text{II}$ is obtained to be 
\begin{align}
\label{eq:vaceta2}
&\eta_\text{II}= \mp\frac{u^H_\text{II} L_\text{I} \nu}{2} \, \frac{ L^2_\text{I} \nu^2 ((u^H_\text{I})^2-(u^H_\text{II})^2) + w (u^H_\text{I})^2(u^H_\text{II})^2 (1+(-1+L^2_\text{I} T^2)\nu^2) }{\sqrt{w(L^2_\text{I} \nu^2 + w (u^H_\text{II})^2 )}\,\sqrt{A (w-w_+)(w-w_-)} }\,,
\end{align} 
with the $w_{\pm}, A$ defined in \eqref{eq:wpm} and \eqref{eq:abc}.

Therefore, with a consistent solution, i.e. \eqref{eq:Trangevac}, we always have $AC\leq 0$ and  $\omega_+\geq 0$. In this case where $u_\text{I}^H\neq u_\text{II}^H$, we have $AC< 0$, leading to $\omega_+> 0$. At first glance, one might infer that $\omega_+> 0$ implies the brane does not intersect with the horizon. However, \eqref{eq:xIxII} reveals finite value for $x_\text{I}$ and $x_\text{II}$ when $w_0= w_+$, indicating that the brane terminates at a specific point. This behavior suggests that the interface brane does not effectively divide the bulk into two distinct regions, making it seem unphysical. Therefore, a static solution does not exist when the temperatures  $\Theta_\text{I}$ and $\Theta_\text{II}$ differ. To address the issue, one has to consider a dynamical brane solution, e.g. the holographic NESS state \cite{Bhaseen:2013ypa, Chang:2013gba}, or possibly a stationary state as described in \cite{Bachas:2021tnp}. 
It is noteworthy that this situation differs from the compact case, where the brane can curve at this ending point and intersects with the boundary again \cite{Bachas:2021fqo}. 

Therefore, the only allowed static solution exists when $u_\text{I}^H=u_\text{II}^H$, which implies $\Theta_\text{I}=\Theta_\text{II}$.  In this scenario, the only allowed configuration is [H2, H2]. The solutions \eqref{eq:vaceta1} and \eqref{eq:vaceta2} can be further simplified. Using \eqref{eq:xIxII} we have 
\begin{align}
\begin{split}
    x_\text{I} &= u^H_\text{I} \,\text{arctanh} \left( \frac{u_\text{I} (-1+(1+L^2_\text{I} T^2)\nu^2) }{\sqrt{u^2_\text{I} (-1+(1+L^2_\text{I} T^2)\nu^2)^2 + (u^H_\text{I})^2 (-1+(1+L^2_\text{I} T^2)\nu^2 - (-1+L^2_\text{I} T^2)^2 \nu^4)}} \right)\,,\\
    x_\text{II} &= u^H_\text{I} \,\text{arctanh} \left( \frac{u_\text{II} (-1+(1-L^2_\text{I} T^2)\nu^2) }{\sqrt{u^2_\text{II} (1+(-1+L^2_\text{I} T^2)\nu^2)^2+  (u^H_\text{I})^2 (-1+(1+L^2_\text{I} T^2)\nu^2 - (-1+L^2_\text{I} T^2)^2 \nu^4)} }\right)\,.
    \end{split}
\end{align}

\begin{figure}[h!]
\begin{center}
\includegraphics[width=0.4\textwidth]{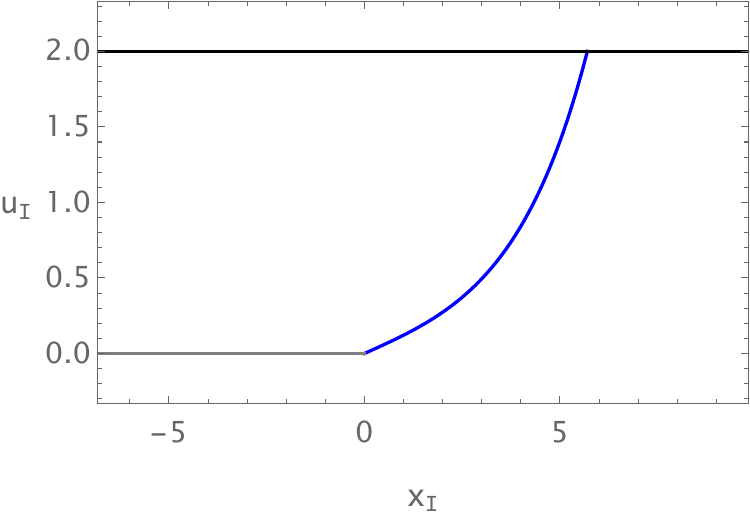}
~~
\includegraphics[width=0.4\textwidth]{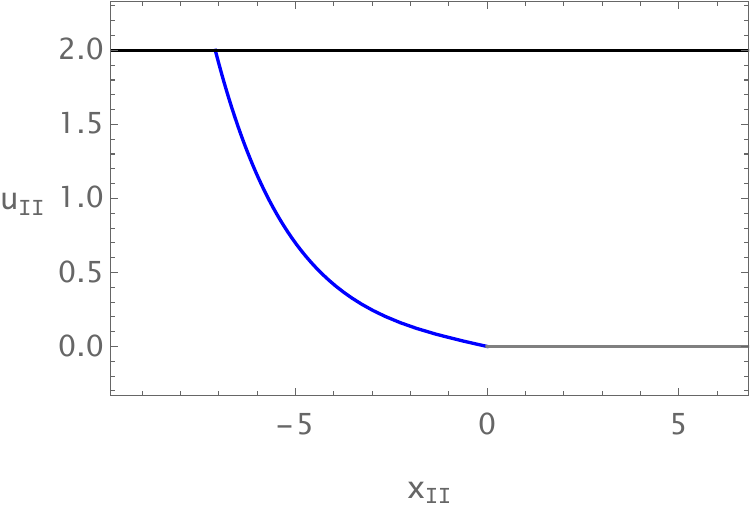}
\end{center}
\vspace{-0.3cm}
\caption{\small An example of the configurations [H2,H2] for trivial scalar field. Here we set $L_\text{I}=1, \nu=0.5, u_\text{I}^H=u_\text{II}^H=2$ and $T=2.99$. 
}
\label{fig:psi1bx-h2h2}
\end{figure}

In Fig. \ref{fig:psi1bx-h2h2}, we show an example of the profiles when $L_\text{I}=1, \nu=0.5, u_\text{I}^H=u_\text{II}^H=2$ and $T=2.99$. We see that the black hole horizon attracts the interface brane to a curved shape.  

Note that from \eqref{eq:lamfin}, we have 
\begin{align}
\lambda(w)= -\frac{1}{4T^2}
\left[
\left(
\frac{1}{L_\text{I}\nu}+\frac{1}{L_\text{I}}
\right)^2-T^2
\right]
\left[
T^2-
\left(
\frac{1}{L_\text{I}\nu}-\frac{1}{L_\text{I}}
\right)^2
\right]
\, .
\end{align}
This indicates that the induced metric on the interface brane 
is AdS$_2$.

\subsubsection{Solution with nontrivial scalar field: [H2,H2]}

We have shown that the [H2,H2] configuration is the only possibility in the trivial scalar field case. Here we will explore if such configuration exists when there is a non-trivial scalar field. 

To obtain an [H2,H2] configuration, we set
\begin{align}
\begin{split}
\eta_\text{I}=\frac{1}{2L_\text{I}}
\frac{1}{\sqrt{a+b w^2}}\,,
\end{split}
\end{align}
with $a>0\,, b>0$. Obviously it is a profile of type H2. The NEC in \eqref{eq:nech1h1} constraints  
$2a(u^H_\text{I})^4-1 \leq 0$. 

\begin{figure}[h!]
\begin{center}
\includegraphics[width=0.4\textwidth]{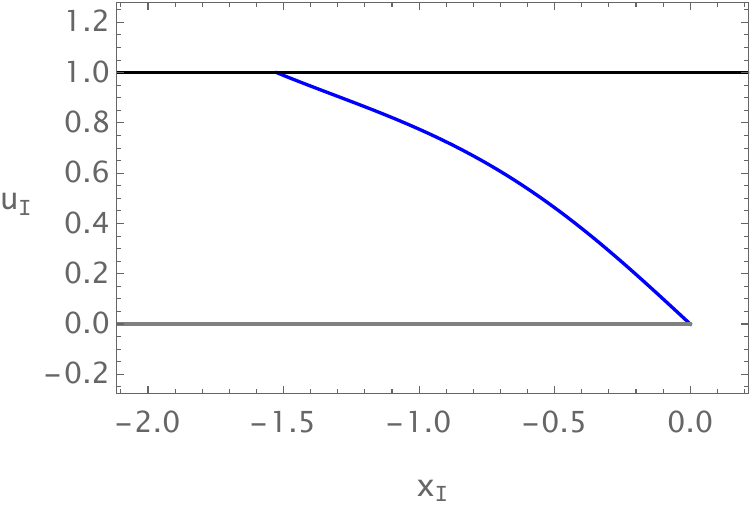}
~~
\includegraphics[width=0.4\textwidth]{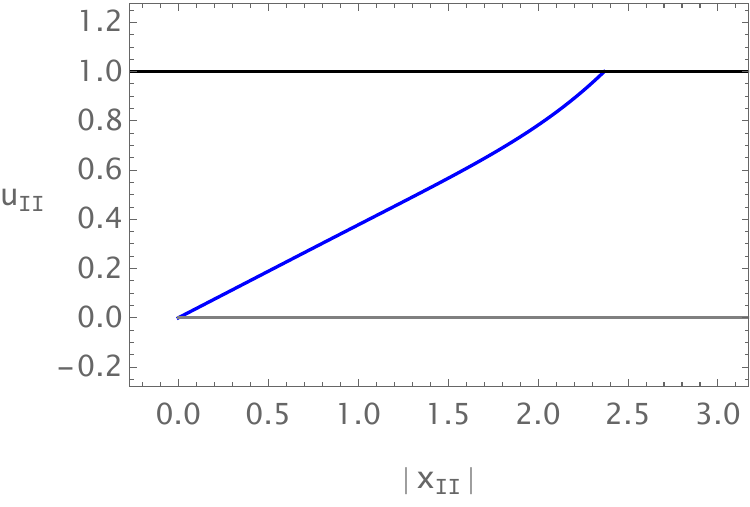}
\end{center}
\vspace{-0.3cm}
\caption{\small An example of the configurations [H2,H2] at finite temperature with dynamical scalar field. Here we set $a=1/5, b=1, L_\text{I}=1, \nu=0.5, u_\text{I}^H=1$. 
}
\label{fig:psi1bx-h2h2-scalar}
\end{figure}

We can solve the first equation in \eqref{eq:q-finiteT} to derive
\begin{align}
    \eta_\text{II}=\pm\frac{1}{2L_\text{I}}
\sqrt{
\frac{1}{a+b w^2}+
\frac{
L^4_\text{I}(u^H_\text{I})^4(1-\nu^2)
}{
((u^H_\text{I})^2w+L^2_\text{I}
)
((u^H_\text{I})^2w+L^2_\text{I}
\nu^2)
}
}\,,
\end{align}
where we have set $u^H_\text{II}=u^H_\text{I}$. With the condition $a>0, b>0$, the NEC is satisfied. The brane can be found by solving the equation
\begin{align}
\frac{dx_\text{I}}{du_\text{I}}=-
\frac{
(u^H_\text{I})^2
}{
\sqrt{
a u^4_\text{I}(u^H_\text{I})^4+b L^4_\text{I}
((u^H_\text{I})^2-u^2_\text{I}
)^2
}
}\, ,
\end{align}
with the boundary condition
$
x_\text{I}(u_\text{I}=0)=0\,.
$
Similarly we have an differential equation for $x_\text{II}(u_\text{II})$ which we do not write out here. 
We find that the functions $x_\text{I}(w),x_\text{II}(w)$ are finite in the limit $w\to 0$ and vanish in the limit $w\to +\infty$ thus this solution is a [H2,H2] solution. A typical example for the profile is shown in Fig. \ref{fig:psi1bx-h2h2-scalar}. 

\begin{figure}[h!]
\begin{center}
\includegraphics[width=0.4\textwidth]{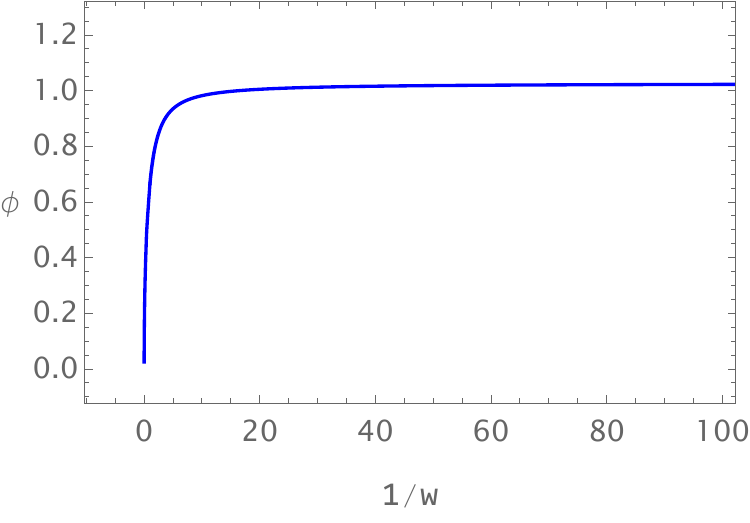}
~~
\includegraphics[width=0.4\textwidth]{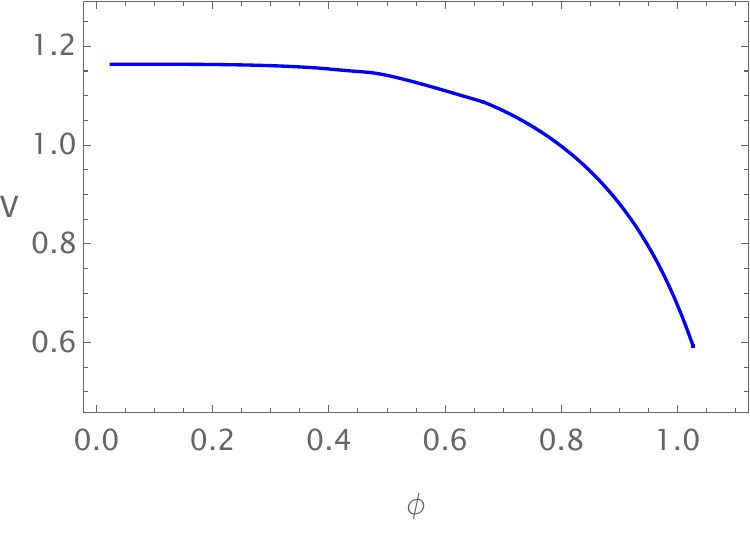}
\end{center}
\vspace{-0.3cm}
\caption{\small  The scalar field and its potential for the configuration [H2,H2] when $a=1/5, b=1, L_\text{I}=1, \nu=0.5, u_\text{I}^H=1$.
}
\label{fig:u1u2-h2h2}
\end{figure}

The scalar field $\phi(w)$ and the potential $V(\phi)$ are very complicated and we just show a numerical result in Fig. \ref{fig:u1u2-h2h2}. In the left plot, $1/w\to 0$ (or $\infty$) is the UV (or IR) limit.  We find that the potential also has a local maximal in the UV limit while a local minimal in the IR limit.

Evaluating \eqref{eq:lamfin}, we obtain the asymptotic behaviours
\begin{align}
\lambda(w)\simeq 
-\frac{b L^2_\text{I}}{1+b L^4_\text{I}}
+\mathcal{O}\left(\frac{1}{w^2}\right)\,, \quad w\to\infty\,,
\end{align}
i.e. the brane is asymptotically AdS$_2$ in the  UV limit,
while 
\begin{align}
\lambda(w)\simeq 
\frac{1-a (u^H_\text{I})^4}{a L^2_\text{I} (u^H_\text{I})^4}
+\mathcal{O}\left(w\right)\,, \quad w\to 0\,,
\end{align}
i.e. the brane is asymptotically dS$_2$ in the IR limit for the example shown in Fig. \ref{fig:psi1bx-h2h2}. Similar structure of spacetime has been constructed in e.g. \cite{Anninos:2017hhn}. It would be interesting to study our setup from the perspective of double holography \cite{Almheiri:2019hni}. 

\subsubsection{Solution with nontrivial scalar field: [E,E] and [E,H1]}
\label{subsec:EE-gbb}

The presence of a nontrivial scalar field enables various types of configurations. This subsection will present an example of the [E,E] and [E,H1] configurations and in the next subsection we will show an example of the 
[H1,E] configuration.

As we have shown in \eqref{eq:uvassumption} and \eqref{eq:ireh1}, a type E or type H1 brane should have a special behavior in the UV and IR limits, respectively. A simple example can be constructed as follows
\begin{align}
\label{eq:eta1e}
\eta_\text{I}=\frac{a}{w- b L^2_\text{I}}\,,
\end{align}
where $a$ and $b$ are constants. Solving the first equation in \eqref{eq:q-finiteT} yields the solution for $\eta_\text{II}$
\begin{align}
\label{eq:fTsetaIIee}
\eta_\text{II}=\pm
\sqrt{
\frac{a^2}{(w- b L^2_\text{I})^2}+L^2_\text{I} \left( \frac{(u^H_\text{I})^2}{4 L^2_\text{I} w + 4 (u^H_\text{I})^2 w^2} - \frac{(u^H_\text{II})^2 \nu^2}{4 L^2_\text{I} \nu^2 w + 4 (u^H_\text{II})^2 w^2} \right)
}\,.
\end{align}

The NEC \eqref{eq:necfiniteT} further constraints the parameters in a complicated way, e.g. when $u_\text{I}^H=u_\text{II}^H$ we have $a>0$. We further impose $b>0$, then the brane is of type E. 
After substituting \eqref{eq:eta1e} into \eqref{eq:dxdu} and performing the integration, we obtain the profile of the brane $Q$
\bea
x_\text{I}=-\frac{2au^H_\text{I}}{L_\text{I} \sqrt{1+b u^H_\text{I}}}
~\text{arctanh} \frac{u_\text{I} \sqrt{1+b u^H_\text{I}}}{u^H_\text{I}}\,,
\eea
where we have used $x_\text{I}(u_\text{I}=0)=0$. 
When $u_\text{I} \to \frac{u^H_\text{I}}{\sqrt{1+b u^H_\text{I}}}$, we have  $x_\text{I} \to -\infty$. 
This indicates that the left spacetime is of type E, i.e. there is no horizon in left spacetime.

An analytical solution for $x_\text{II}$ does not exist. Nevertheless, its derivative with respect to $u_\text{II}$ can be derived from \eqref{eq:dxdu}, 
from which 
we know that $x_\text{II}$ is a monotonic increasing or decreasing function of $u_\text{II}$. And there is a pole $u_\text{II}^p$ where $x_\text{II} \to \pm\infty$, as can be seen from 
\begin{align}
    \frac{dx_\text{II}}{du_\text{II}} \propto \pm \frac{1}{u_\text{II}-u_\text{II}^p}\,, \quad \text{with} \quad
    u_\text{II}^p = \frac{u^H_\text{II} \nu}{\sqrt{b(u^H_\text{II})^2+\nu^2} }\,.
\end{align}

\begin{figure}[h!]
\begin{center}
\includegraphics[width=0.4\textwidth]{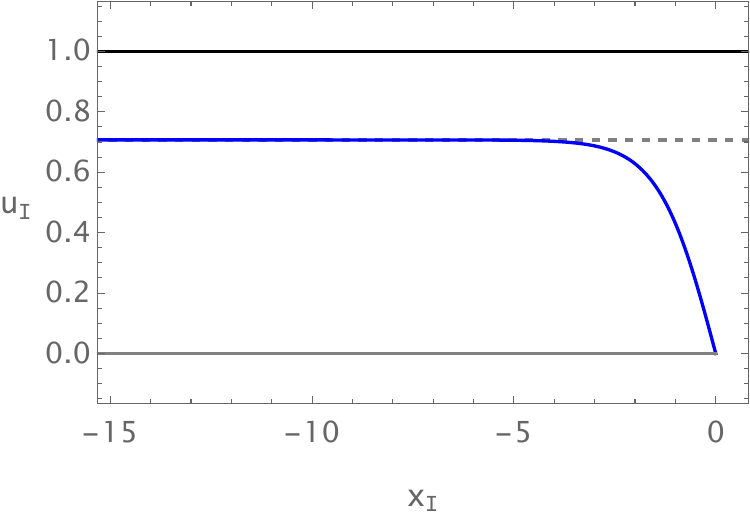}
~~
\includegraphics[width=0.41\textwidth]{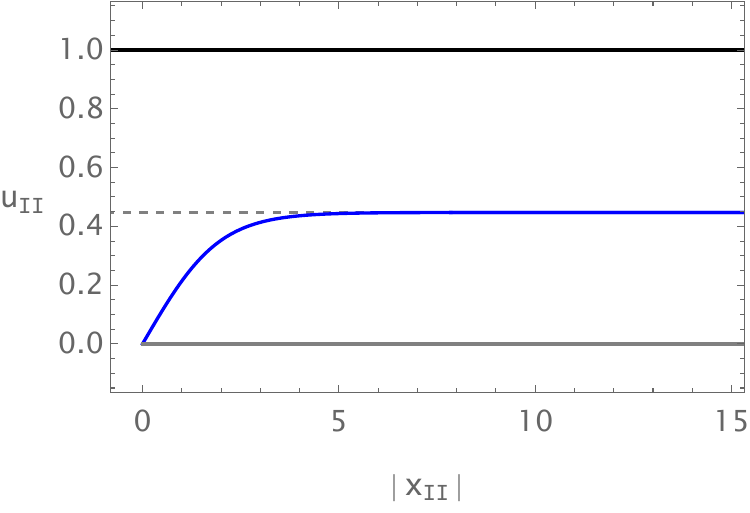}
\end{center}
\vspace{-0.3cm}
\caption{\small An example of the configurations [E,E] or [E, H1] at finite temperature with dynamical scalar field. Here we set $a= b=L_\text{I}=1, \nu=0.5, u_\text{I}^H=u_\text{II}^H=1$. 
}
\label{fig:u1u2-EEEH1}
\end{figure}

When we choose $\eta_\text{II}>0$ in \eqref{eq:fTsetaIIee}, the right part of the bulk contains a black hole. This is the configuration [E,H1].
The inertial observer in the left bulk will hit the brane and the inertial observer in the right bulk will hit the horizon. When we choose $\eta_\text{II}<0$ in \eqref{eq:fTsetaIIee}, the right part of the bulk does not contain a horizon of the black hole. Now the configuration is 
[E,E]. In Fig. \ref{fig:u1u2-EEEH1}, we show the typical profiles of [E,E] or [E,H1], depending on the sign we chose in \eqref{eq:fTsetaIIee}. These profiles are allowed only when we have a dynamical scalar field on the interface brane. 

\begin{figure}[h!]
\begin{center}
\includegraphics[width=0.375\textwidth]{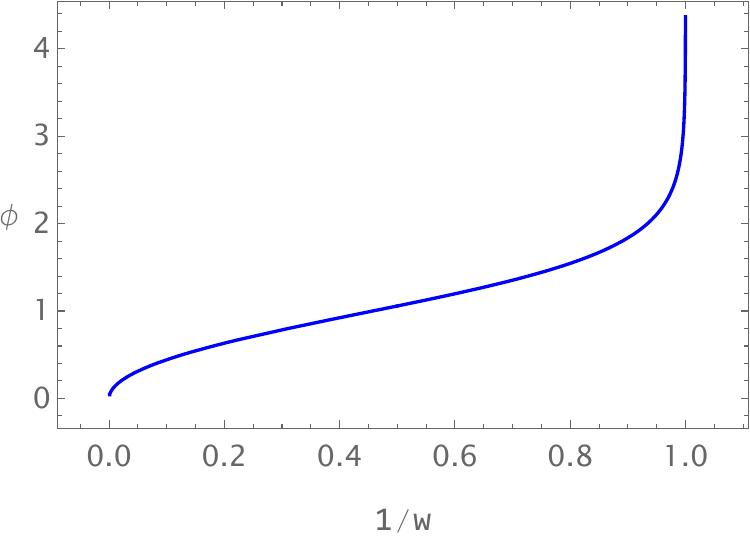}
~~
\includegraphics[width=0.4\textwidth]{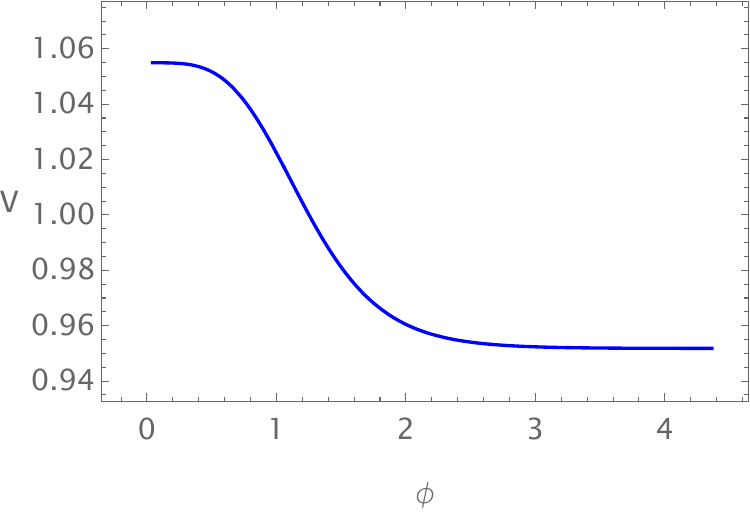}
\end{center}
\vspace{-0.3cm}
\caption{\small The scalar field and its potential for the configuration [E,H1] when $a= b=L_\text{I}=1, \nu=0.5, u_\text{I}^H=u_\text{II}^H=1$. A similar profile for the configuration [E,E] can be obtained and we do not show it here. 
}
\label{fig:sp-EEEH1}
\end{figure}

Similar to the previous discussions, we can show the profile of the scalar field and its potential for the configuration, e.g. [E,H1], as shown in Fig. \ref{fig:sp-EEEH1}. On the brane, the potential also has a local maximal in the UV limit and a local minimum in the IR limit. Moreover, the scalar field diverges in the deep IR. The profiles of the configuration [E, E] have similar behavior. 

The induced metric on the interface brane is
\begin{align}
ds^2_Q=-wdt^2+
\left(
\frac{a^2}{(w-bL^2_\text{I})}+
\frac{
L^2_\text{I}(u^H_\text{I})^2
}{
4L^2_\text{I}
w+
4(u^H_\text{I})^2w^2
}
\right)
dw^2\,.
\end{align}
In the IR limit $w\to bL^2_\text{I}$, the induced metric becomes
\begin{align}
ds^2_Q\simeq -bL^2_\text{I} dt^2+
\frac{a^2}{(w-bL^2_\text{I})}
dw^2\,,
\end{align}
which implies the brane is not a black hole.

Evaluating \eqref{eq:lamfin}, we obtain the asymptotic behaviours
\begin{align}
\lambda(w)\simeq 
-\frac{1}{4a^2+L^2_\text{I}}+\mathcal{O}\left(
\frac{1}{w^2}
\right)\,, ~~~~\quad w\to \infty\,,
\end{align}
i.e. the interface brane is asymptotically AdS in the UV limit, 
while 
\begin{align}
\lambda(w)\simeq 
-\frac{w-b L^2_\text{I}}{2 a^2 b L^2_\text{I}}+\mathcal{O}\left((w-b L^2_\text{I})^2
\right)\,,~~~ w\to b L^2_\text{I}\,,
\end{align}
i.e. the interface brane is asymptotically flat in IR limit.

\subsubsection{Solution with nontrivial scalar field: [H1,E]}

\begin{figure}[h!]
\begin{center}
\includegraphics[width=0.4\textwidth]{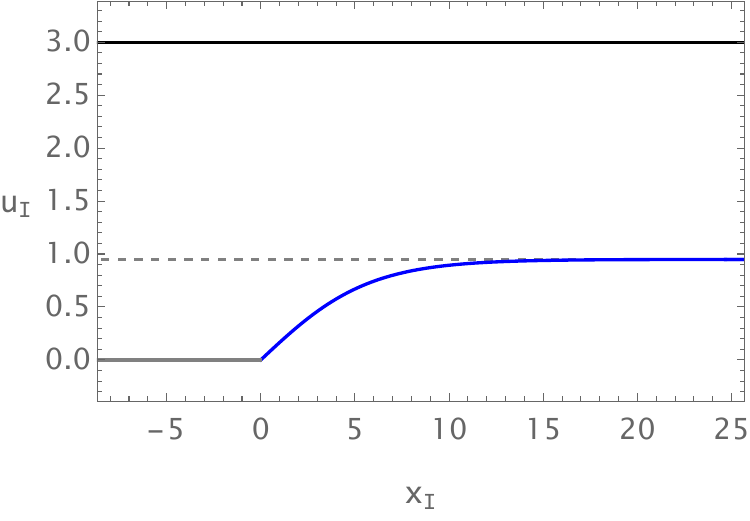}
~~
\includegraphics[width=0.415\textwidth]{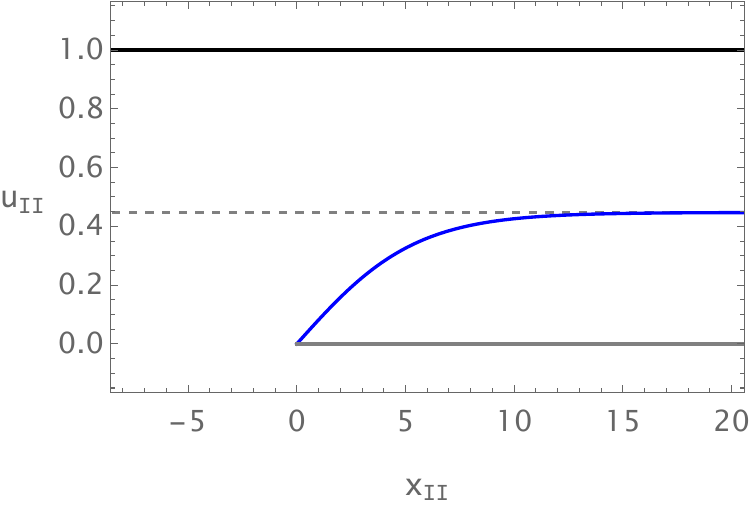}
\end{center}
\vspace{-0.3cm}
\caption{\small Plot of the configuration [H1, E] for $a=3,\, b=1,\, L_\text{I}=1,\, \nu=0.5,\, u_\text{I}^H=3$ and $u_\text{II}^H=1$. 
}
\label{fig:u1u2-H1E}
\end{figure}

In this subsection we show an example of the configuration [H1,E]. 
Motivated by the previous subsection, we consider the following configuration 
\begin{align}
\begin{split}
\label{eq:finiteTH1E}
\eta_\text{I}=&-\frac{a}{w- b L^2_\text{I}}\, ,\\
\eta_\text{II}=&-
\sqrt{
\frac{a^2}{(w- b L^2_\text{I})^2}+L^2_\text{I} \left( \frac{(u^H_\text{I})^2}{4 L^2_\text{I} w + 4 (u^H_\text{I})^2 w^2} - \frac{(u^H_\text{II})^2 \nu^2}{4 L^2_\text{I} \nu^2 w + 4 (u^H_\text{II})^2 w^2} \right)
}\,.
\end{split}
\end{align}
The NEC \eqref{eq:necfiniteT} constraints the choices of the parameters. 
We find that under the choice of parameters
$
a=3,\, b=1, \ L_\text{I}=1,\ 
u^H_\text{I}=3,\ 
u^H_\text{II}=1,\ 
\nu=0.5,
$ the NEC is satisfied. This case provides an example of configuration [H1,E]. The profiles of the configuration is shown in Fig. \ref{fig:u1u2-H1E}. 

\begin{figure}[h!]
\begin{center}
\includegraphics[width=0.375\textwidth]{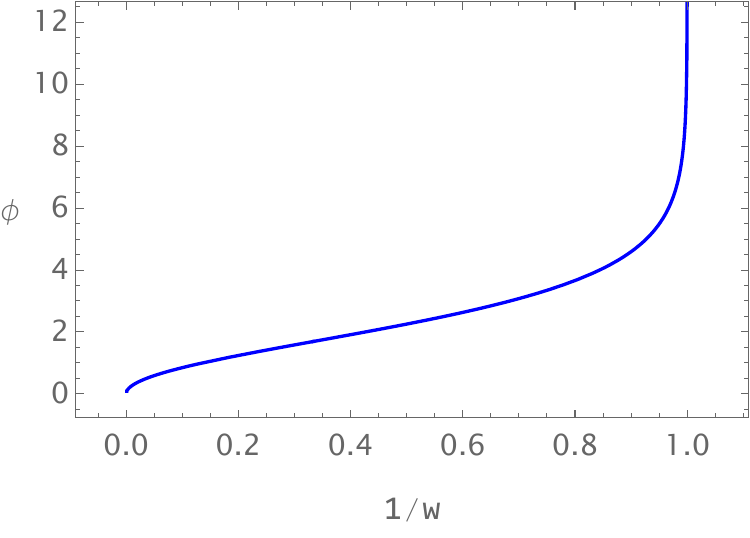}
~~
\includegraphics[width=0.415\textwidth]{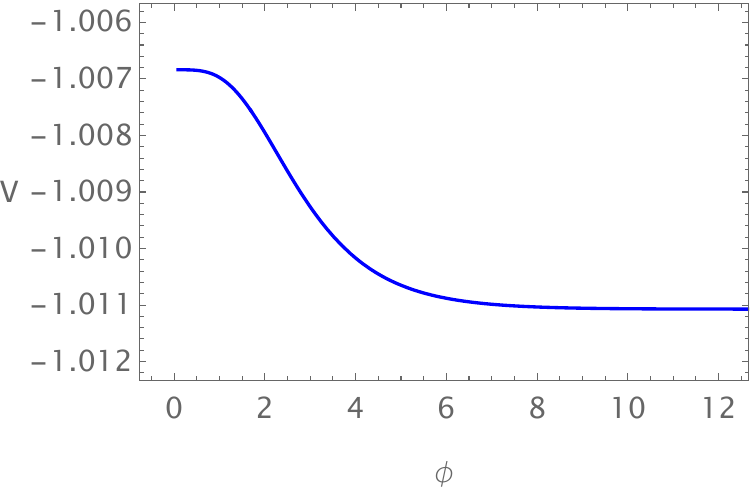}
\end{center}
\vspace{-0.3cm}
\caption{\small The scalar field and its potential for the configuration [H1, E] when $a=3,\, b=1,\, L_\text{I}=1,\, \nu=0.5,\, u_\text{I}^H=3$ and $u_\text{II}^H=1$.
}
\label{fig:sp-H1E}
\end{figure}

In Fig. \ref{fig:sp-H1E} we show the profiles of the scalar field and its potential. Similar to the previous examples, the potential evolves from a local  maximum in UV to a global minimum in IR. In the deep IR, the induced metric on $Q$ is asymptotically flat, which is again associated with a divergent scalar field in IR. 

\subsubsection{Solution with nontrivial scalar field: no [H1,H1]}
\label{sec:noh1h1}

We have demonstrated examples of the [E,E], [E,H1] and [H2,H2] configurations. The nonexistence of the [E,H2], [H1,H2], [H2,E] and [H2,H1] configurations has been mentioned below \eqref{eq:xIxII}. It can be seen as follows: Suppose there exists a solution for [E,H2] or [H1,H2]. For E or H1 configuration, there exists a lower bound at $w=w_0$ where the value of $x_\text{I}$ approach $\infty$ or $-\infty$. However, for the other side, from 
\eqref{eq:xIxII}, $u_\text{II}$ can not approach the horizon when $w\to w_0$. Therefore the configurations [E,H2] and [H1,H2] are not allowed. Similar arguments apply for the [H2,E] and [H2,H1] configurations. 
In the following, we will prove that the [H1,H1] configuration cannot exist neither from the NEC. 

If there exists an [H1,H1] solution, i.e. we can find a function $\eta_\text{I}(w)$. Let us  assume that $w_0>0$ is the lower bound of $w$ where the brane goes to $x_\text{I}\to \infty$. From \eqref{eq:dxdu}, we have 
$ x'_\text{I}(u_\text{I})\sim -\eta_\text{I}(w)\to \infty \, 
$ when $w\to w_{0+}$. We assume this solution satisfies the NEC \eqref{eq:nech1h1} which is only valid when we impose $u^H_\text{I}=u^H_\text{II}$.\footnote{For the configurations [H1,E] and [E, H1] we do not have such constraint and therefore the following argument does not apply for these two cases. } 
Then from the NEC, when $w$ decreases to $w_0$, we should have $\eta'_\text{I}\to -\infty$. 
This implies that when $w$ decreases to $w_0$, we have 
$\eta_\text{I}$ increases which is inconsistent with $\eta_\text{I}(w\to w_{0+})\to -\infty$. 
One possibility is that we have a turning point for the non-monotonic profile of the interface brane. However, in this case one could repeat the aforementioned reasoning near the turning point, where $u_\text{I}'(x_\text{I})=0$. By doing so, we arrive at the conclusion that it is indeed impossible to obtain the profile [H1, H1].

\subsubsection{BCFT limit}

Here we briefly discuss the BCFT limits of the system, following the same method for the zero temperature case.  The first BCFT limit is $\nu\to 0$, 
the equation of motion in \eqref{eq:q-finiteT} becomes
\begin{align}
\label{bcft:ft1}
\begin{split}
&\eta_\text{II}=\pm
\frac{1}{2}\sqrt{
\frac{
L_\text{I}(u^H_\text{I})^2
}{
w(L^2_\text{I}+(u^H_\text{I})^2w)
}
+4\eta^2_\text{I}
}+\mathcal{O}(\nu^2)  \,  ,\\
&\phi'^2(w)=-
\frac{
L_\text{I}\left(
(u^H_\text{I})^4
\eta_\text{I}
+(L^2_\text{I}+(u^H_\text{I})^2w)
((u^H_\text{I})^2\eta'_\text{I}-2\eta^3_\text{I}) \right)
}{
2u^H_\text{I}(L^2_\text{I}+(u^H_\text{I})^2w)
\sqrt{
L^2_\text{I}(u^H_\text{I})^2+4w (L^2_\text{I}+w (u^H_\text{I})^2)
\eta^2_\text{I}
}
}+\mathcal{O}(\nu)   \,   ,\\
&V(\phi)=\pm
\frac{1}{L_\text{I}\nu}  \\
&-\frac{2 (L^2_\text{I}+2w (u^H_\text{I})^2) \eta_\text{I} (L^2_\text{I} (u^H_\text{I})^2+2w (L^2_\text{I}+w (u^H_\text{I})^2) \eta^2_\text{I})
+2 L^2_\text{I} (u^H_\text{I})^2 w
(L^2_\text{I}+(u^H_\text{I})^2 w)\eta'_\text{I}
}{
L_\text{I}u^H_\text{I}
(
L^2_\text{I}(u^H_\text{I})^2+
4w(L^2_\text{I}+(u^H_\text{I})^2w)\eta^2_\text{I}
)^{\frac{3}{2}}
}\\
&~~~~~
+\mathcal{O}(\nu)  \,  .
\end{split}
\end{align}
Similar to the zero temperature case \eqref{eq:bcftlimit2}, there is a divergent term $\frac{1}{L_\text{I}\nu}$ in the expression of $V(\phi)$.
In above equations, the parameter $u^H_\text{II}$ appears in higher order terms. 

The second way to obtain a BCFT is to consder the limit
$\nu= 1$, $u^H_\text{I}=u^H_\text{II}$ and  $\eta_\text{II}=-\eta_\text{I}$ , where we can perform the folding trick. In this case,
the nontrivial equations of motion are
\begin{align}
\begin{split}
\label{bcft:ft2}
&\phi'^2(w)=-
\frac{
L_\text{I}\left(
(u^H_\text{I})^4
\eta_\text{I}
+(L^2_\text{I}+(u^H_\text{I})^2w)
((u^H_\text{I})^2\eta'_\text{I}-2\eta^3_\text{I}) \right)
}{
u^H_\text{I}(L^2_\text{I}+(u^H_\text{I})^2w)
\sqrt{
L^2_\text{I}(u^H_\text{I})^2+4w (L^2_\text{I}+w (u^H_\text{I})^2)
\eta^2_\text{I}
}
} \,  ,\\
&V(\phi)=\\
&
-\frac{4 (L^2_\text{I}+2w (u^H_\text{I})^2) \eta_\text{I} (L^2_\text{I} (u^H_\text{I})^2+2w (L^2_\text{I}+w (u^H_\text{I})^2) \eta^2_\text{I})
+4 L^2_\text{I} (u^H_\text{I})^2 w
(L^2_\text{I}+(u^H_\text{I})^2 w)\eta'_\text{I}
}{
L_\text{I}u^H_\text{I}
(
L^2_\text{I}(u^H_\text{I})^2+
4w(L^2_\text{I}+(u^H_\text{I})^2w)\eta^2_\text{I}
)^{\frac{3}{2}}
} \,   ,
\end{split}
\end{align}
where the quantities $\phi'^2(w),V(\phi)$ are twice of those in \eqref{bcft:ft1}.

\subsection{Gluing thermal AdS$_3$ and BTZ black hole}
\label{sec:gab}

In this subsection we will glue a thermal AdS$_3$ with a BTZ black hole along the interface brane. The permissible configuration in thermal AdS$_3$ spacetime is expected to be of type empty, denoted as  E$_\text{tAdS}$, while the allowed configurations for a BTZ black hole can be type E, H1 or H2 as illustrated in Fig. \ref{fig:Qconfig}. 
In the absence of a scalar field, we will show that finding a solution is impossible. However, with the inclusion of a brane-located scalar field,  we will demonstrate that the permissible configurations only include  [E$_\text{tAdS}$, E], 
and [E, E$_\text{tAdS}$]. 

The metric for thermal AdS$_3$ is \eqref{eq:poincare}, and the metric for BTZ is \eqref{eq:btzbh}. Note that one can set $u_A^H\to\infty$ with $A=\text{I}$ or $\text{II}$ in \eqref{eq:btzbh}
to make it a thermal AdS$_3$. As an example, we take the left portion of the bulk as thermal AdS$_3$ while the right portion of spacetime as BTZ black hole. The parameterization is 
\bea
\label{eq:para-gtb}
\begin{split}
&u_\text{I}(w)=\frac{L_\text{I}}{
\sqrt{w}}\,,~~~~~~
~~~~~~~~~~~~~~
u_\text{II}(w)=\frac{u_\text{II}^HL_\text{II}}{
\sqrt{L_\text{II}^2+(u_\text{II}^H)^2 w}}\,,~~~~
\\
& x_\text{I}(w)=
-\int^{+\infty}_w
\frac{
\eta_\text{I}(\sigma)}{
\sqrt{
\sigma}
}d\sigma\,,~~~~
x_\text{II}(w)=
-\int^{+\infty}_w
\frac{
u^H_\text{II}\eta_\text{II}(\sigma)}{
\sqrt{
L_\text{II}^2+(u^H_\text{II})^2\sigma}
}d\sigma\,.
\end{split}
\eea
It is straightforward to generalize following analysis to the case that the left portion is BTZ black hole while the right portion is a thermal AdS spacetime.  

Plugging the above parameterization into the equations in Sec. \ref{sec2} we can obtain the equations for this case. Another consistent way is to take the limit $u_\text{I}^H\to \infty$ to the equations in Sec. \ref{sec:g2b}. The independent equations of motion are 
\begin{align}
\begin{split}
\label{eq:q-finiteT-tb}
& 0=\frac{L^2_\text{I}}
{
4w^2
}+\eta^{2}_\text{I}  
- \frac{L^2_\text{II}(u^H_\text{II})^2}
{
4 L^2_\text{II} w+4(u^H_\text{II})^2w^2
}-\eta^{2}_\text{II}   \, ,\\
&\frac{V(\phi)}{2}= -
\frac{ 2w  \eta_\text{I} (L^2_\text{I} +2w^2 \eta^2_\text{I})
+ L^2_\text{I}  w^2 \eta'_\text{I}
}{
L_\text{I}
(L^2_\text{I}+4w^2\eta^2_\text{I})^{\frac{3}{2}}
}  \\
&~~+
\frac{
 (L^2_\text{II}+2w (u^H_\text{II})^2) \eta_\text{II} (L^2_\text{II} (u^H_\text{II})^2+2w (L^2_\text{II}+w (u^H_\text{II})^2) \eta^2_\text{II})
+ L^2_\text{II} (u^H_\text{II})^2 w
(L^2_\text{II}+(u^H_\text{II})^2 w)\eta'_\text{II}
}{
L_\text{II}u^H_\text{II}
(L^2_\text{II}(u^H_\text{II})^2+4w(L^2_\text{II}+w (u^H_\text{II})^2)\eta^2_\text{II})^{\frac{3}{2}}
}
\, ,\\
&\phi'^2(w)=
-\frac{
L^2_\text{I}+
4 w^2 \eta^2_\text{I}
}{
4w^3
}\,\bigg{[}
-wV(\phi)  \\
&~~-
\frac{
2w^2\eta_\text{I}
}{
L_\text{I}
\sqrt{
L^2_\text{I}+
4 w^2 \eta^2_\text{I}
}
}+
\frac{
2w(L^2_\text{II}+(u^H_\text{II})^2w)\eta_\text{II}
}{
L_\text{II}u^H_\text{II}
\sqrt{
L^2_\text{II}(u^H_\text{II})^2+
4(L^2_\text{II}+(u^H_\text{II})^2w) w \eta^2_\text{II}
}
}
\bigg{]}
\, .
\end{split}
\end{align}

The NEC becomes 
\begin{align}
\label{eq:necfiniteT-ads-btz}
\begin{split}
-\frac{
\,\eta_\text{I}
+w\eta'_\text{I}
}
{
\nu 
(L^2_\text{I}+4w^2\eta^2_\text{I})^{\frac{3}{2}}
}
+
\frac{
(u^H_\text{II})^4\,\eta_\text{II}
+(L^2_\text{I}\nu^2+(u^H_\text{II})^2w)
(-2\eta^3_\text{II}+(u^H_\text{II})^2\eta'_\text{II})
}
{
u^H_\text{II}
(L^2_\text{I}(u^H_\text{I})^2\nu^2+4w(L^2_\text{I}\nu^2+(u^H_\text{II})^2w)\eta^2_\text{II})^{\frac{3}{2}}
}\ge 0
\,,
\end{split}
\end{align}
which is again consistent with the condition $\phi'^2\geq0$.  

One observation from the continuous equation for the metric field is that the profile [E$_\text{tAdS}$, H2] does not exist. If such configuration exists, the parameter regimes for $w$ should be from zero to infinity. Then from the first equation in \eqref{eq:q-finiteT-tb}, we have $\eta_\text{II}\to \frac{1}{w}$ when $w\to 0$. However, from \eqref{eq:q-finiteT-tb} we know $x_\text{II}\to \infty$. Therefore it is not possible to have the profile [E$_\text{tAdS}$, H2]. Note that this observation does not depend on whether there is any scalar field located on the interface brane. 

When the scalar field is trivial, i.e. $\phi=0, V=T$, one can confirm that there is no solution for this case as follows. Solving \eqref{eq:q-finiteT-tb}, one obtains 
\be
\label{eq:etaI-gtb}
\eta_\text{I}=\pm\frac{L_\text{I}}{2w}\frac{L_\text{I}^2\nu^2+(u_\text{II}^H)^2  (1-(1+L_\text{I}^2 T^2) \nu^2)w}{\sqrt{\left(1-(1-L_\text{I} T)^2\nu^2\right)(u_\text{II}^H)^2\,w+L_\text{I}^2\nu^2}\sqrt{\left((1+L_\text{I} T)^2\nu^2-1\right)(u_\text{II}^H)^2\,w-L_\text{I}^2\nu^2}}\,.
\ee
Similarly one can obtain a solution of $\eta_\text{II}$ from the above expression. From \eqref{eq:etaI-gtb} one finds that the tension has to be in the regime \eqref{eq:tensionreg0T}. Furthermore, the parameter $w$ takes value from $w_0=L_\text{I}^2\nu^2/(\left((1+L_\text{I} T)^2\nu^2-1\right)(u_\text{II}^H)^2)$ to infinity which means it can not be the profile of H2 on the BTZ side. Close to $w_0$ we find $\eta_\text{I}\propto 1/\sqrt{w-w_0}$ which indicates $x_\text{II}$ is finite. This is inconsistent with the profile E and H1. Therefore we do not have any consistent solution for this case of trivial scalar field with constant tension on the interface brane.  

\begin{figure}[h!]
\begin{center}
\includegraphics[width=0.39\textwidth]{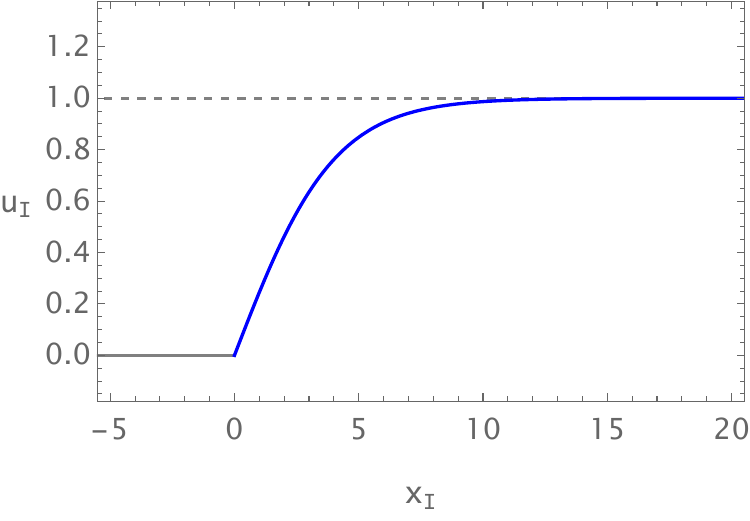}
~~
\includegraphics[width=0.402\textwidth]{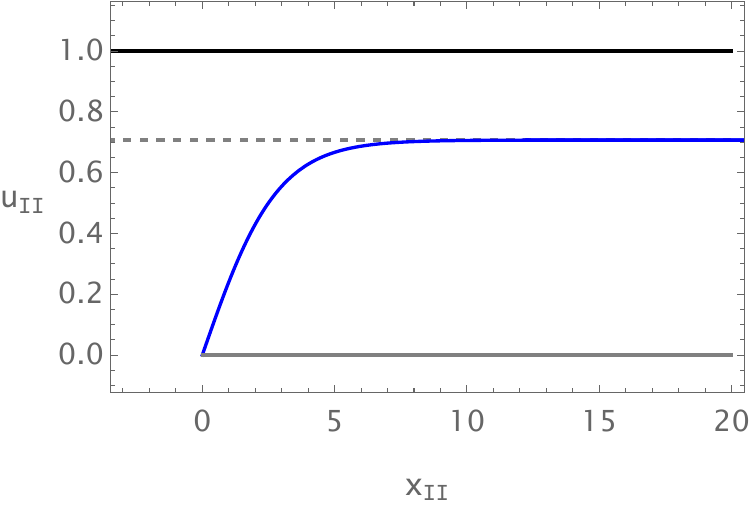}\\
\includegraphics[width=0.38\textwidth]{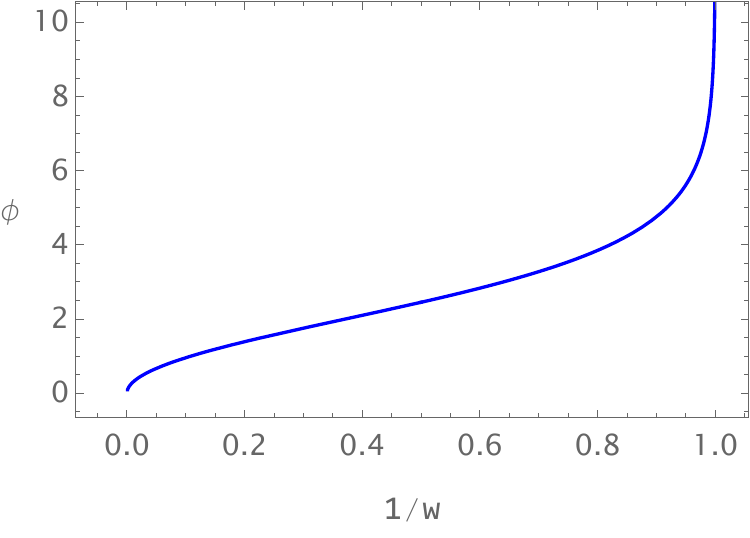}
~~
\includegraphics[width=0.41\textwidth]{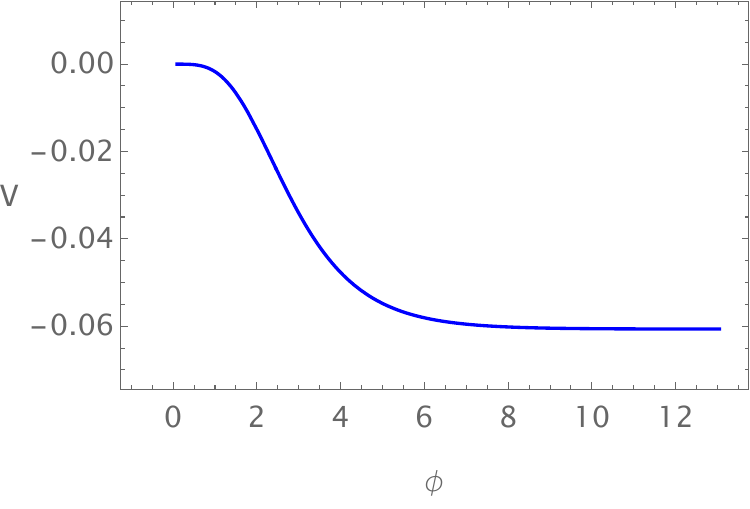}
\end{center}
\vspace{-0.3cm}
    \caption{\small Top: Plots of configurations $x_\text{I}$ and $x_\text{II}$. Bottom:  Plots of the scalar field $\phi$ as a function of $1/w$ and its potential as a function of $\phi$. The parameters are chosen to be $a=2, b= u^H_\text{II}= L_\text{I}=\nu=1$.
}
\label{fig:adsEE}
\end{figure}

We find \eqref{eq:finiteTH1E} with $\nu=1$ and $u_\text{I}^H\to\infty$ satisfies all the above equations, 
\begin{align}
\begin{split}
\label{eq:finiteTH1E-gtb}
\eta_\text{I}=&-\frac{a}{w- b L^2_\text{I}}\, ,\\
\eta_\text{II}=&-
\sqrt{
\frac{a^2}{(w- b L^2_\text{I})^2}+L^2_\text{I} \left( \frac{1}{4 w^2} - \frac{(u^H_\text{II})^2 \nu^2}{4 L^2_\text{I} \nu^2 w + 4 (u^H_\text{II})^2 w^2} \right)
}\,.
\end{split}
\end{align}
This can give the solution of [E$_\text{tAdS}$, E]. With the choice of $a=2, b= u^H_\text{II}= L_\text{I}=\nu=1$, we show the profiles as well as the scalar potential in Fig. \ref{fig:adsEE}. Here again the potential evolves from a locally maximal in UV to a global minimal in IR. Note that when $a<0$ in \eqref{eq:finiteTH1E-gtb}, e.g. 
the choice of 
$a=-2,b= u^H_\text{II}= L_\text{I}=\nu=1$, we also have consistent solutions. Now the brane in the left bulk curves towards $x_\text{I}\to -\infty$, while in the right bulk it curves towards $x_\text{II}\to +\infty$. The scalar field and its potential exhibit similar behaviors to the one shown in Fig. \ref{fig:adsEE}. We will not show the plots here. 

Suppose there exists a configuration [E$_\text{tAdS}$, H1], from \eqref{eq:irh2} we should have when $w\to w_0$
\be\label{eq:etaIIasy-gtb}
\eta_\text{II}\to \frac{B}{(w-w_0)^b}
\ee
with $b\geq 1$, $w_0>0$ and $B>0$. From the first equation in \eqref{eq:q-finiteT-tb}, we should have 
$
\eta_\text{I}\to \pm \frac{B}{(w-w_0)^b}
$
when $w\to w_0$. 
From the last equation in \eqref{eq:q-finiteT-tb}, we find for both plus and minus sectors in $\eta_\text{I}$,
\be
\phi'^2\to -\frac{ B^2 L_\text{I}\nu}{2 u_\text{II}^H\sqrt{L_\text{I}^2\nu^2w_0+(u_\text{II}^H)^2w_0^2}} \,(w-w_0)^{-2b}\,.
\ee
Therefore it is impossible to find a consistent solution of [E$_\text{tAdS}$, H1] as above behavior. Note that when $B<0$ in \eqref{eq:etaIIasy-gtb}, we have a consistent solution, i.e. $\phi'^2>0$ as expected.

When the left side is a BTZ black hole while the right side is thermal AdS$_3$, by exchange $\eta_\text{I}$ with $-\eta_\text{II}$ as well as other geometry quantities, we obtain the same results as above. Namely, only configuration of type [E, E$_\text{tAdS}$] is permissible, while all other configurations are not allowed.

Therefore, with a dynamical scalar field, solutions of type [E$_\text{tAdS}$, E], [E, E$_\text{tAdS}$] are permissible. However, the solutions [E$_\text{tAdS}$, H1], [H1, E$_\text{tAdS}$], [E$_\text{tAdS}$, H2] and [H2, E$_\text{tAdS}$] do not exist due to the NEC or the requirement of metric compatibility on the interface brane.

Finally, let us briefly discuss the above configuration. This category of intriguing configurations includes  [E$_{\text{tAdS}}$,E], [E,E$_{\text{tAdS}}$], [E$_{\text{AdS}}$,E$_{\text{AdS}}$] and [E,E]. The latter two type of solutions were previously discussed in Sec. \ref{subsec:solVI}, 
and Sec. \ref{subsec:EE-gbb}. These configurations share the common feature of covering only an ``empty" portion of the spacetime, devoid of horizons. 
The metrics of E$_{\text{AdS}}$ and E$_{\text{tAdS}}$ are the same in real time but differ solely in the periodicity of Euclidean time. These two solutions are quite different from the type E solution of the BTZ black hole, as no regular coordinate transformation can make them the same. Additionally, the properties of the dual field theory for the bulk AdS and type E solution of BTZ black hole differ, exhibiting different vacuum expectation value (VEV) of energy densities  \cite{Kraus:2006wn}.\footnote{It is crucial to keep in mind that the vacuum expectation value of the operators in the dual field theory might be influenced by the special dynamics occurring on the brane. For instance, a discussion on the VEV of scalar operator in AdS/BCFT can be found in \cite{Fujita:2011fp}. It would indeed be intriguing to explicitly calculate the VEV of the energy momentum tensor in these various scenarios.} Note that both the left and right portions of the bulk do not involve horizons, allowing for arbitrary periodicity of Euclidean time, independent of the Hawking temperature in black hole case. While a conifold singularity may exist for type E in the Euclidean BTZ black hole, it is hidden behind the interface brane. Considering both the similarities and differences among these configurations, it would be interesting to understand the detailed emergence of specific bulk geometries for a given ICFT.

\section{Conclusions and open questions}
\label{sec:cd}

We have constructed a holographic model of two CFTs that could have different central charges, contacting at an interface with an interface field theory away from a fixed point. In this AdS$_3$/ICFT$_2$ system, we introduced a dynamical scalar field residing on the interface brane. 
This bulk scalar field serves as a distinguishing feature, characterizing nontrivial intrinsic dynamics within the interface field theory.  

At zero temperature, we show the typical profiles of the system,  highlighting scenarios where the interface field theory deviates from the fixed point. Without the scalar field, the interface brane is limited to a linear configuration. However, when a nontrivial scalar field is present on the brane, a variety of monotonic configurations become possible, 
as illustrated in Fig. \ref{fig:brane pair}, while considering the null energy condition for the scalar field on the brane. Then we define an interface entropy from the holographic entanglement entropy, to characterize the effective number of degrees of freedom for the interface field theory.
Through the examination of several illustrative examples, we consistently observe that the $g$-theorem is upheld whenever the null energy condition is satisfied. Our study extends the existing $g$-theorem in BCFT to ICFT with two different central charges. We also studied the BCFT limits of the system, including the properties of $g$-functions, which are consistent with the BCFT results. Through concrete examples, summarized at the beginning of Sec. \ref{sec:heeexample}, we identified  several relationships between the profile of the brane and the RG flow properties of the interface field theory. We established concrete connections between aspects of interface entropy and the scalar potential on the interface brane. Among all the concrete examples, one particularly interesting solution shows that the brane can terminate at a finite value of $z$ along the radial direction, as discussed in Sec. \ref{subsec:solVI}. A similar configuration was obtained at finite temperature and this kind of configuration is only possible due to the presence of a scalar field on the interface brane. Additionally, we noted  that when the scalar potential is non-monotonic, the $g$-function for the interface field theory becomes non-smooth, indicating a ``second order phase transition" in the degrees of freedom of the interface field theory along the RG flow. 

We have also studied the configurations at finite temperature. 
We first glue together two BTZ black holes along an interface brane. In the absence of a scalar field the only permissible profiles are two segments of BTZ black hole spacetime with the interface brane intersecting the horizon. However, with the inclusion of a brane-localized scalar field, there are more allowed profiles, as summarized in Table  \ref{tab:allowedprofile}. Moreover, we also glue together a thermal AdS$_3$ spacetime with a BTZ black hole, further exploring the rich configurations enabled by the brane-localized scalar field. This reveals the interesting properties of the interface field theory away from the fixed point at finite temperature.

There are several potential extensions of our work that deserve further exploration. While we have primarily focused on static systems, where a global time exists and the interface brane remains static, it would be intriguing to generalize our analysis to the dynamical interface branes. Such a generalization would enable us to investigate the diverse configurations and  entanglement structures that may arise, providing deeper insights into the behavior of the $g$-theorem when the interface field theory deviates from the fixed point. 

Quantum complexity is an important quantum information quantity 
 which plays a crucial role in holography \cite{Susskind:2018pmk}.  Holographic complexity was studied in the context of AdS/ICFT 
\cite{Chapman:2018bqj, Braccia:2019xxi,Sato:2019kik,Auzzi:2021nrj, Auzzi:2021ozb}.  
 It would be interesting to study the properties of quantum complexity for the interface field theory in the setup considered in this paper.

It is also interesting to generalize our study to scenarios involving  multiple interfaces in ICFTs.  The emergence of diverse interface brane topologies and the exploration of phase transitions among them offer rich physics for exploration. In the case of AdS/ICFT with CFTs living in compact spaces, it was shown that there are also multiple phases \cite{Bachas:2021fqo}, with the phase transition identified as manifestations of the ER=EPR hypothesis \cite{Maldacena:2013xja, ruan}. However, within our current setup, the absence of additional dimensional quantities beyond temperature precludes such phase transitions. Nevertheless, in systems involving CFTs living in non-compact spaces, the presence of multiple defects introduces extra dimensional quantities  associated with the finite interval between the defects. Exploring phase transitions in such settings would provide valuable insights into the holographic dual of an ICFT. 

We introduced a dynamical brane-localized scalar field to break the scaling symmetry of the interface field theory. It would be worthwhile to explore the impact of incorporating other matter fields or higher derivative gravitational effects on the interface brane. Subsequently, studying the coefficients of energy transport and entanglement structure within these models would be intriguing. 
It would be interesting to calculate the transmission/reflection coefficient of energy flux  following both the linear response method   \cite{Bachas:2020yxv, Bachas:2022etu} and the heat flow approach in non-equilibrium-steady-states (NESS) configurations \cite{Bachas:2021tnp}. 
In particular, the breaking of scaling symmetries due to the presence of the scalar on the brane could 
lead to the violation of the universality of these energy transport coefficients, as proposed in \cite{Meineri:2019ycm, Bachas:2020yxv, Bachas:2022etu}.

Another natural extension of our work would be to explore higher-dimensional AdS/ICFT scenarios. Additionally, investigating the stability of the system through the analysis of its perturbations presents an intriguing direction. Lastly, delving deeper into the understanding of AdS/ICFT from the perspective of bulk reconstruction  \cite{Leutheusser:2022bgi} would be a highly intriguing avenue for future research.

\vspace{.3cm}
\subsection*{Acknowledgments}
 We thank 
 Shan-Ming Ruan and Ya-Wen Sun for useful discussions. This work is supported by the National Natural Science Foundation of China grant No. 11875083,
12375041. 

\vspace{.6 cm}
\appendix
\section{Configurations of non-monotonic profile at zero temperature}
\label{app:a}
In this appendix we show an example of non-monotonic profile for the interface brane. The interface brane profile in $N_\text{I}$ is chosen as 
\begin{align}
\psi_\text{I}(z)=&-az(z-b)\, . \label{eq:nm1}
\end{align} 
The NEC constrains $a\ge 0$ and
we will focus on the case $a>0,b>0$ where $\psi_\text{I}(z)$ is non-monotonic clearly.  
This brane is asymptotically AdS in UV. 
From \eqref{eq:zrelation}, the solution of $\psi_\text{II}(z)$ is 
\bea
\begin{split}
&\psi_\text{II}(z)=\pm
\frac{
(2z-b)}{4}
\sqrt{\frac{
 1-\nu^2+a^2(2z-b)^2\nu
}
{
\nu
}}\\
&~~~~~~~~~~~\mp\frac{1-\nu^2}{2a\nu}
\tanh^{-1}
\frac{
2az\sqrt{\nu}
}{
\sqrt{1-\nu^2+a^2b^2\nu}
-\sqrt{1-\nu^2+a^2(2z-b)^2\nu}
}   \\
&~~~~~~~~~\mp
\frac{b}{4}
\sqrt{\frac{
 1+a^2b^2\nu-\nu^2
}
{
\nu
}}
\mp \frac{\nu^2-1}{2a\nu}
\tanh^{-1}
\frac{
\sqrt{a^2b^2\nu+1-\nu^2}
}{
ab\sqrt{\nu}
}\,.
\end{split}  \label{eq:nm2}
\eea
It is monotonic in $N_\text{II}$.

\begin{figure}[h!]
\begin{center}
\includegraphics[width=0.4\textwidth]{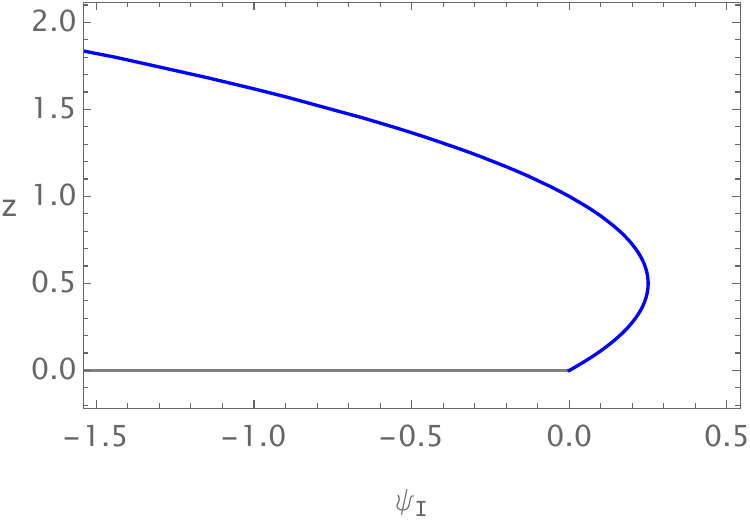}
~~
\includegraphics[width=0.4\textwidth]{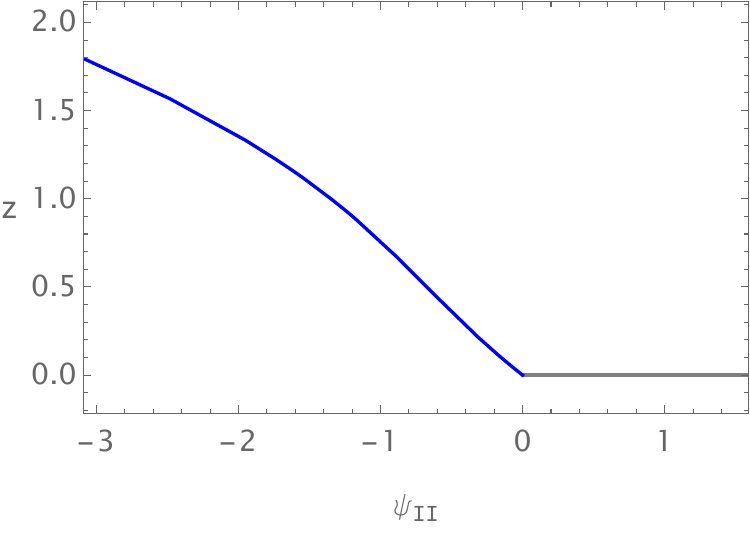}
\end{center}
\vspace{-0.3cm}
\caption{\small Typical configurations of $\psi_\text{I}$ and $\psi_\text{II}$ when $a=b=L_\text{I}=1,\nu=0.5$. 
We have chosen the minus sector for $\psi_\text{II}$ in \eqref{eq:nm2}. 
}
\label{fig:nm br}
\end{figure}

An example of the configuration \eqref{eq:nm1} and \eqref{eq:nm2} is shown in Fig. \ref{fig:nm br} when $a=b=L_\text{I}=1, \nu=0.5$.

In the UV limit $z\to 0$, the scalar and its potential have following asymptotic behaviour 
\begin{align}
\begin{split}
|\phi(z)|=&~ 
2\left[
aL_\text{I}
\sqrt{
\frac{\nu}{1+a^2(b-2z)^2\nu}
}
\left(
1\mp
\frac{
a(b-2z)\nu^{\frac{3}{2}}
}{
\sqrt{{1+a^2(b-2z)^2\nu-\nu^2}}
}
\right)
\right]^{\frac{1}{2}}
\sqrt{z}+...
\,  ,\\
V(\phi)=&~-
\frac{
\sqrt{\nu}\left(
-ab\pm (a^2b^2+\frac{1}{\nu}-\nu)^{\frac{3}{2}}
-a^3b^3\nu+
\sqrt{
\nu (1-\nu+a^2b^2\nu)
}
\right)
}{
L_\text{I}(1+a^2b^2\nu)^{\frac{3}{2}}
}
+m^2\phi^2+...
\,  ,
 \end{split}
\end{align}
and
\begin{align}
m^2=
\frac{
\pm ab\nu^2-
\sqrt{
\nu (1-\nu+a^2b^2\nu)
}
}{
4L^2_\text{I} (1+a^2b^2\nu)
(\sqrt{
\nu (1-\nu+a^2b^2\nu)
}+ab\nu^2 )
}
 \,  .
\end{align}
In the IR limit $z\to +\infty$, the scalar and its potential behave as 
\begin{align}
\begin{split}
|\phi(z)|=&~
\sqrt{
\frac{L_\text{I}(1\pm\nu)}{2}
}\log z+...\, , \\
V(\phi)=&~
\frac{\mp 1-\nu}{L_\text{I}\nu}
+\frac{1\mp \nu}{4a^2L_\text{I}\nu}\frac{1}{z^2}
+\mathcal{O}\left(\frac{1}{z^3}\right)  \,  .
\end{split}
\end{align}

\begin{figure}[h!]
\begin{center}
\includegraphics[width=0.4\textwidth]{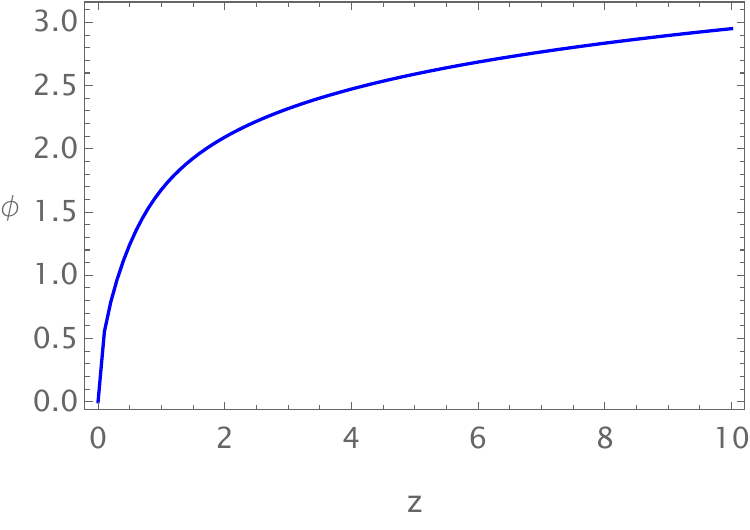}
~~
\includegraphics[width=0.4\textwidth]{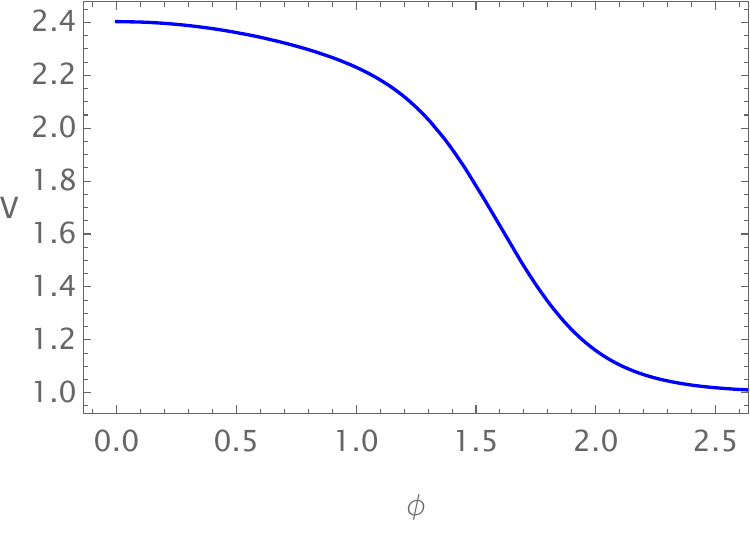}
\end{center}
\vspace{-0.3cm}
\caption{\small Typical configurations of the scalar field and its potential when $a=b=L_\text{I}=1,\nu=0.5$. In the limit $z\to \infty$, $\phi(z)$ diverges and $V(\phi)$ approaches a constant value $V_0=1$. 
}
\label{fig:nm scalar}
\end{figure}

In Fig. \ref{fig:nm scalar}, we show an example of the profiles of the scalar field and its potential where we choose $\psi'_\text{II}\le 0$. The scalar field goes to infinity in the deep IR resulting in a flat induced metric. Along the RG flow, the monotonic potential is a  maximum in UV and minimal in IR. 

\begin{figure}[h!]
\begin{center}
\includegraphics[width=0.4\textwidth]{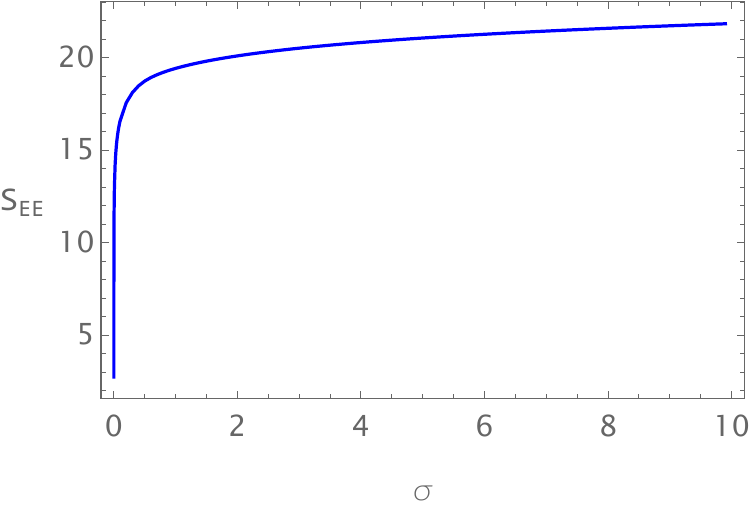}
~~
\includegraphics[width=0.4\textwidth]{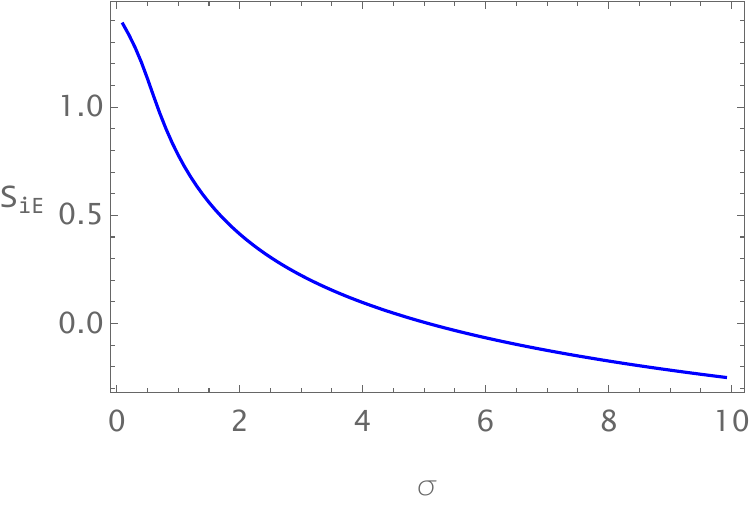}
\end{center}
\vspace{-0.3cm}
\caption{\small Typical configurations of the scalar field and its potential when $a=b=L_\text{I}=1,\nu=0.5$. 
}
\label{fig:nm ee iee}
\end{figure}

In Fig. \ref{fig:nm ee iee}, we show an example of the entanglement entropy and the interface entropy, where we choose $\psi'_\text{II}\le 0$. We see that the interface entropy goes to $-\infty$ in the deep IR and the $g$-theorem holds. 

Evaluating \eqref{eq:lam0} we obtain 
\begin{align}
    \lambda(z)=&~-\frac{1}{L^2_\text{I}(\nu^2+a^2b^2)} \,, \qquad (z\to 0)    \,,\\
\lambda(z)=&~0\,,~~~~~~~~~~~~~~~~~~~\qquad (z\to\infty)\,.
\end{align}
This implies that the brane asymptotically approaches AdS$_2$ in the UV regime. However, the condition $\lambda(z)\to 0$ as $z\to\infty$ indicates that it becomes flat in the deep IR.

\section{UV and IR limits of the scalar field and its potential}
\label{app:b}

In this appendix we list the UV and IR limits of the scalar field and its potential for the solutions in Sec. \ref{sec:heeexample}.  For the solution I in Sec. \ref{subsub:solI}, the scalar field vanishes and its potential is a constant. For the solution V in Sec. \ref{subsub:sol5}, we do not have an analytical solution of the expansions. Here we list the expressions for other solutions discussed in Sec. \ref{sec:heeexample}. 

\subsection{For solution II}

For the solution discussed in Sec. \ref{subsub:sol2}, we show the analytical expansions of the scalar field and its potential in the UV and IR limits. 

In the UV limit $z\to 0$, the scalar field and its potential exhibit the following asymptotic behavior 
\begin{align}
\begin{split}
\phi^2(z)=&~
\frac{
4(a-b)L_\text{I}\sqrt{\nu}(\sqrt{1+a^2\nu-\nu^2}\mp a\nu^{3/2})
}{
\sqrt{1+a^2\nu}
\sqrt{1+a^2\nu-\nu^2}
}z+\cdots\,,\\
V(\phi)=&~
\frac{a \nu^{3/2}\mp\sqrt{1+a^2\nu-\nu^2}}{L_{\text{I}}\nu \sqrt{1+a^2\nu} } -\frac{1}{4L_{\text{I}}^2(1+a^2\nu^2)}
\phi^2+\cdots\,.
\end{split}
\end{align}
In the IR limit as $z \to \infty$, the scalar field and its potential demonstrate the following asymptotic behavior
\begin{align}
\begin{split}
\label{eq:solIIphi}
(\phi(z)-\phi_0)^2=&~
\frac{
(a-b)L_\text{I} \sqrt{\nu}
(\sqrt{1+b^2\nu-\nu^2}\mp b\nu^{3/2})
}{
\sqrt{1+b^2\nu}
\sqrt{1+b^2\nu-\nu^2}
}\frac{1}{z^2}+\cdots\,,\\
V(\phi)=&~
\frac{b\nu^{3/2}\mp
\sqrt{1+b^2\nu-\nu^2}
}{
L_\text{I}\nu
\sqrt{1+b^2\nu}
}+\frac{2}{
L_\text{I}^2(1+b^2\nu)
}(\phi-\phi_0)^2+\cdots\,.
\end{split}
\end{align}
where $\phi_0$ is a constant of integration.
From the above expressions, the potential $V$ exhibits a maximum at $\phi=0$ and a minimum at $\phi=\phi_0$. These points respectively correspond to the UV and IR fixed points. For the plot shown in Fig. \ref{fig:case2-2}, the integration constant $\phi_0$ in \eqref{eq:solIIphi} is $1.62$.

\subsection{For solution III}

For the solution discussed in Sec. \ref{subsub:sol3}, in the UV limit $z\to 0$, the scalar field and its potential behave as follows, 
\begin{align}
\begin{split}
\phi^2(z)=&~ \frac{ab^3L_{\text{I}}\sqrt{\nu} (\sqrt{1+((ab+c)^2-\nu)\nu} \mp (ab+c)\nu^{3/2})}{\sqrt{1+(ab+c)^2\nu} \sqrt{1+((ab+c)^2-\nu)\nu}}  z^2+\dots\,,\\
V(\phi)=&~ \frac{(ab+c)\nu^{3/2} \mp \sqrt{1+((ab+c)^2-\nu)\nu}}{L_{\text{I}}\nu \sqrt{1+(ab+c)^2\nu}} +C_1\cdot\phi^4+\dots\,,
\end{split}
\end{align}
where $C_1$ are complicated functions depending on $a, b, c$ and $\nu$. It is interesting to see that the effective mass of the scalar field vanishes in the UV limit which means that the scalar field is sourceless on the brane. 
In the IR limit $z\to\infty$, the scalar field and its potential have following asymptotic behaviour,
\begin{align}
\begin{split}
(\phi(z)-\phi_0)^2=&~\frac{
aL_\text{I}\sqrt{\nu} (\sqrt{1+c^2\nu-\nu^2} \mp c\nu^{3/2})
}{
b\sqrt{1+c^2\nu} \sqrt{1+c^2\nu-\nu^2}
}\frac{1}{z^2}+\dots\,,\\
V(\phi)=&~\frac{c\nu^{3/2} \mp
\sqrt{1+c^2\nu-\nu^2}
}{
L_\text{I}\nu\sqrt{1+c^2\nu}
}+\frac{2}{L_\text{I}^2(1+c^2\nu)}(\phi-\phi_0)^2+\dots\,.
\end{split}
\end{align}
For the plot shown in Fig. \ref{fig:Stan},  we have $\phi_0\simeq 1.28$. 

\subsection{For solution IV}

For the solution discussed in Sec. \ref{subsub:sol4}, the behaviors of the scalar field and its potential at the UV limit ($z\to 0$) are 
\begin{align}
\begin{split}
\label{eq:scalarinsol4}
    \phi^2&= -\frac{2n}{n-1} L_{\text{I}} \sqrt{\nu} \gamma\  z^{n-1}+\cdots\,,\\
    V&= \mp\frac{\sqrt{1-\nu^2}}{\nu L_{\text{I}}} + \frac{(n-1)(n-3)}{4 L^2_{\text{I}}}\ \phi^2 + \cdots\,,
\end{split}
\end{align}
where we have assumed $\phi(0)=0$, i.e. the source of the scalar field on $Q$ in UV is set to be zero. From \eqref{eq:scalarinsol4}, we observe that $V'(0)=0$ and $V''(0)>0$ 
and $<0$ when $n<3$ 
and $n>3$ respectively. Here, the prime $'$ denotes the derivative with respect to $\phi$. This suggests that $\phi=0$ (i.e. UV) corresponds to a local maximum, 
and a local minimum for $n<3$, 
and $n>3$, respectively.

The behaviors of the scalar field and its potential at the IR limit ($z\to \infty$) are
\begin{align}
\label{eq:appB-solIV}
\begin{split}
    |\phi|&=  \sqrt{\frac{(n-1)L_{\text{I}}(1\pm\nu)}{2}} \log(z)+ \cdots\,,\\
    V&= \frac{\mp 1-\nu}{\nu L_{\text{I}}}+ \cdots\,.
    \end{split}
\end{align}
In contrast to the previous examples, the scalar field $\phi$ undergoes a  monotonic increase until infinity
while the potential $V$ converges towards a constant.

\subsection{For solution VI}

For the solution discussed in Sec. \ref{subsec:solVI}, in the UV limit $z\to 0$, the scalar and its potential exhibit the following asymptotic behavior
\begin{align}
\begin{split}
    \phi^2 &= \frac{4a L_\text{I} \sqrt{\nu} (\sqrt{b^4+a^2\nu-b^4\nu^2}\pm a \nu^{3/2})}{b\sqrt{(b^4+a^2\nu)(b^4+a^2\nu-b^4\nu^2)}}\ z+\cdots\,,\\
    V &= \frac{-a\nu^{3/2}\mp\sqrt{b^4+a^2\nu-b^4\nu^2}}{L_\text{I} \nu \sqrt{b^4+a^2\nu}} - \frac{b^4}{4 L^2_\text{I}(b^4+a^2\nu)} \phi^2+\cdots\,,
\end{split}
\end{align}
where we have assumed that $\phi(z=0)=0$. Near the $z=b$, the behaviors become
\begin{align}
\begin{split}
    (\phi-\phi_0)^2 &= \frac{4L_\text{I} (1\pm\nu)}{b} (b-z)+\cdots\,,\\
    V&= \frac{\mp 1-\nu}{L_\text{I}\nu} + \frac{b^4}{64a^2 L^4_\text{I} \nu (\pm1+\nu)^2}(\phi-\phi_0)^6 +\cdots\,,
    \end{split}
\end{align}
where $\phi_0$ is the constant of integration and denotes the position at which the potential $V$ attains its minimum value.


\end{document}